\pdfoutput=1

\documentclass[11pt,twoside,a4paper,cmspaper,final,collab]{cms-tdr}
\def\svnVersion{257071}\def\cmsCernNo{2013-037}\def\cmsCernDate{\today}\def\cmsMessage{Published in the Journal of Instrumentation as \href{http://dx.doi.org/10.1088/1748-0221/9/06/P06009}{\doi{10.1088/1748-0221/9/06/P06009.}}}

\begin{document}\cmsNoteHeader{TRK-11-002}

\hyphenation{had-ron-i-za-tion}
\hyphenation{cal-or-i-me-ter}
\hyphenation{de-vices}

\RCS$Revision: 230857 $
\RCS$HeadURL: svn+ssh://alverson@svn.cern.ch/reps/tdr2/papers/TRK-11-002/trunk/TRK-11-002.tex $
\RCS$Id: TRK-11-002.tex 230857 2014-03-10 15:51:29Z alverson $
\cmsNoteHeader{TRK-11-002} 
\title{Alignment of the CMS tracker with LHC and cosmic ray data}

\date{\today}

\abstract{
The central component of the CMS detector is the largest silicon
tracker ever built. The precise alignment of this complex device is a
formidable challenge, and only achievable with a significant extension
of the technologies routinely used for tracking detectors in the
past. This article describes the full-scale alignment procedure as it
is used during LHC operations. Among the specific features of the
method are the simultaneous determination of up to 200\,000 alignment
parameters with tracks, the measurement of individual sensor curvature
parameters, the control of systematic misalignment effects, and the
implementation of the whole procedure in a multi-processor environment
for high execution speed. Overall, the achieved statistical accuracy
on the module alignment is found to be significantly better than 10\mum.}

\hypersetup{%
pdfauthor={CMS Collaboration},%
pdftitle={Alignment of the CMS tracker with LHC and cosmic ray data},%
pdfsubject={CMS},%
pdfkeywords={CMS, physics, software, computing, tracking, alignment, Millepede}}

\maketitle 

\newcommand{\mptwo}{{\scshape Millepede~II}\xspace}
\newcommand{\minres}{{\scshape Minres}\xspace}
\newcommand{\chisq}{$\chi^{2}$\xspace}
\newcommand{\murad}{\ensuremath{\,\mu\text{rad}}\xspace}
\newcommand{\mrad} {mrad\xspace} 
\newcommand{\rphi}{$r\varphi$\xspace}

\tableofcontents 
\section{Introduction} \label{Sec:Intro}

The scientific programme of the Compact Muon Solenoid (CMS) experiment
\cite{CMS:2008zzk} at the Large Hadron Collider (LHC) \cite{LHC:2008}
covers a very broad spectrum of physics and focuses on the search for
new phenomena in the TeV range. Excellent tracking performance is
crucial for reaching these goals, which requires high precision of the
calibration and alignment of the tracking devices.  The task of the
CMS tracker \cite{Trk:TDR, Trk:TDRadd1} is to measure the trajectories
of charged particles (tracks) with very high momentum, angle, and
position resolutions, in combination with high reconstruction
efficiency \cite{CMS:TDRvol1}. According to design specifications, the
tracking should reach a resolution on the transverse momentum, \PT, of
1.5\% (10\%) for 100\GeVc (1000\GeVc) momentum
muons~\cite{CMS:TDRvol1}. This is made possible by the precise
single-hit resolution of the silicon detectors of the tracker and the
high-intensity magnetic field provided by the CMS solenoid.

The complete set of parameters describing the geometrical properties
of the modules composing the tracker is called the \textit{tracker
geometry} and is one of the most important inputs used in the
reconstruction of tracks. Misalignment of the tracker geometry is a
potentially limiting factor for its performance. The large number of
individual tracker modules and their arrangement over a large volume
with some sensors as far as $\approx$6\unit{m} apart takes the alignment
challenge to a new stage compared to earlier experiments. Because of
the limited accessibility of the tracker inside CMS and the high level
of precision required, the alignment technique is based on the tracks
reconstructed by the tracker in situ.  The statistical accuracy of the
alignment should remain significantly below the typical intrinsic
silicon hit resolution of between 10 and 30\mum
\cite{Trk:StripRecoCraftRef, Trk:PixelRecoCraftRef}. Another important
aspect is the efficient control of systematic biases in the alignment
of the tracking modules, which might degrade the physics performance
of the experiment. Systematic distortions of the tracker geometry
could potentially be introduced by biases in the hit and track
reconstruction, by inaccurate treatment of material effects or
imprecise estimation of the magnetic field, or by the lack of
sensitivity of the alignment procedure itself to such degrees of
freedom. Large samples of events with different track topologies are
needed to identify and suppress such distortions, representing a
particularly challenging aspect of the alignment effort.

The control of both the statistical and systematic uncertainties on
the tracker geometry is crucial for reaching the design physics
performance of CMS. As an example, the b-tagging performance can
significantly deteriorate in the presence of large misalignments
\cite{CMS:TDRvol1,CMS:BTV-12-001}. Electroweak precision measurements
are also sensitive to systematic misalignments. Likewise, the
imperfect knowledge of the tracker geometry is one of the dominant
sources of systematic uncertainty in the measurement of the weak
mixing angle,
$\sin^{2}(\theta_{\text{eff}})$~\cite{CMS:EWK11003}.

The methodology of the tracker alignment in CMS builds on past
experience, which was instrumental for the fast start-up of the
tracking at the beginning of LHC operations. Following simulation
studies~\cite{ALI:Stoye_MC_2008}, the alignment at the tracker
integration facility~\cite{ALI:TIF} demonstrated the readiness of the
alignment framework prior to the installation of the tracker in CMS by
aligning a setup with approximately 15\% of the silicon modules by using
cosmic ray tracks. Before the first proton-proton collisions at the
LHC, cosmic ray muons were recorded by CMS in a dedicated run known as
``cosmic run at four Tesla'' (CRAFT)~\cite{Craft08Ref} with the magnetic
field at the nominal value. The CRAFT data were used to align and
calibrate the various subdetectors.  The complete alignment of the
tracker with the CRAFT data involved 3.2 million cosmic ray tracks
passing stringent quality requirements, as well as optical survey
measurements carried out before the final installation of the tracker
\cite{ALI:TkAlCraft08}. The achieved statistical accuracy on the
position of the modules with respect to the cosmic ray tracks was 3--4\mum
and 3--14\mum in the central and forward regions, respectively. The
performance of the tracking at CMS has been studied during the first
period of proton-proton collisions at the LHC and proven to be very
good already at the start of the operations
\cite{CMS-PAS-TRK-10-001,Trk:TRK-FirstColl, Muo:MUO-Performance}.

While the alignment obtained from CRAFT was essential for the early
physics programme of CMS, its quality was still statistically limited by
the available number of cosmic ray tracks, mainly in the forward
region of the pixel detector, and systematically limited by the lack of 
kinematic diversity in the track sample. In order to achieve the ultimate
accuracy, the large track sample accumulated in the proton-proton physics
run must be included.
This article describes the full alignment procedure for the modules of
the CMS tracker as applied during the commissioning campaign of 2011, and
its validation.  The procedure uses tracks from cosmic ray muons and
proton-proton collisions recorded in 2011. The final positions and
shapes of the tracker modules have been used for the reprocessing of
the 2011 data and the start of the 2012 data taking. A similar
procedure has later been applied to the 2012 data.

The structure of the paper is the following: in
section~\ref{Sec:TKLayout} the tracker layout and coordinate system
are introduced.  Section~\ref{Sec:GlobalTilt} describes the alignment
of the tracker with respect to the magnetic field of CMS.  In
section~\ref{Sec:Method} the algorithm used for the internal alignment
is detailed with emphasis on the topic of systematic
distortions of the tracker geometry and the possible ways for
controlling them. The strategy pursued for the alignment with tracks
together with the datasets and selections used are presented in
section~\ref{Sec:Strategy}.  Section~\ref{Sec:Stability} discusses the
stability of the tracker geometry as a function of
time. Section~\ref{Sec:StatAccuracy} presents the evaluation of the
performance of the aligned tracker in terms of statistical
accuracy. The measurement of the parameters describing the
non-planarity of the sensors is presented in
section~\ref{Sec:SensorShape}.  Section~\ref{Sec:WeakModes} describes
the techniques adopted for controlling the presence of systematic
distortions of the tracker geometry and the sensitivity of the
alignment strategy to systematic effects.


\section{Tracker layout and coordinate system}\label{Sec:TKLayout}

The CMS experiment uses a right-handed coordinate system, with the
origin at the nominal collision point, the $x$-axis pointing to the
centre of the LHC ring, the $y$-axis pointing up (perpendicular to the LHC
plane), and the $z$-axis along the anticlockwise beam direction. The
polar angle ($\theta$) is measured from the positive $z$-axis and the
azimuthal angle ($\varphi$) is measured from the positive $x$-axis in
the $x$-$y$ plane. The radius ($r$) denotes the distance from
the $z$-axis and the pseudorapidity ($\eta$) is defined as $\eta = - \ln
[\tan(\theta / 2)]$.

A detailed description of the CMS detector can be found in~\cite{CMS:2008zzk}.
 The central feature of the CMS apparatus is a
3.8\unit{T} superconducting solenoid of 6\unit{m} internal diameter.

Starting from the smallest radius, the silicon tracker, the crystal electromagnetic
calorimeter (ECAL), and the brass/scintillator hadron calorimeter (HCAL) are located within the inner field volume. The muon system is installed outside the solenoid
and is embedded in the steel return yoke of the magnet.

\begin{figure*}[hbtp]
  \begin{center}
    \includegraphics[width=0.7\textwidth]{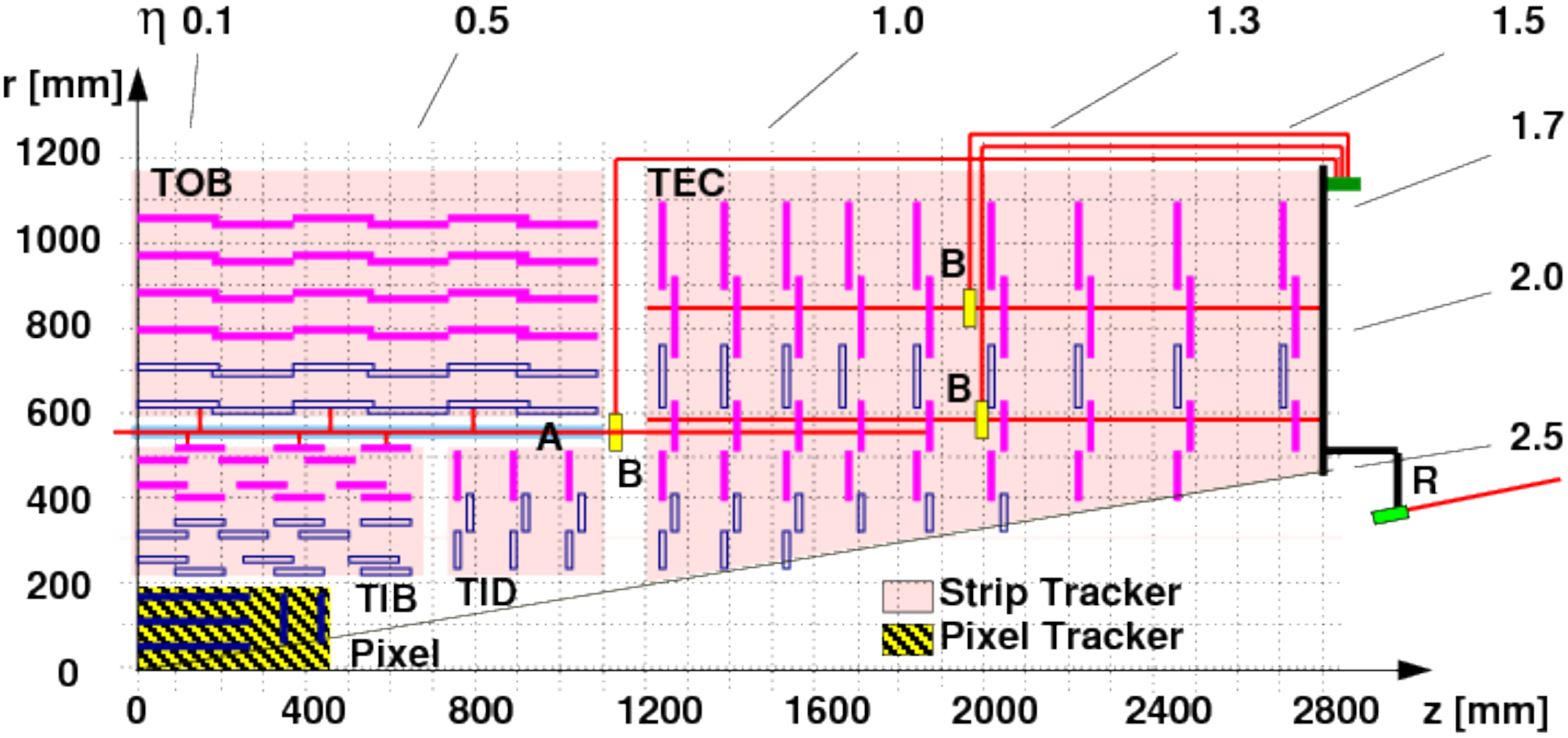}
    \caption{Schematic view of one quarter of the silicon tracker in the $r$-$z$ plane.
      The positions of the pixel modules are indicated within the hatched area. At
      larger radii within the lightly shaded areas, solid rectangles represent single
      strip modules, while hollow rectangles indicate pairs of strip
      modules mounted back-to-back with a relative stereo angle.
      The figure also illustrates the paths of the laser rays (R), the
      alignment tubes (A) and the beam splitters (B) of the laser alignment system.}
    \label{fig:TrackerLayout}
  \end{center}
\end{figure*}

The CMS tracker is composed of 1440 silicon pixel and 15\,148 silicon
microstrip modules organised in six sub-assemblies, as shown in figure
\ref{fig:TrackerLayout}.  Pixel modules of the CMS tracker are grouped
into the barrel pixel (BPIX) and the forward pixel (FPIX) in the
endcap regions. Strip modules in the central pseudorapidity region are
divided into the tracker inner barrel (TIB) and the tracker
outer barrel (TOB) at smaller and larger radii respectively. Similarly, strip modules in the
endcap regions are arranged in the tracker inner disks (TID) and tracker
endcaps (TEC) at smaller and larger values of $z$-coordinate, respectively.

The BPIX system is divided into two semi-cylindrical half-shells along the $y$-$z$
plane. The TIB and TOB are both divided into two half-barrels at positive and
negative $z$ values, respectively. The pixel modules composing the BPIX half-shells are
mechanically assembled in three concentric layers, with radial positions at 4\cm,
7\cm, and 11\cm in the design layout. Similarly, four and six layers of
microstrip modules compose the TIB and TOB half-barrels, respectively. The FPIX, TID,
and TEC are all divided into two symmetrical parts in the forward ($z>0$) and backward
($z<0$) regions. Each of these halves is composed of a series of disks arranged at
different $z$, with the FPIX, TID, and TEC having two, three, and nine such disks,
respectively.  Each FPIX disk is subdivided into two mechanically independent
half-disks.  The modules on the TID and TEC disks are further arranged in concentric
rings numbered from 1 (innermost) to 3 (outermost) in TID and from 1
up to 7 in TEC.

Pixel modules provide a two-dimensional measurement of the hit position. The pixels
composing the modules in BPIX have a rectangular shape, with the narrower side in the
direction of the ``global''  \rphi and the larger one along the "global"
$z$-coordinate, where ``global'' refers to the overall CMS coordinate system
introduced at the beginning of this section. Pixel modules in the FPIX have a finer
segmentation along global $r$ and a coarser one roughly along global \rphi,
but they are tilted by 20$^\circ$ around $r$. Strip modules
positioned at a given $r$ in the barrel generally measure the global \rphi coordinate
of the hit. Similarly, strip modules in the endcaps measure the global $\varphi$
coordinate of the hit.  The two layers of the TIB and TOB at smaller radii, rings 1 and 2
in the TID, and rings 1, 2, and 5 in the TEC are instrumented with pairs of microstrip
modules mounted back-to-back, referred to as ``\rphi'' and ``stereo'' modules,
respectively.  The strip direction of the stereo modules is tilted by 100\unit{mrad}
relative to that of the \rphi modules, which provides a measurement component in
the $z$-direction in the barrel and in the $r$-direction in the endcaps.  The modules
in the TOB and in rings 5--7 of the TEC contain pairs of sensors with strips
connected in series.

The strip modules have the possibility to take data in two different configurations,
called ``peak'' and ``deconvolution'' modes~\cite{French:2001xb,Raymond:2005qa}. The peak mode
uses directly the signals from the analogue pipeline, which stores the amplified and
shaped signals every 25\unit{ns}. In the deconvolution mode, a weighted sum of three
consecutive samples is formed, which effectively reduces the rise time and contains
the whole signal in 25\unit{ns}. Peak mode is characterised by a better signal-over-noise
ratio and a longer integration time, ideal for collecting cosmic ray tracks that
appear at random times, but not suitable for the high bunch-crossing frequency of the
LHC. Therefore, the strip tracker is operated in deconvolution mode when recording
data during the LHC operation. The calibration and time synchronization of the strip
modules are optimized for the two operation modes separately.

\begin{figure*}[htbp]
  \begin{center}
    \includegraphics[width=0.52\textwidth]{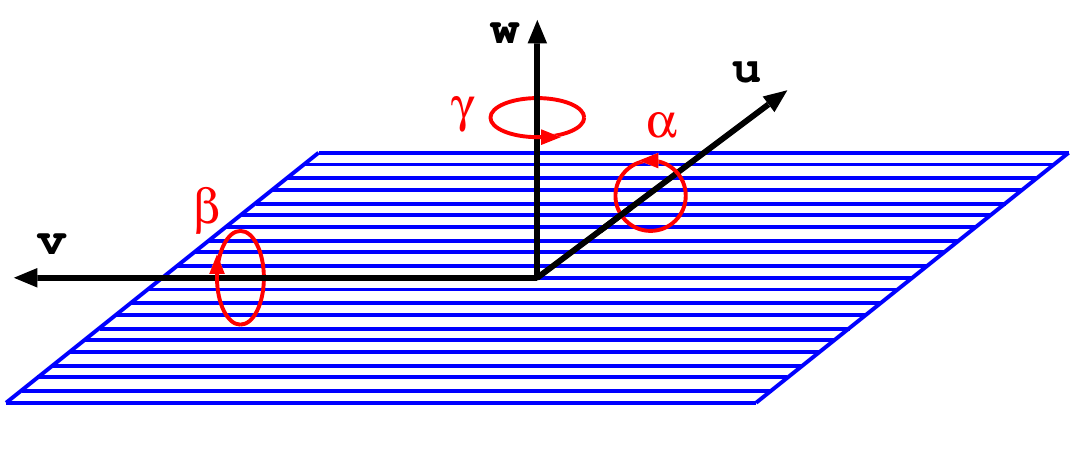}
    \includegraphics[height=0.2\textwidth]{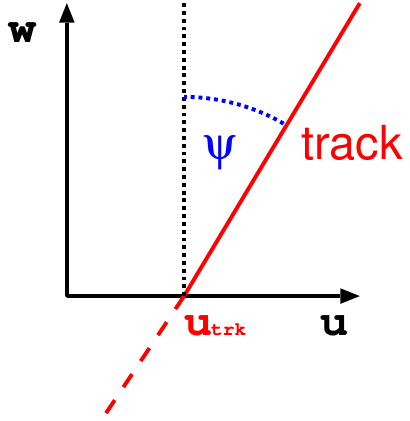}
    \includegraphics[height=0.2\textwidth]{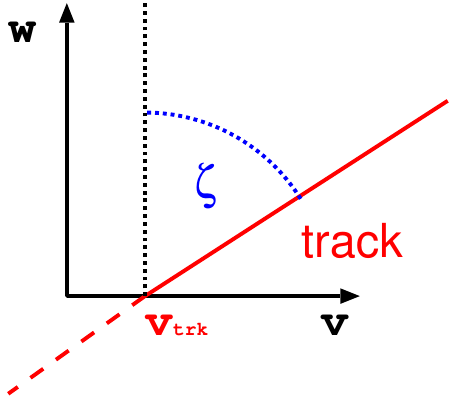} 
    \caption{Sketch of a silicon strip module showing the axes of its local coordinate
      system, $u$, $v$, and $w$, and the respective local rotations $\alpha$, $\beta$,
      $\gamma$ (left), together with illustrations of the local track angles $\psi$ and
      $\zeta$ (right).}
    \label{fig:LocalCoord}
  \end{center}
\end{figure*}

A local right-handed coordinate system is defined for each module with
the origin at the geometric centre of the active area of the
module. As illustrated in the left panel of figure~\ref{fig:LocalCoord}, the
$u$-axis is defined along the more precisely measured coordinate of
the module (typically along the azimuthal direction in the global
system), the $v$-axis orthogonal to the $u$-axis and in the module
plane, pointing away from the readout electronics, and the $w$-axis
normal to the module plane.  The origin of the $w$-axis is in the
middle of the module thickness.  For the pixel system, $u$ is chosen
orthogonal to the magnetic field, \ie in the global \rphi direction
in the BPIX and in the radial direction in the FPIX. The $v$-coordinate is perpendicular to $u$ in the sensor plane, \ie along
global $z$ in the BPIX and at a small angle to the global \rphi
direction in the FPIX. The angles $\alpha$, $\beta$, and $\gamma$
indicate right-handed rotations about the $u$-, $v$-, and $w$-axes,
respectively.  As illustrated in the right panel of
figure~\ref{fig:LocalCoord}, the local track angle $\psi$ ($\zeta$)
with respect to the module normal is defined in the $u$-$w$ ($v$-$w$) plane.

\section{Global position and orientation of the tracker}\label{Sec:GlobalTilt}
While the track based internal alignment (see section~\ref{sec:methodology})
mainly adjusts the positions
and angles of the tracker modules relative to each other, it cannot ascertain
the absolute position and orientation of the tracker.
Survey measurements
of the TOB, as the largest single sub-component, are thus used to determine
its shift and rotation around the beam axis relative to the design values.
The other sub-components are then aligned relative to the TOB
by means of the track based internal alignment procedure.
The magnetic field of the CMS solenoid is to good approximation parallel to the $z$-axis.
The
orientation of the tracker relative to the magnetic field is of
special importance,
since the correct parameterisation of the trajectory in the
reconstruction depends on it. This global orientation is described by
the angles $\theta_x$ and $\theta_y$, which correspond to rotations of
the whole tracker around the $x$- and $y$-axes defined in the previous section.
Uncorrected overall
tilts of the tracker relative to the magnetic field could result in
biases of the reconstructed parameters of the tracks and the measured masses of
resonances inferred from their charged daughter particles. Such biases would be hard to
disentangle from the other systematic effects addressed in section~\ref{subsec:weakmodes}.
 It is therefore essential to determine the
global tracker tilt angles prior to the overall alignment
correction, because the latter might be affected by a wrong assumption on the magnetic field orientation. It is not expected that tilt angles will change
significantly with time, hence one measurement should be sufficient for many years of operation.
The tilt angles have been determined with the 2010 CMS
data, and they have been used as the input for the internal alignment
detailed in subsequent sections of this article.
A repetition of the procedure with 2011 data led to compatible results.

The measurement of the tilt angles is based on the study of overall track quality
as a function of the $\theta_x$ and $\theta_y$ angles. Any non-optimal
setting of the tilt angles will result, for example, in incorrect assumptions on
the transverse field components relative to the tracker axis. This may degrade the observed track
quality. The tilt angles $\theta_x$ and $\theta_y$ are
scanned in appropriate intervals centred around zero. For each set of values, the standard
CMS track fit is applied to the whole set of tracks, and an overall track quality
estimator is determined. The track quality is estimated by the total $\chi^2$ of all the fitted tracks, $\sum \chi^2$.
As cross-checks, two other track quality estimators are also studied:
the mean normalised track $\chi^2$ per degree of freedom, $\langle\chi^2/{N_\mathrm{dof}}\rangle$,
and the mean p value, $\langle\mathrm{P}(\chi^2,N_\mathrm{dof})\rangle$,
which is the probability of the $\chi^2$ to exceed the value obtained in the track fit,
assuming a $\chi^2$ distribution with the appropriate number of degrees of freedom.
All methods yield similar results; remaining small differences are attributed to the different
relative weight of tracks with varying number of hits and the effect of any remaining outliers.

Events are considered if they have exactly one primary vertex
reconstructed by using at least four tracks,
and a reconstructed position along the beam line within $\pm 24$\unit{cm} of the nominal centre of the CMS detector.
Tracks are required to have at least ten reconstructed hits and a pseudorapidity
of $\abs{\eta}<2.5$. The track impact parameter with respect to the
primary vertex must be less than 0.15\unit{cm} (2\unit{cm}) in the transverse (longitudinal)
direction. For the baseline analysis that provides the central values, the transverse momentum threshold
is set to 1\GeVc; alternative values of 0.5 and 2\GeVc are used to check the stability of the results.
Only tracks with $\chi^2/N_\mathrm{dof}<4$ are selected
in order to reject those with wrongly associated hits. For each setting of the tilt angles, each track is refitted
by using a full 3D magnetic field model~\cite{Maroussov:2008zz,Chatrchyan:2009si}
that also takes tangential field components into account.
This field model is based on measurements obtained during a dedicated
mapping operation with Hall and NMR probes~\cite{Klyukhin:2011xq}.

\begin{figure*}[hbt]
  \begin{center}
    \includegraphics[width=0.45\textwidth,angle=0]{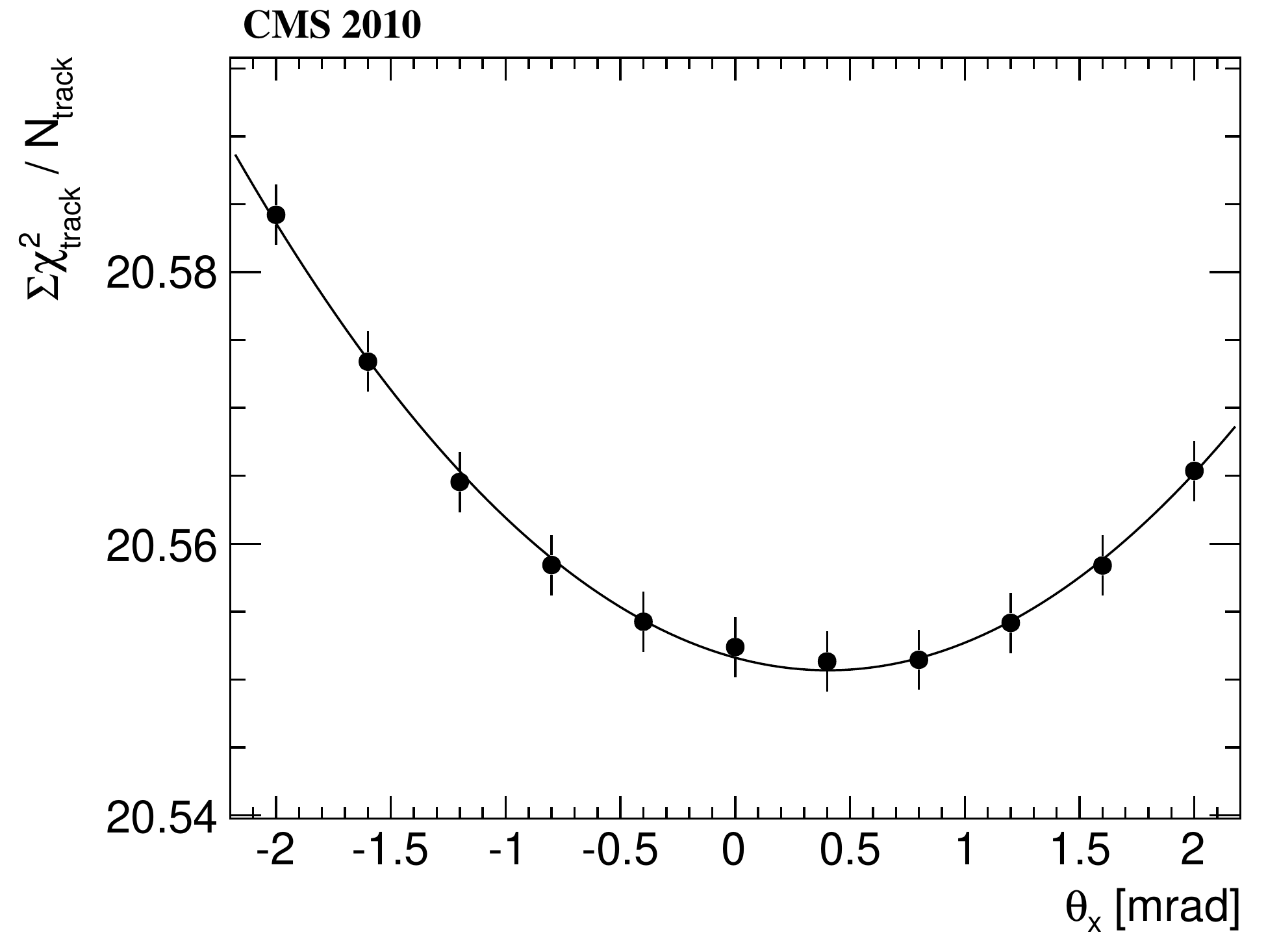}\hspace{0.1cm}\includegraphics[width=0.45\textwidth,angle=0]{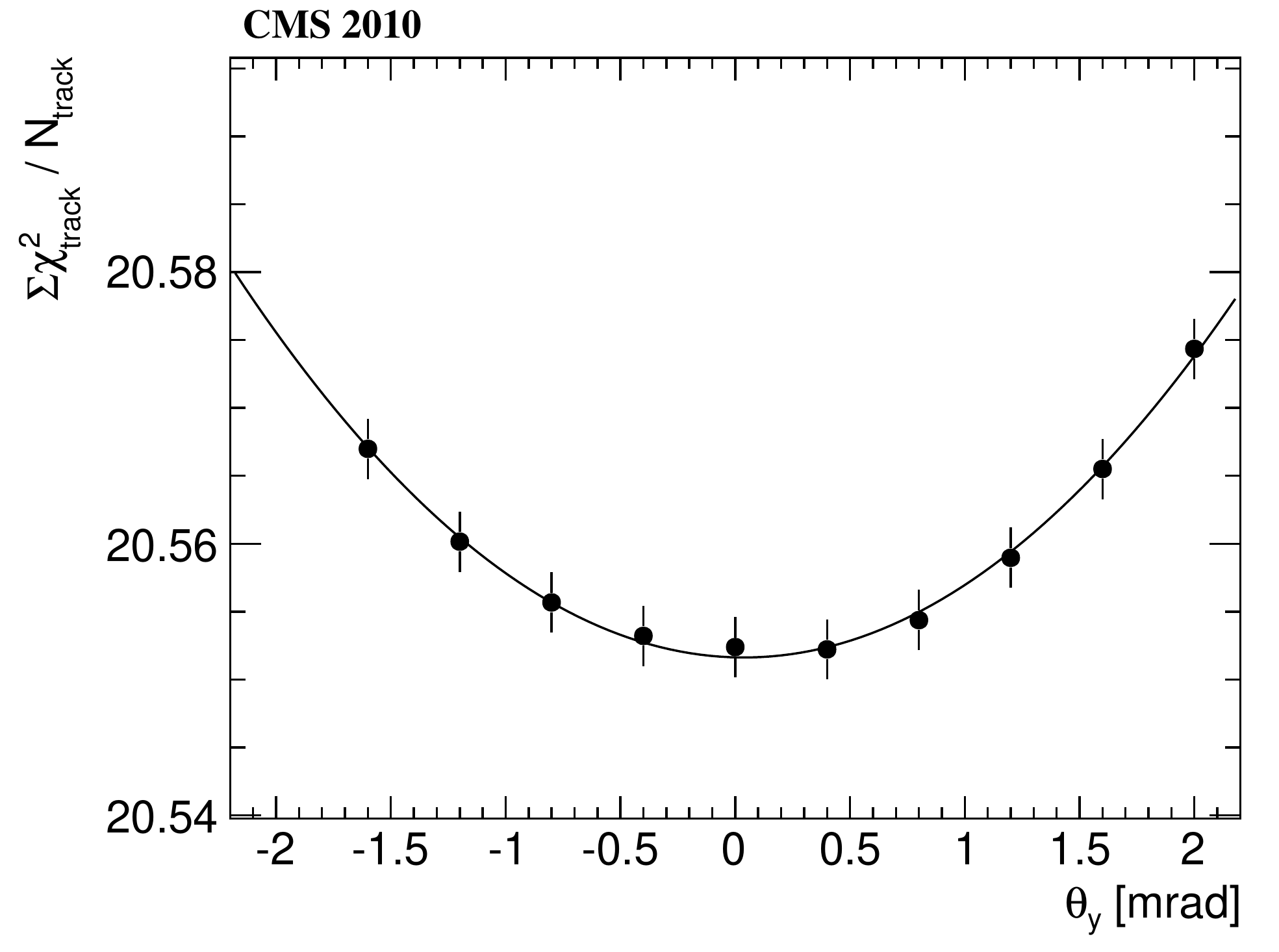}
    \caption{Dependence of the total $\chi^2$ of the track
    fits, divided by the number of tracks, on the
    assumed $\theta_x$ (left) and $\theta_y$ (right) tilt
    angles for $\abs{\eta}<2.5$ and \pt $>1$\GeVc. The error bars are purely statistical and
    correlated point-to-point because the same tracks are used for each point.}
    \label{fig:sumt}
  \end{center}
\end{figure*}

Each tilt angle is scanned at eleven
settings in the range $\pm$2\unit{mrad}. The angle of correct alignment is derived as
the point of maximum track quality, corresponding to minimum total $\chi^2$, determined by a least squares fit with a
second-order polynomial function.
The dependence of the total $\chi^2$ divided by the number of tracks
on the tilt angles $\theta_x$ and $\theta_y$ is
shown in figure~\ref{fig:sumt}.

In each plot, only one angle is varied, while the other remains fixed at 0. The second-order
polynomial fit describes the functional dependence very well. There is no result for the scan
point at $\theta_y=-2$\unit{mrad}, because this setting is outside the range
allowed by the track reconstruction programme.
While the $\theta_y$ dependence is symmetric
with a maximum near $\theta_y \approx 0$, the $\theta_x$ dependence
is shifted towards positive values, indicating a noticeable vertical downward
tilt of the tracker around the $x$-axis with respect to the magnetic
field. On an absolute scale, the tilt is small, but nevertheless visible within
the resolution of beam line tilt
measurements~\cite{Trk:TRK-FirstColl}. 

\begin{figure*}[hbtp]
  \begin{center}
    \includegraphics[width=0.45\textwidth,angle=0]{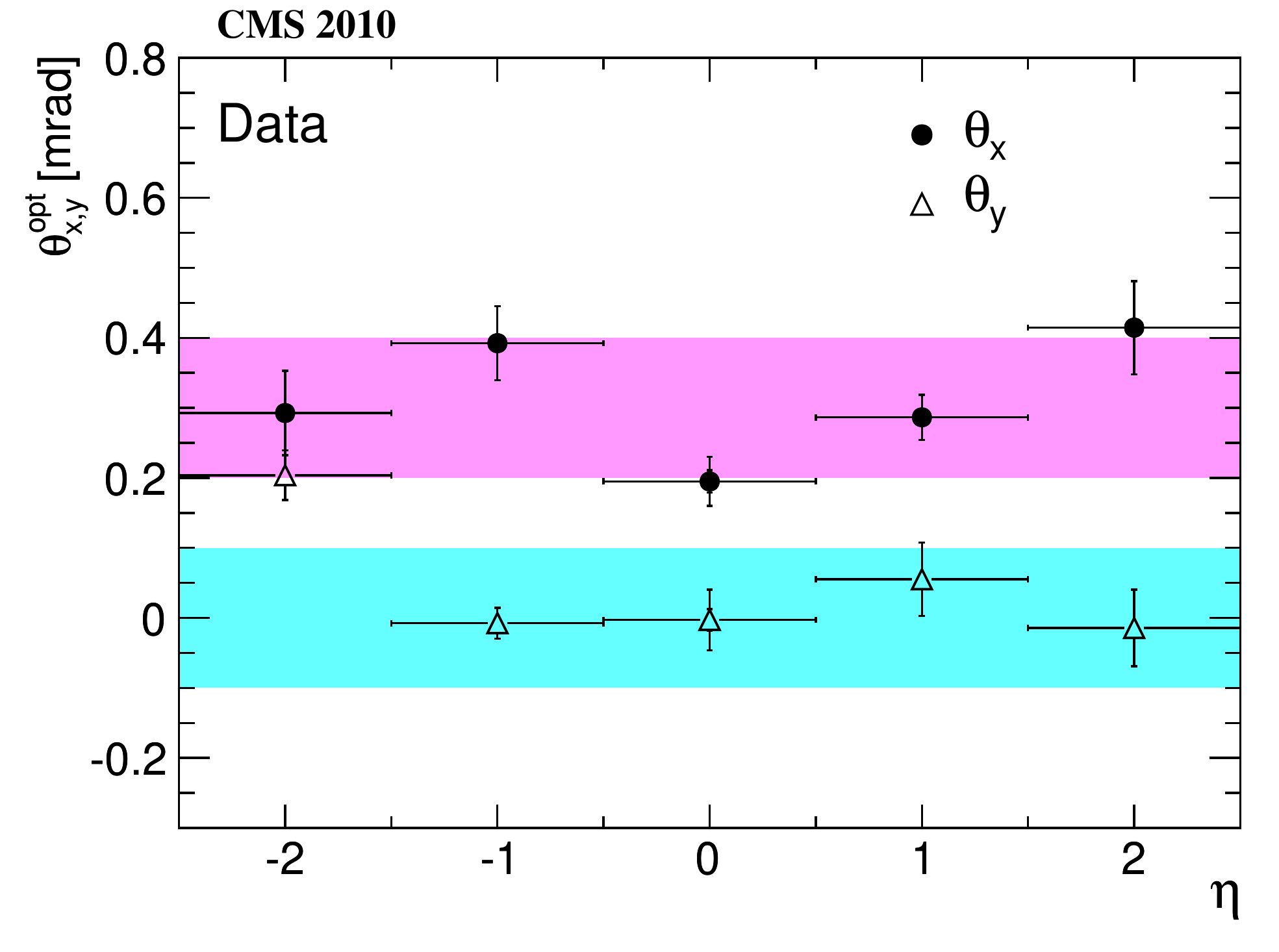}\hspace{0.1cm}\includegraphics[width=0.45\textwidth,angle=0]{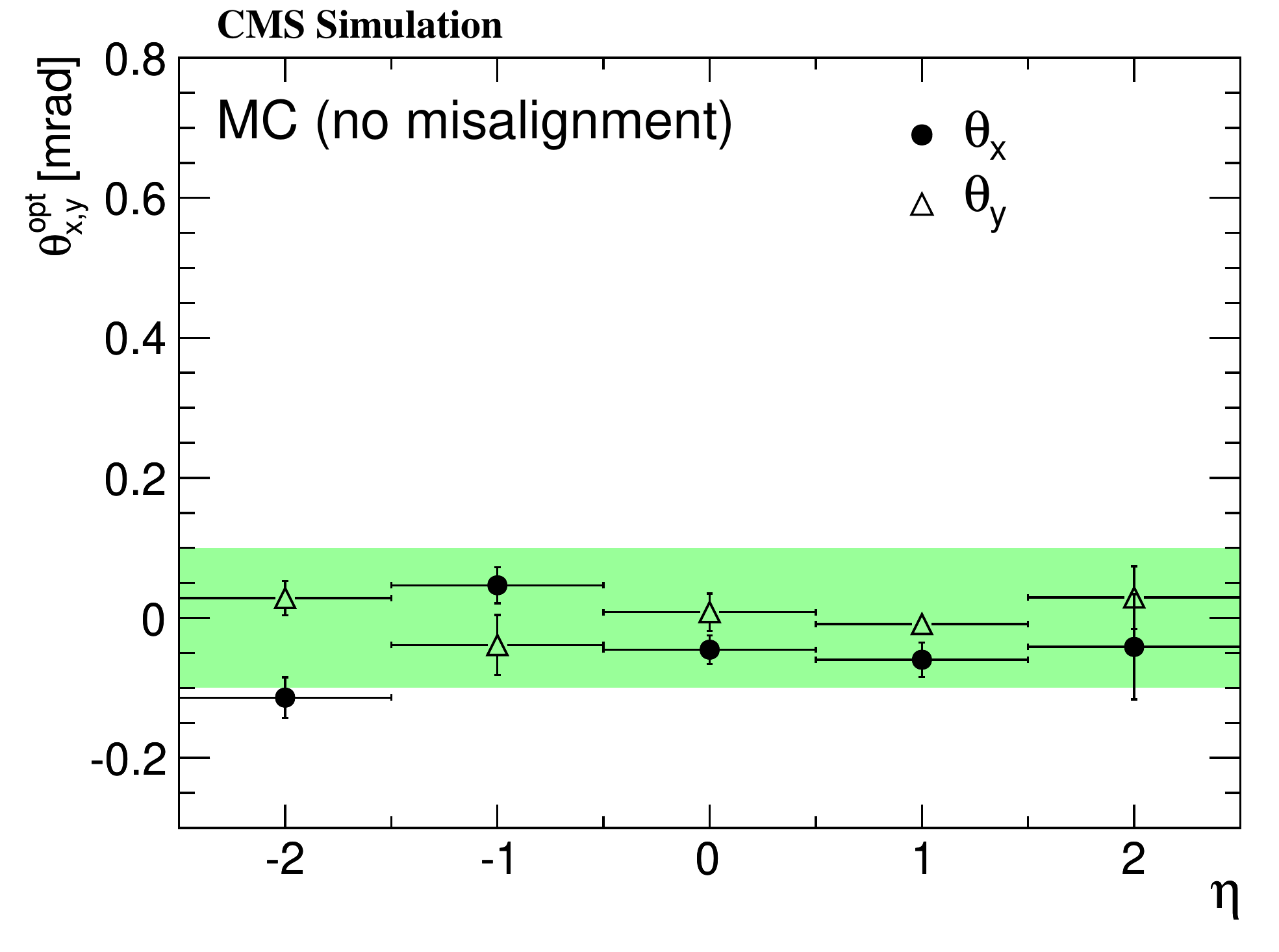}
    \caption{Tracker tilt angles $\theta_x$ (filled circles)
    and $\theta_y$ (hollow triangles) as a function of track
    pseudorapidity. The left plot shows the values measured with the
    data collected in 2010; the right plot has been obtained from simulated
    events without tracker misalignment. The statistical uncertainty
    is typically smaller than the symbol size and mostly invisible. The outer error bars indicate
    the RMS of the variations which are observed when varying
    several parameters of the tilt angle determination. The shaded bands indicate the
    margins of $\pm$0.1 discussed in the text.}
    \label{fig:tiltband}
  \end{center}
\end{figure*}

Figure~\ref{fig:tiltband} shows the resulting values of $\theta_x$ and $\theta_y$ for
five intervals of track pseudorapidity, for tracks with $\pt >1\GeVc$ for the total $\chi^2$ estimator.
The statistical uncertainties are taken as the distance corresponding to a one unit increase
of the total $\chi^2$; they are at most of the order of
the symbol size and thus hardly visible. The outer error bars show
the root mean square (RMS) of the
changes observed in the tilt angle estimates
when changing the track quality estimators and the \pt threshold.
The right plot shows the results of the method applied to simulated events
without any tracker misalignment. They are consistent
with zero tilt within the systematic uncertainty. The variations are smaller in the central
region within $\abs{\eta}<1.5$, and they are well contained within a margin of $\pm$0.1\unit{mrad}, which is used as a rough estimate of the systematic uncertainty of this method.
The left plot in figure~\ref{fig:tiltband} shows the result obtained from data; the $\theta_x$ values are systematically shifted
by $\approx$0.3\unit{mrad}, while the $\theta_y$ values are close to zero. The nominal tilt angle
values used as alignment constants are extracted from the central $\eta$
region of figure~\ref{fig:tiltband} (left) as
$\theta_x=(0.3 \pm 0.1)$\unit{mrad} and $\theta_y=(0 \pm 0.1)$\unit{mrad}, thus eliminating an important potential
source of systematic alignment uncertainty. These results represent
an important complementary step to the internal alignment procedure
described in the following sections.

\section{Methodology of track based internal alignment} \label{Sec:Method}
\label{sec:methodology}

Track-hit residual distributions are generally broadened if the
assumed positions 
of the silicon modules
used in track reconstruction differ from the true positions.
Therefore standard alignment algorithms
follow the least squares approach and minimise
the sum of squares of normalised residuals from many tracks.
Assuming the 
measurements $m_{ij}$, \ie usually the reconstructed hit positions on the modules,
with uncertainties $\sigma_{ij}$
are independent, the minimised objective function is
\begin{equation}
  \chi^2(\mathbf{p}, \mathbf{q}) =
  \sum_j^{\text{tracks}}\,\sum_i^{\text{measurements}}
  \left(\frac{m_{ij} - f_{ij}\left(\mathbf{p}, \mathbf{q}_j\right)}
       {\sigma_{ij}}\right)^2,
       \label{eq:globalChi2}
\end{equation}
where $f_{ij}$ is the trajectory prediction of the track model at the position of
the measurement,
depending on the geometry ($\mathbf{p}$) and track ($\mathbf{q}_j$) parameters.
An initial geometry description $\mathbf{p}_0$ is usually available from
design drawings, survey measurements, or previous alignment results.
This can be used to determine approximate track parameters $\mathbf{q}_{j0}$.
Since alignment corrections can be assumed to be small,
$f_{ij}$ can be linearised around these initial values. 
Minimising \chisq after the linearisation
leads to the normal equations of least squares.
These can be expressed as a linear equation system
$\mathbf{C} \mathbf{a} = \mathbf{b}$
with $\mathbf{a}^T = (\mathbf{\Delta p},\mathbf{\Delta q})$,
\ie the alignment parameters $\mathbf{\Delta p}$
and corrections to all parameters of all $n$ used tracks
$\mathbf{\Delta q}^T = (\mathbf{\Delta q}_1, \dots , \mathbf{\Delta  q}_n)$.
If the alignment corrections are not small, the linear approximation
is of limited precision and the procedure has to be iterated.

For the alignment of the CMS tracker a global-fit approach~\cite{ALI:MP1}
is applied, by using the \mptwo program~\cite{ALI:MPII}.
It makes use of
the special structure of $\mathbf{C}$ that facilitates,
by means of block matrix algebra, the reduction of the large system
of equations $\mathbf{C} \mathbf{a} = \mathbf{b}$ to a smaller one for the
alignment parameters only,
\begin{equation}
  \mathbf{C'} \mathbf{\Delta p} = \mathbf{b'}.
  \label{eq:reducedNormal}
\end{equation}
Here $\mathbf{C'}$ and
$\mathbf{b'}$ sum contributions from all tracks.
To derive $\mathbf{b'}$, the solutions $\mathbf{\Delta q}_j$ of the track fit
equations
$\mathbf{C}_j \mathbf{\Delta q}_j = \mathbf{b}_j$ are needed.
For $\mathbf{C'}$, also the covariance matrices $\mathbf{C}^{-1}_j$ have
to be calculated.
The reduction of the matrix size from $\mathbf{C}$ to $\mathbf{C'}$ is
dramatic. For $10^7$ tracks with on average 20 parameters and $10^5$
alignment parameters, the number of matrix elements is reduced by
a factor larger than $4 \times 10^6$. Nevertheless, no information is lost
for the determination of the alignment parameters $\mathbf{\Delta p}$.

The following subsections explain the track and alignment parameters
$\mathbf{\Delta q}$ and $\mathbf{\Delta p}$ that are used for
the CMS tracker alignment. Then the concept of a hierarchical and differential
alignment by using equality constraints is introduced, followed by a
discussion of ``weak modes'' and how they
can be avoided.
The section closes with the computing optimisations needed to make
\mptwo a fast tool with modest computer memory needs, even for the alignment
of the CMS tracker with its unprecedented complexity.

\subsection{Track parameterisation}
\label{sec:method_trackparametrisation}
In the absence of material effects, five parameters are needed to describe
the trajectory of a charged particle in a magnetic field.
Traversing material, the particle experiences
multiple scattering, mainly due to Coulomb interactions with
atomic nuclei. 
These effects are significant in the CMS tracker, \ie the particle trajectory
cannot be well described without taking them into account in the track model.
This is achieved in a rigorous and efficient way as explained
below in this section, representing
an improvement compared to previous \mptwo alignment
procedures~\cite{ALI:Stoye_MC_2008,ALI:TIF,ALI:TkAlCraft08} for the CMS silicon tracker,
which ignored correlations induced by multiple scattering.

A rigorous treatment of multiple scattering can be
achieved by increasing the number of track parameters to
$n_{\text{par}} = 5 + 2 n_{\text{scat}}$, \eg
by adding two deflection angles for each of the
$n_{\text{scat}}$ thin scatterers
traversed by the particle. For thin scatterers, the trajectory offsets induced by multiple scattering
  can be ignored. If a scatterer is thick, it can be approximately
  treated as two thin scatterers.
The distributions of these deflection angles have mean values of zero.
Their standard deviations can be estimated by using preliminary knowledge
of the particle momentum and of the amount of material crossed.
This theoretical knowledge is used to extend the list of measurements,
originally containing all the track hits, by ``virtual measurements''.
Each scattering angle is virtually measured to be zero with an uncertainty
corresponding to the respective standard deviation.
These virtual measurements compensate for the degrees of freedom introduced
in the track model by increasing its number of parameters.
For cosmic ray tracks this complete parameterisation often leads
to $n_{\text{par}} > 50$.
Since in the general case the effort to calculate $\mathbf{C}^{-1}_j$ is
proportional to $n_{\text{par}}^3$, a significant amount of computing time
would be spent to calculate $\mathbf{C}^{-1}_j$ and thus
$\mathbf{C'}$ and $\mathbf{b'}$.
The progressive Kalman filter fit, as used in the CMS track
reconstruction, avoids
the $n_{\text{par}}^3$ scaling by a sequential fit procedure,
determining five track parameters
at each measurement.
However, the Kalman filter does not provide the full (singular) covariance matrix $\mathbf{C}^{-1}_j$
of these parameters as needed in a global-fit alignment approach.
As shown in~\cite{ALI:Hulsbergen2009},
the Kalman filter fit procedure can be extended to provide this covariance matrix,
but since \mptwo is designed for a simultaneous fit of all measurements, another
approach is followed here.

The general broken lines (GBL) track refit~\cite{GBL2010_CR2010_89,GBL_Kleinwort2012} 
that generalises the algorithm described in Ref.~\cite{BrokenLines_Blobel},
avoids the $n_{\text{par}}^3$ scaling
for calculating $\mathbf{C}^{-1}_j$
by defining a custom track parameterisation. The parameters are
$\mathbf{q}_j = (\Delta \frac{q}{p},\allowbreak \mathbf{u}_1, \dots,\allowbreak
\mathbf{u}_{\left(n_{\text{scat}}+2\right)})$,
where $\Delta \frac{q}{p}$ is the change of the
inverse momentum multiplied by the particle charge 
and $\mathbf{u}_i$ are the two-dimensional offsets to an initial
reference trajectory
in local systems at each scatterer and at the first and last measurement.
All parameters except $\Delta \frac{q}{p}$ influence only a small part of
the track trajectory.
This locality of all track parameters (but one) results in
$\mathbf{C}_j$ being a bordered band matrix with band width
$m \leq 5$ and border size $b = 1$, \ie
the matrix elements $c_j^{kl}$ are non-zero only for
$k \leq b$, $l \leq b$ or $\abs{k - l} \le m$.
By means of root-free Cholesky decomposition
($\mathbf{C}_j^{\mathrm{\scriptsize band}} = \mathbf{LDL}^T$) of the band part
$\mathbf{C}_j^{\mathrm{\scriptsize band}}$
into a diagonal matrix $\mathbf{D}$ and a unit left triangular band matrix
$\mathbf{L}$,
the effort to calculate $\mathbf{C}^{-1}_j$ and $\mathbf{q}_j$
is reduced to
$\propto n_{\text{par}}^2\cdot (m+b)$ and $\propto n_{\text{par}}\cdot (m+b)^2$,
respectively.
This approach saves a factor of 6.5 in CPU time for track refitting in \mptwo for
isolated muons (see section~\ref{Sec:Strategy}) and of 8.4 for cosmic ray
tracks in comparison with an (equivalent) linear equation system with a
dense matrix solved by inversion.

The implementation of the GBL refit
used for the \mptwo alignment of the CMS tracker 
is based on a seed trajectory derived from the position and direction
of the track at its first hit as resulting from the standard Kalman filter
track fit. From the first hit, the trajectory is propagated taking into account
magnetic-field inhomogeneities, by using the Runge--Kutta technique,
and the average energy loss in the material as for muons. As in the CMS
Kalman filter track
fit, all traversed material is assumed to coincide with the silicon
measurement planes that are treated as thin scatterers.
The curvilinear frame defined in~\cite{Strandlie_Wittek_Jacobians2006}
is chosen for the local coordinate systems at these scatterers.
Parameter propagation along the trajectory needed to link the local systems
uses Jacobians assuming a locally constant magnetic
field between them~\cite{Strandlie_Wittek_Jacobians2006}.
To further reduce the computing time,
two approximations are used in the standard processing: material
assigned to stereo and \rphi modules that are mounted together is
treated as a single thin scatterer, and the Jacobians are calculated
assuming the magnetic field $\vec{B}$ to be parallel to the $z$-axis
in the limit of weak deflection, $\frac{\abs{\vec{B}}}{p} \to 0$. This
leads to a band width of $m = 4$ in the matrix $\mathbf{C}_j$.

\subsection{Alignment parameterisation}
\label{sec:method_alignmentparametrisation}
\begin{figure}[tb]
  \centering
      \includegraphics[width=.3\textwidth]{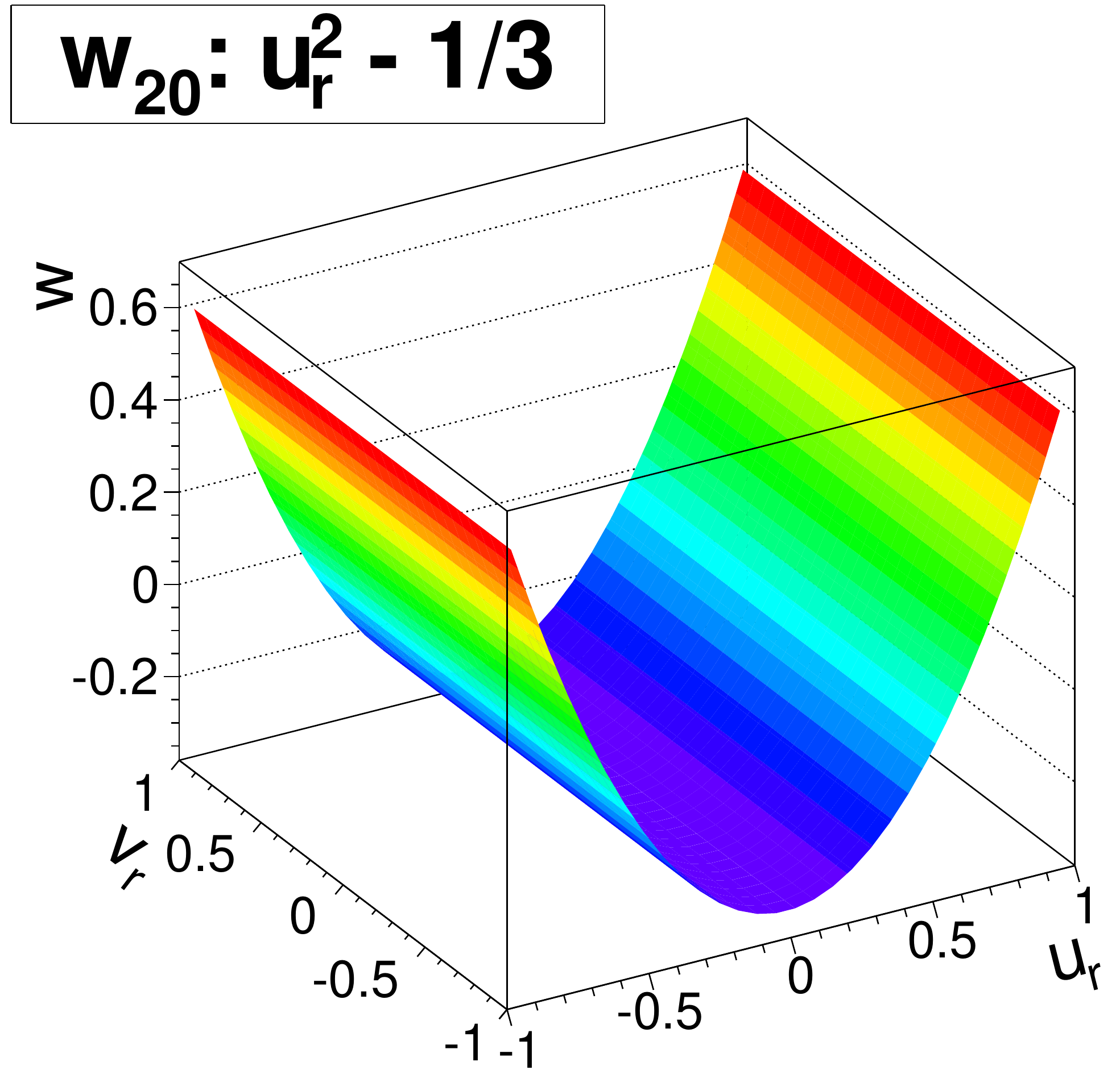}  \hfill
      \includegraphics[width=.3\textwidth]{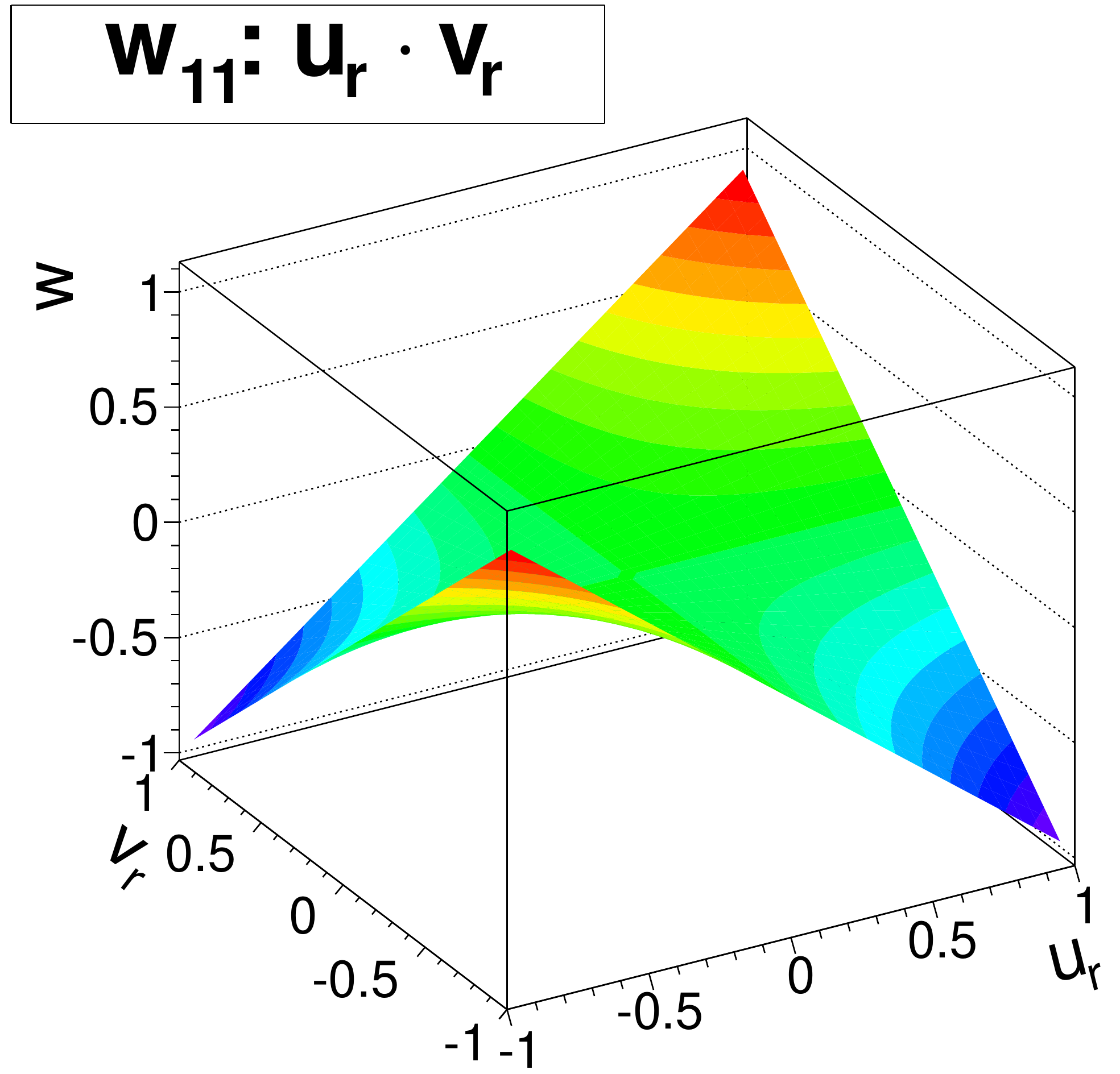}  \hfill
      \includegraphics[width=.3\textwidth]{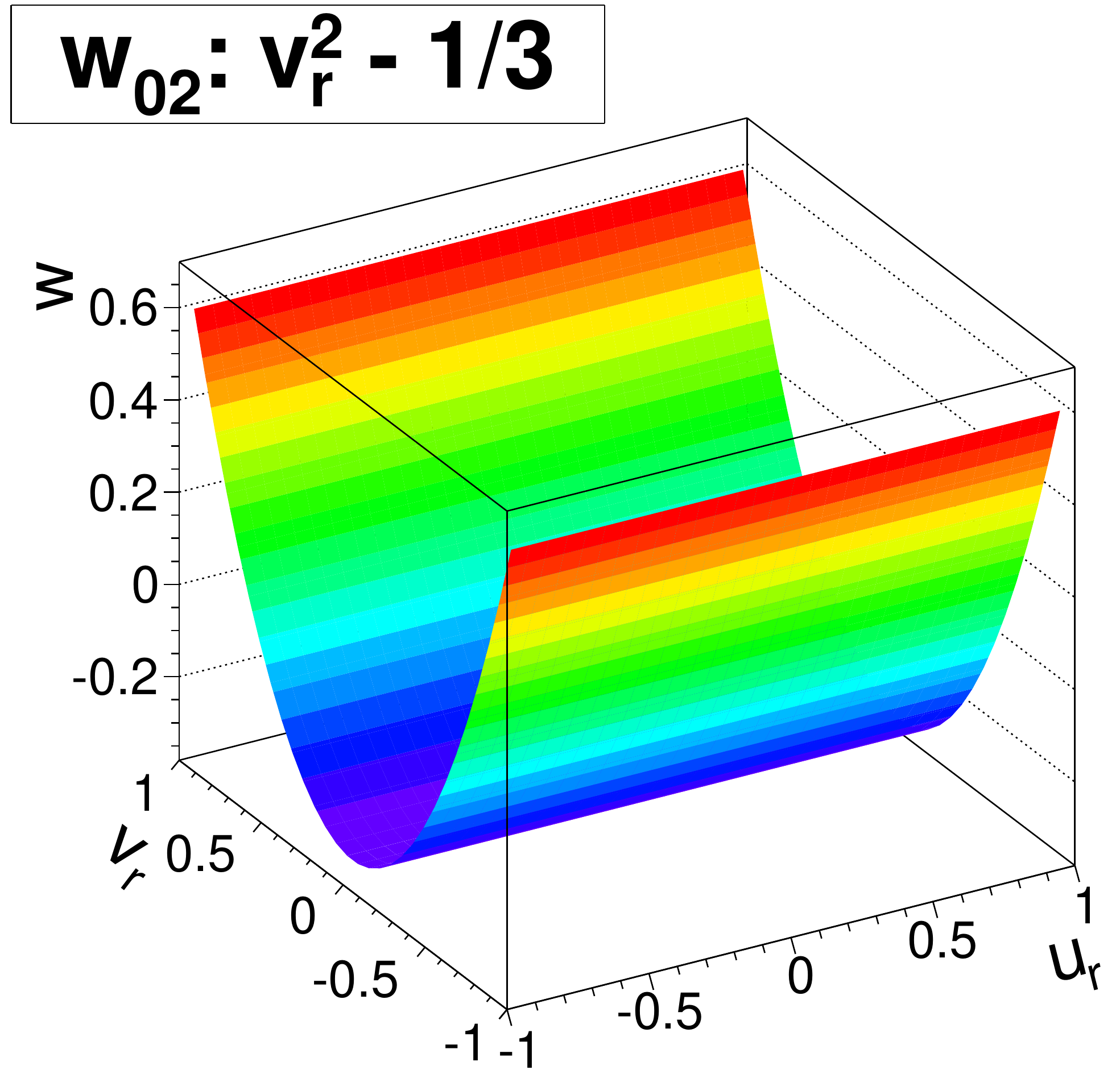}
  \caption{
    The three two-dimensional second-order polynomials to describe sensor
    deviations from the flat plane, illustrated for sagittae
    $w_{20} = w_{11} = w_{02} = 1$.
    \label{fig:bow_parameters}
  }
\end{figure}

To first approximation, the CMS silicon modules are flat planes.
Previous alignment approaches in CMS ignored possible deviations from
this approximation and determined only corrections to the
initial module positions,
\ie up to three shifts ($u$, $v$, $w$) and three rotations
($\alpha$, $\beta$, $\gamma$). 
However, tracks
with large angles of incidence relative to the silicon module normal
are highly sensitive to the exact positions of the modules along their $w$
directions and therefore also to local $w$ variations if the modules
are not flat.
These local variations can arise from
possible curvatures of silicon sensors and, for strip modules with two sensors in a
chain, from their relative misalignment.
In fact, sensor curvatures can be expected because of tensions after mounting
or because of single-sided silicon processing as for the strip sensors.
The specifications for the construction of the sensors
required the deviation from perfect planarity to be less than 100\mum~\cite{CMS:2008zzk}.
To take into account such deviations,
the vector of alignment parameters $\mathbf{\Delta p}$ is
extended to up to nine degrees of freedom per sensor instead of six per module.
The sensor shape is parameterised as a sum of products of modified (orthogonal)
Legendre polynomials up to the second order where the constant and linear
terms are equivalent to the rigid body parameters $w$, $\alpha$ and $\beta$:
\begin{equation}
        \begin{array}{rlll}
          w(u_{r},v_{r})  = & \hphantom{+}\;w&&\\ 
          & + \; w_{10} \cdot u_{r} &+\; w_{01} \cdot v_{r}& \\
          & + \; w_{20} \cdot (u_{r}^2 - 1/3) &+\; w_{11} \cdot (u_{r} \cdot v_{r}) &+\; w_{02} \cdot (v_{r}^2 - 1/3). \\
        \end{array}
\end{equation}
Here $u_r \in [-1, 1]$ ($v_r \in [-1, 1]$) is the 
position on the sensor in the $u$- ($v$-) direction,
normalised to its width $l_u$ (length $l_v$).
The coefficients $w_{20}$, $w_{11}$ and $w_{02}$ quantify the sagittae of the
sensor curvature as illustrated in figure~\ref{fig:bow_parameters}.
The CMS track reconstruction algorithm treats the hits under the assumption
of a flat module surface.
To take into account the determined sensor shapes,
the reconstructed hit positions
in $u$ and (for pixel modules) $v$ are corrected by
$\left(-w\left(u_r, v_r\right) \cdot \tan\psi \right)$ and
$\left(-w\left(u_r, v_r\right) \cdot \tan\zeta\right)$, respectively.
Here the track angle from the sensor normal $\psi$ ($\zeta$) is defined in the
$u$-$w$ ($v$-$w$) plane (figure~\ref{fig:LocalCoord}),
and the track predictions are used for $u_r$ and $v_r$.

To linearise the track-model prediction $f_{ij}$, derivatives with respect
to the alignment parameters have to be calculated.
If $f_{ij}$ is in the local $u$ ($v$) direction,
denoted as $f_u$ ($f_v$) in the following, the derivatives are
\begin{equation}
  \renewcommand{\arraystretch}{1.35} 
  \left(
  \begin{array}{cc}
    \frac{\partial f_u}{\partial u}     & \frac{\partial f_v}{\partial u}\\
    \frac{\partial f_u}{\partial v}     & \frac{\partial f_v}{\partial v}\\
    \frac{\partial f_u}{\partial w}     & \frac{\partial f_v}{\partial w}\\
    \frac{\partial f_u}{\partial w_{10}}& \frac{\partial f_v}{\partial w_{10}}\\
    \frac{\partial f_u}{\partial w_{01}}& \frac{\partial f_v}{\partial w_{01}}\\
    \frac{\partial f_u}{\partial\gamma'}& \frac{\partial f_v}{\partial\gamma'}\\
    \frac{\partial f_u}{\partial w_{20}} & \frac{\partial f_v}{\partial w_{20}}\\
    \frac{\partial f_u}{\partial w_{11}} & \frac{\partial f_v}{\partial w_{11}}\\
    \frac{\partial f_u}{\partial w_{02}} & \frac{\partial f_v}{\partial w_{02}}\\
  \end{array}
  \right) = \left(
  \begin{array}{rr}
    \multicolumn{1}{c}{-1} &  \multicolumn{1}{c}{0} \\
    \multicolumn{1}{c}{0}  & \multicolumn{1}{c}{-1} \\
    \tan\psi               & \tan\zeta \\
    u_{r} \cdot \tan\psi   & u_{r} \cdot \tan\zeta \\
    v_{r} \cdot \tan\psi   & v_{r} \cdot \tan\zeta \\
    \multicolumn{1}{c} {v_r l_v/(2\, s)}& \multicolumn{1}{c}{-u_r l_u/(2\, s)}\\
    (u_{r}^2 -1/3) \cdot \tan\psi   & (u_{r}^2 -1/3) \cdot \tan\zeta \\
    u_{r} \cdot v_{r} \cdot \tan\psi& u_{r}\cdot v_{r} \cdot \tan\zeta\\
    (v_{r}^2 -1/3) \cdot \tan\psi   & (v_{r}^2 -1/3) \cdot \tan\zeta \\
  \end{array}
  \right).
  \label{eq:bowedDerivs}
\end{equation}

Unlike the parameterisation used in previous CMS alignment
procedures~\cite{KarimakiDerivs2003},
the coefficients of the first order polynomials
$w_{01} = \frac{ l_v}{2} \cdot \tan\alpha$ and
$w_{10} = \frac{-l_u}{2} \cdot \tan\beta $
are used as alignment parameters
instead of the angles.
This
ensures the orthogonality of the sensor surface parameterisation.
The in-plane rotation $\gamma$ is replaced by $\gamma' = s \cdot \gamma$ with
$s = \frac{l_u+l_v}{2}$. This has the advantage that all parameters
have a length scale and their derivatives have similar numerical size.

The pixel modules provide uncorrelated measurements in both $u$ and $v$
directions.
The strips of the modules in the TIB and TOB are parallel along $v$, so
the modules provide measurements only in the $u$ direction.
For TID and TEC modules, where the strips are not parallel, the hit
reconstruction provides highly correlated two-dimensional measurements
in $u$ and $v$.
Their covariance matrix is
diagonalised and the corresponding transformation applied
to the derivatives and residuals as well. The measurement in the less precise
direction, after the diagonalisation, is not used for the alignment.

\subsection{Hierarchical and differential alignment by using equality constraints}
\label{subsec:hierarchDiffAlignment_constraints}
The CMS tracker is built in a hierarchical way from mechanical substructures,
\eg three BPIX half-layers form each of the two BPIX half-shells.
To treat translations and rotations of these substructures as a whole,
six alignment parameters $\mathbf{\Delta p}_{l}$ for each of the considered
substructures can be introduced.
The derivatives of the track prediction with respect to these parameters,
$\rd f_{u\!/\!v}/\rd\mathbf{\Delta p}_{l}$, are obtained from the six
translational and rotational parameters of the hit sensor
$\mathbf{\Delta p}_{s}$ 
by coordinate transformation with the chain rule
\begin{equation}
  \frac{\rd f_{u\!/\!v}}{\rd\mathbf{\Delta p}_{l}} =
  \frac{\rd\mathbf{\Delta p}_{s}}{\rd\mathbf{\Delta p}_{l}} \cdot
  \frac{\rd f_{u\!/\!v}}{\rd\mathbf{\Delta p}_{s}},
\end{equation}
where $\frac{\rd\mathbf{\Delta p}_{s}}{\rd\mathbf{\Delta p}_{l}}$
is the $6\times 6H$ Jacobian matrix expressing the effect of translations and
rotations of the large structure on the position
of the sensor.

These large-substructure parameters are useful in two different cases.
If the track sample is too small for the determination of the large number of alignment parameters at module level,
the alignment can be restricted to the much smaller set of parameters of these substructures.
In addition they can be used in a hierarchical alignment approach,
simultaneously with the alignment parameters of the sensors. This has the
advantage that coherent misplacements of large structures in directions of the
non-sensitive coordinate $v$ of strip sensors can be taken into account.

This hierarchical approach introduces redundant degrees of freedom since
movements of the large structures can be expressed either by their alignment
parameters or by combinations of the parameters of their components.
These degrees of freedom 
are eliminated by using linear equality constraints. In general,
such constraints can be formulated as
\begin{equation}
  \sum_i c_i \, {\Delta p}_i = s,
\end{equation}
where the index $i$ runs over all the alignment parameters.
In \mptwo these constraints are implemented by extending the matrix
equation~(\ref{eq:reducedNormal}) by means of Lagrangian multipliers.
In the hierarchical approach,
for each parameter $\Delta p_l$ of the larger structure one constraint
with $s=0$ has to be applied and then all constraints for one large structure
form a matrix equation,
\begin{equation}
  \sum_i^{\text{components}}
  \left[\frac{\rd\mathbf{\Delta p}_{s,i}}{\rd\mathbf{\Delta p}_{l}}\right]^{-1}
  \cdot \mathbf{\Delta p}_{i}
  = \mathbf{0},
\end{equation}
where $\mathbf{\Delta p}_{s,i}$ are the shift and rotation parameters of
component $i$ of the large substructure.
Similarly, the technique of equality constraints is used to fix the six
undefined overall shifts
and rotations of the complete tracker.

The concept of ``differential alignment'' means that in one alignment step
some parameters are treated as time-dependent while the majority of the
parameters stays time-independent. 
Time dependence is achieved by replacing an alignment parameter
$\Delta p_t$ in the linearised form of equation~(\ref{eq:globalChi2})
by several parameters, each to be used for one period of time only.
This method allows the use of the full statistical power of the whole
dataset for the determination of parameters that are stable with time,
without neglecting the time dependence of others.
This is especially useful in conjunction with a hierarchical alignment:
the parameters of larger structures can vary with time, but the sensors
therein are kept stable relative to their large structure.

\subsection{Weak modes}
\label{subsec:weakmodes}
\label{sec:resonanceTreatment}
A major difficulty of track based alignment arises if the
matrix~$\mathbf{C'}$ in equation~(\ref{eq:reducedNormal}) is
ill-con\-di\-tioned,
\ie singular or numerically close to singular. This can result from linear
combinations of the alignment parameters that do not (or only slightly) change
the track-hit residuals and thus the overall
$\chi^2(\mathbf{\Delta p}, \mathbf{\Delta q})$ in
equation~(\ref{eq:globalChi2}), after linearisation of the track model $f_{ij}$.
These linear combinations are called ``weak modes'' since the amplitudes of their
contributions to the solution are either not determinable or only barely so.

Weak modes 
can emerge if certain coherent changes of alignment parameters
$\mathbf{\Delta p}$ can be compensated by changes of the track parameters
$\mathbf{\Delta q}$. 
The simplest example is an overall shift of the tracker that would be
compensated by changes of the impact parameters of the tracks. For that reason
the overall shift has to be fixed by using constraints as mentioned above.
Other weak modes discussed below influence especially the transverse
momenta of the tracks.
A specific problem is that even very small biases in the track model $f_{ij}$
can lead to a significant distortion of the tracker if a 
linear
combination of the alignment parameters is not well determined by the data
used in equation~(\ref{eq:globalChi2}).
As a result, weak modes contribute significantly to the systematic
uncertainty of kinematic properties determined from the track fit.

The range of possible weak modes depends largely on the geometry and
segmentation of the detector, the topology of the tracks used for
alignment, and on the alignment and track parameters.
The CMS tracker has a highly segmented detector geometry with a cylindrical
layout within a solenoidal magnetic field. If aligned only with tracks
passing through the beam line, the characteristic weak modes can be
classified in cylindrical coordinates, \ie by
module displacements $\Delta r$, $\Delta z$, and $\Delta \varphi$ as functions
of $r$, $z$, and $\varphi$ \cite{ALI:Babar}. 
To control these weak modes it is crucial to
include additional information in equation~(\ref{eq:globalChi2}), \eg
by combining
track sets of different topological variety and
different physics constraints by means of
\begin{itemize}
\item cosmic ray tracks that break the cylindrical symmetry,
\item straight tracks without curvature, recorded when the magnetic field
  is off,
\item knowledge about the production vertex of tracks,
\item knowledge about the invariant mass of a resonance
  whose decay products are observed as tracks.
\end{itemize}

Earlier alignment studies~\cite{ALI:TkAlCraft08} have shown that the
usage of cosmic ray tracks is quite effective in controlling several classes of
weak modes.
However, for some types of coherent deformations
of the tracker the sensitivity of an alignment based on cosmic ray tracks
is limited. A prominent example biasing the track curvature
$\kappa \propto \frac{q}{\pt}$ (with $q$ being the track charge) is a
\textit{twist} deformation of the tracker, in which the modules are moved coherently in
$\varphi$ by an amount directly proportional to their longitudinal
position ($\Delta\varphi = \tau \cdot z$).
This has been studied extensively in \cite{ALI:draegerThesis2011}.
Other potential weak modes
are the off-centring of
the barrel layers and endcap rings (\textit{sagitta}), described by
$(\Delta x, \Delta y) = \sigma \cdot r \cdot (\sin\varphi_\sigma, \cos\varphi_\sigma)$,
and a \textit{skew}, parameterised as
$\Delta z = \omega \cdot \sin(\varphi + \varphi_\omega)$.
Here $\sigma$ and $\omega$ denote the amplitudes of the sagitta and
skew weak modes,
whereas $\varphi_\sigma$ and $\varphi_\omega$ are their azimuthal phases.

As a measure against weak modes that influence the track momenta, such as a twist
deformation, information on the mass of a resonance decaying into two
charged particles is included in the alignment fit with the following
method. A common parameterisation for the two trajectories of the
particles produced in the decay is defined as in~\cite{TBDparams_2007_widl}. Instead of
$2\cdot 5$ parameters (plus those accounting for multiple
scattering), the nine common parameters are the position of the decay
vertex, the momentum of the resonance candidate, two angles defining
the direction of the decay products in the rest-frame of the
resonance, and the mass of the resonance.  The mean mass of the resonance
is added as a virtual measurement with an uncertainty equal to
the standard deviation of its invariant 
mass distribution.
Mean and standard deviation are estimated from the distribution of the invariant
mass in simulated decays, calculated from the decay particles after final-state
radiation.
In the sum on
the right hand side of equation~(\ref{eq:globalChi2}), the two
individual tracks are replaced by the common fit object.  With the
broken lines parameterisation the corresponding $\mathbf{C}_j$ has the
border size $b=9$.
This approach to include resonance mass information in the alignment fit
implies an implementation of a vertex constraint as well,
since the coordinates of the decay vertex are parameters of the combined
fit object and thus force the tracks to a common vertex.

The dependence of the reconstructed resonance mass $M$ on the size $\tau$
of a twist deformation can be shown to follow
\begin{equation}
  \frac{\partial M^2}{\partial \tau} =
  \left(
  \frac{M^2}{p^+}\frac{\partial p^+}{\partial\tau} +
  \frac{M^2}{p^-}\frac{\partial p^-}{\partial\tau}
  \right) =
  \frac{2M^2}{B_z} \left(p_z^+ - p_z^-\right).
  \label{eq:dmass2_twist}
\end{equation}
Here $B_z$ denotes the strength of the solenoidal magnetic field along
the $z$-axis, $p^+$ ($p^-$) and $p^+_z$ ($p^-_z$) are the momentum and
its longitudinal component of the positively (negatively) charged
particle, respectively.  Equation~(\ref{eq:dmass2_twist}) shows that the inclusion of a
heavy resonance such as the \Z boson in the alignment procedure is
more effective for controlling the twist than the \JPsi and \PgU\
quarkonia, since at the LHC the decay products of the latter are
usually boosted within a narrow cone, and the difference of their
longitudinal momenta is small. The decay channel of \Z to muons is
particularly useful because the high-\pt muons are measured precisely
and with high efficiency by the CMS detector. The properties of the
\Z boson are predicted by the Standard Model and have been
characterised experimentally very well at the LHC
\cite{CMS:ZBoson1,CMS:ZBoson2} and in past experiments
\cite{LEP:ZBoson1}. This allows the muonic decay of the \Z boson to be used as a
standard reference
to improve the ability of the alignment procedure to resolve systematic
distortions, and to verify the absence of any bias
on the track reconstruction.

Under certain conditions, equality constraints can be utilised against
a distortion in the starting geometry $\mathbf{p}_0$
induced by a weak mode in a previous alignment attempt.
The linear combination of the alignment parameters corresponding to the weak mode
and the amplitude of the distortion in the starting geometry have to be known.
In this case, a constraint used in a further alignment step can
remove the distortion, even if the data used in the alignment cannot determine
the amplitude.
If, for example,  each aligned object $i$ is, compared to its true position,
misplaced in $\varphi$
according to a twist $\tau$ with reference point $z_0$
($\Delta\varphi_i = \tau\cdot (z_i - z_0)$),
this constraint takes the form $\sum_i\sum_j
\frac{\partial \Delta\varphi_i}{\partial
\Delta p_{ij}} \frac{(z_i - z_0)}{\sum_k (z_k-z_0)^2} \, \Delta p_{ij}
= -\tau$, where the sums over $i$ and $k$ include the aligned objects
and the sum over $j$ includes their active alignment parameters $\Delta
p_{ij}$.

\subsection{Computing optimisation}
\label{sec:ComputingOptimisations}
The \mptwo program proceeds in a two-step approach. First, the standard
CMS software environment~\cite{CMS:TDRvol1}
is used to produce binary files containing the residuals $m_{ij}-f_{ij}$,
their dependence on the parameters $\mathbf{\Delta p}$ and $\mathbf{\Delta q}$
of the linearised track model,
the uncertainties $\sigma_{ij}$,
and labels identifying the fit parameters. Second, these binary files are
read by an experiment-independent program that sets up
equation~(\ref{eq:reducedNormal}), extends it to incorporate the Lagrangian
multipliers to implement constraints, and solves it, \eg by the iterative
\minres algorithm~\cite{minres}. 
In contrast to other fast algorithms for solving large matrix
equations,
\minres does not require a positive definite matrix, and
because of the Lagrangian multipliers $\mathbf{C'}$ is indefinite.
Since the convergence speed of \minres depends on the eigenvalue spectrum of the matrix
$\mathbf{C}'$, preconditioning is used by multiplying
equation~(\ref{eq:reducedNormal}) by the inverse of the diagonal of the matrix.
The elements of the symmetric matrix $\mathbf{C}'$ in general require storage
in double precision while they are summed.
For the 200\,000 alignment parameters
used in this study, this would require 160\unit{GB} of RAM.
Although the matrix is rather sparse and only non-zero elements are stored,
the reduction is not sufficient.
High alignment precision also requires the use of many millions
of tracks of different topologies that are fitted several times
within \mptwo, leading to
a significant contribution to the CPU time.
To cope with the needs of the CMS tracker alignment described in this article,
the \mptwo program has been further developed, in particular to reduce the
computer memory needs, to enlarge the number of alignment parameters
beyond what was used 
in~\cite{ALI:TkAlCraft08}, and to reduce the processing time.
Details are described in the following.

Since the non-zero matrix elements are usually close to each other in the matrix,
further reduction of memory needs is reached by bit-packed addressing of non-zero
blocks in a row.
In addition, some matrix elements sum contributions of only a few tracks,
\eg cosmic ray tracks from rare directions. For these elements, single
precision storage is sufficient.

Processing time is highly reduced in \mptwo by shared-memory parallelisation
by means of the Open Multi-Processing (\textsc{OpenMP}$^{\mbox{\scriptsize\textregistered}}$)
package~\cite{openMP}
for the most computing intensive parts like the product of the huge matrix
$\mathbf{C'}$ with a vector for \minres, the track fits for the calculation of
$\mathbf{\Delta q}_j$ and $\mathbf{C}^{-1}_j$, and the construction
of $\mathbf{C'}$ from those.
Furthermore, bordered band matrices $\mathbf{C}_j$
are automatically detected and root-free Cholesky decomposition
is applied subsequently (see section~\ref{sec:method_trackparametrisation}).

Reading data from local disk and memory access are further potential
bottlenecks. The band structure of $\mathbf{C}_j$ that is due to the GBL refit
and the approximations in the track model (see
section~\ref{sec:method_trackparametrisation}) also aim to alleviate the binary file size.
To further reduce the time needed for reading, \mptwo
reads compressed input and caches the information of many tracks
to reduce the number of disk accesses.


\section{Strategy of the internal alignment of the CMS tracker}
\label{Sec:Strategy}
In general, the tracker has been
sufficiently stable throughout 2011 to treat alignment parameters as
constant in time. The stability of large structures has been
checked as described in section~\ref{Sec:Stability}.
An exception to this stability is the pixel
detector whose movements have been carefully monitored
and are then treated as described below.  Validating the statistical
alignment precision by means of the methods of
section~\ref{Sec:StatAccuracy} shows no need to have a further time
dependent alignment at the single module parameter level.
Also, calibration
parameters 
influence the reconstructed position of a hit on a module.
These parameters account for the Lorentz drift of the charge carriers
in the silicon due to the magnetic field and for the inefficient collection
of charge generated near the back-plane of strip sensors if these are operated in
deconvolution mode.
Nevertheless, for 2011 data there is no need to integrate the determination
of calibration parameters into the alignment procedure. The hit
position effect of any Lorentz drift miscalibration is compensated by the
alignment corrections and as long as the Lorentz drift is stable with time,
the exact miscalibration has no influence on the statistical alignment precision.
Also the back-plane correction has only a very minor influence.
No significant degradation of the statistical alignment precision with
time has been observed.

Given this stability, the 2011 alignment strategy of the CMS tracker consists
of two steps;
both apply the techniques and tools described in section~\ref{Sec:Method}.
The first step uses data collected in 2011 up to the
end of June, corresponding to an integrated luminosity
of about 1\fbinv. This step is based on the full
exploitation of different track topologies, making use of resonance
mass and vertex information. The details are described in the rest of this
section.
The second step treats the four relevant movements of the pixel detector
after the end of June, identified with the methods of section~\ref{sec:monitorPixel}.
Six alignment parameters for each BPIX layer and FPIX half-disk are
redetermined by a stand-alone
alignment procedure, keeping their internal structures unchanged and the
positions of the strip modules constant.

Tracks from several data sets are used simultaneously in the alignment
procedure.  The following selection criteria are
applied:
\begin {itemize}

\item \textbf{ Isolated muons}: \textit{Global muons}~\cite{Muo:MUO-Performance}
  are reconstructed in both the tracker and the muon
  system. They 
  are selected if their number of hits $N_{\text{hit}}$ in the tracker
  exceeds nine (at least one of which is in the pixel detector,
  $N_{\text{hit}}(\text{pixel}) \ge 1$),
  their momenta $p$ are above 8\GeVc (in order to minimise the effects of
  Multiple Coulomb scattering), 
  and their transverse momenta \pt are above 5\GeVc. 
  Their
  distances $\Delta R = \sqrt{(\Delta \varphi)^2 + (\Delta \eta)^2}$ from the axes
  of jets reconstructed in the calorimeter and fulfilling $\pt > 40\GeVc$
  have to be larger than 0.1. This class of
  events is populated mainly by muons belonging to leptonic $W$-boson
  decays and about 15 million of these tracks are used for the alignment.

\item \textbf{ Tracks from minimum bias events}: a minimum bias data
  sample is selected online with a combination of triggers varying
  with pileup conditions, i.e. the mean number of additional
  collisions, overlapping to the primary one, within the same bunch
  crossing (on average 9.5 for the whole 2011 data sample).
  These triggers are based, for example, on pick-up signals indicating
  the crossing of two filled proton bunches, 
  signals from the Beam Scintillator Counters~\cite{CMS:2008zzk}, 
  or moderate requirements on hit and track
  multiplicity in the pixel detectors. The offline track selection
  requires $N_{\text{hit}} > 7$, $p > 8\GeVc$.  
  Three million of these tracks are used for alignment.
\item \textbf{ Muons from \Z-boson decays}: events passing any trigger filter requesting two
  muons reconstructed online are used for reconstructing \Z-boson candidates. Two muons with
  opposite charge must be identified as \textit{global muons} and fulfil the requirement
  $N_{\text{hit}} > 9$ ($N_{\text{hit}}(\text{pixel})$
  $\ge 1$).
  Their transverse momenta must exceed $\pt > 15\GeVc$
  and their distances to jets reconstructed in the calorimeter $\Delta R > 0.2$.
  The invariant mass of the reconstructed
  dimuon system must lie in the range
  $85.8 < M_{\mu^+\mu^-} < 95.8\GeVcc$, in order to obtain a
  pure sample of \Z-boson candidates.
  The total number of such muon pairs is 375\,000.

\item \textbf{ cosmic ray tracks}: cosmic ray events used in the alignment
  were recorded with the strip tracker operated both in peak and
  deconvolution modes.  Data in peak mode were recorded in a dedicated
  cosmic data taking period before the restart of the LHC operations
  in 2011 and during the beam-free times between successive LHC fills.
  In addition, cosmic ray data were taken in deconvolution mode both
  during and between LHC fills, making use of a dedicated trigger
  selecting cosmic ray tracks passing through the tracker barrel.  In
  total 3.6~million cosmic ray tracks with $p > 4\GeVc$ and
  $N_{\text{hit}} > 7$ are used, where about half of the sample
  has been collected with the strip tracker operating in peak mode, while the
  other half during operations in deconvolution mode.
\end{itemize}

For all the data sets, basic quality criteria are applied on the hits used in
the track fit and on the tracks themselves:
\begin{itemize}
\item the signal-over-noise ratio of the strip hits must be higher
  than 12 (18) when the strip tracker records data in deconvolution
  (peak) mode;
\item for pixel hits, the probability of the hit to match the
  expected shape of the charge cluster for the given track parameters~\cite{Trk:PixelTemplate}
  must be higher than $0.001$ ($0.01$)
  in the $u$ ($v$) direction;
\item for all hits, the angle between the track and the module surface
  must be larger than $10^{\circ}$ ($20^{\circ}$) for tracks from
  proton-proton collisions (cosmic rays) to avoid a region where
  the estimates of the hit position and uncertainty are less reliable;
\item to ensure a reliable determination of the polar track angle, $\theta$,
  tracks have to have at least two hits  in pixel or stereo strip modules;
\item tracks from proton-proton collisions have to satisfy the ``high-purity''
  criteria~\cite{CMS-PAS-TRK-10-001}
  of the CMS track reconstruction code;
\item in the final track fit within \mptwo, tracks are rejected if their
  \chisq value is larger than the 99.87\% quantile
  (corresponding to three standard deviations) of the \chisq
  distribution for the number of degrees of freedom $N_{\mathrm{dof}}$ of the track.
\end{itemize}

The tracker geometry, as determined by the alignment with the 2010
data~\cite{ALI:draegerThesis2011}, is the starting point of the 2011
alignment procedure.  In general, for each sensor all nine parameters
are included in the alignment procedure.  Exceptions are the $v$
coordinate for strip sensors since it is orthogonal to the measurement
direction, and the surface parameterisation parameters $w_{10}$,
$w_{01}$, $w_{20}$, $w_{11}$, $w_{02}$ for the FPIX modules.  The
latter exception is due to their small size and smaller sensitivity
compared to the other subdetectors, caused by the smaller spread of
track angles with respect to the module surface.

The hierarchical alignment approach discussed in
section~\ref{subsec:hierarchDiffAlignment_constraints} is utilised by
introducing parameters for shifts and rotations of
half-barrels and end-caps of the strip tracker and of the BPIX
layers and FPIX half-disks. For the parameters of the BPIX layers and FPIX
half-disks the differential alignment is used as well. The need for nine
time periods (including one for the cosmic ray data before the LHC start)
has been identified with the validation procedure of section~\ref{sec:monitorPixel}.
The parameters for the six degrees of freedom of each of the two TOB
half-barrels are constrained to have opposite sign,
fixing the overall reference system.

Three approaches have been investigated to overcome the twist weak mode
introduced in section~\ref{subsec:weakmodes}.
The first uses tracks from cosmic rays, recorded in 2010 when the magnetic
field was off. This successfully controls the twist, but no equivalent
data were available in 2011.
Second, the twist has been measured in the starting geometry with the method
of section~\ref{subsec:eoverp}. An equality constraint has been introduced
to compensate for it. While this method controls the twist, it does not reduce
the dependence of the muon kinematics on the azimuthal angle
seen in sections~\ref{subsec:zmumu}
and \ref{subsec:eoverp}.
Therefore the final alignment strategy is based on the muons from \Z-boson decays
to include mass information and vertex
constraints in the alignment procedure as described in
section~\ref{sec:resonanceTreatment}, with a virtual mass measurement
of $M_{\mu^+\mu^-}=90.86 \pm 1.86\GeVcc$.

In total, more than 200\,000 alignment parameters are determined in
the common fit, by using 138 constraints.
To perform this fit, 
246 parallel 
jobs produce the compressed input files
containing residuals, uncertainties,
and derivatives for the linearised track model for the \mptwo program.
The total size of these files is 46.5\unit{GB}.
The matrix $\mathbf{C}'$ constructed from this by \mptwo contains 31\% non-zero
off-diagonal elements. With a compression
ratio of 40\% this fits well into an affordable 32\unit{GB} of memory.
The \minres\ algorithm has been run four times with increasingly tighter
rejection of bad tracks.
Since $\mathbf{C}'$ is not significantly changed by this rejection, it does not need to be
recalculated after the first iteration.
By means of eight threads on an
Intel$^{\mbox{\scriptsize \textregistered}}$ Xeon$^{\mbox{\scriptsize\textregistered}}$
L5520 processor with 2.27\unit{GHz}, the CPU usage was 44:30\unit{h} with a wall clock time of only 9:50\unit{h}.
This procedure has been repeated four times to treat effects from
non-linearity: iterating the procedure is particularly important for eliminating the twist weak mode.

The same alignment procedure as used on real data was run on a
sample of simulated tracks, prepared with the same admixture of track
samples as in the recorded data. The geometry obtained in this way is
characterised by a physics performance comparable to the one obtained with recorded data and
can be used for comparisons with simulated events. This geometry is
referred as the \textit{``realistic''} misalignment scenario.

\section{Monitoring of the large structures}\label{Sec:Stability}

A substantial fraction of the analyses in CMS use data reconstructed
immediately after its acquisition (\textit{prompt reconstruction}) for
obtaining preliminary sets of results. Therefore, it is important to
provide to the physics analyses the best possible geometry for use in
the prompt reconstruction, immediately correcting any possible
time-dependent large misalignment.
Specifically, the position of the large structures in the pixel detector is
relevant for the performance of b-tagging algorithms. As described in
\cite{CMS:BTV-12-001}, misalignment at the level of a few tens of
microns can seriously affect the b-tagging performance.

In order to obtain the best possible track reconstruction performance,
the tracker geometry is carefully monitored as a function of time, so
that corrections can be applied upon movements large enough to affect
the reconstruction significantly. The CMS software and reconstruction
framework accommodates time-dependent alignment and calibration
conditions by ``intervals of validity'' (IOV), which are periods during
which a specific set of constants retain the same values
\cite{CMS:TDRvol1}. While the alignment at the level of the single
modules needs data accumulated over substantial periods of time, the
stability of the position of the large structures can be controlled
with relatively small amounts of data or via a system of infrared
lasers. The short data acquisition times required by these monitoring
methods allow fast and frequent feedback to the alignment procedure.
A system of laser beams is able to monitor the positions of a
restricted number of modules in the silicon strip tracker. Movements
of large structures in the pixel tracker can be detected with high
precision with collision tracks by a statistical study of the \textit{primary-vertex residuals}, defined as the distance between the tracks
and the primary vertex at the point of closest approach of the tracks
to the vertex. All these techniques allow the monitoring of the
position of the large structures on a daily basis. This frequent
monitoring, together with the fast turn-around of the alignment with
\mptwo, allows, if needed, the correction of large movements
on the timescale of one day.

\subsection{Monitoring of the strip tracker geometry}

The CMS laser alignment system (LAS)~\cite{Wittmer:2007zz} provides a
source of alignment information independent of tracks. It is based on
40 near-infrared (1075\unit{nm}) laser beams passing through a
subset of the silicon sensors that are used also for the standard
track reconstruction (see figure \ref{fig:TrackerLayout}). The laser
optics are mounted on mechanical structures independent of those used
to support the tracker. With this limited number of laser beams one
can align large-scale structures such as the TOB, TIB, and both TECs. The
mechanical accuracy of LAS components limits the absolute precision of
this alignment method to $\sim$50\mum in comparison to the alignment
with tracks, which reaches better than 10\mum resolution (see
section \ref{Sec:StatAccuracy}), but the response time of the LAS is
at the level of only a few minutes.  Within this margin of accuracy,
the LAS measurement demonstrated very good stability of the strip
detector geometry over the whole 2011 running period. This observation
is confirmed by a dedicated set of alignments with tracks, where the
dataset was divided into different time periods. No significant
movements of the large structures of the silicon strip tracker were
found.

\subsection{Monitoring of the pixel detector geometry with tracks}
\label{sec:monitorPixel}
The large number of tracks produced in a pp collision allows precise
reconstruction of the interaction vertices
\cite{Trk:TRK-FirstColl}. The resolution of the reconstructed vertex
position is driven by the pixel detector since it is the sub-structure
that is closest to the interaction point and has the best hit
resolution.  The primary vertex residual method is based on the study
the distance between the track and the vertex, the latter
reconstructed without the track under scrutiny (\textit{unbiased
track-vertex residual}).  Events used in this analysis are selected
online with minimum bias triggers as mentioned in
section~\ref{Sec:Strategy}. The analysis uses only vertices with
distances from the nominal interaction point $\sqrt{\smash[b]{x_\text{vtx}^{2} +
y_\text{vtx}^{2}}}<2$\unit{cm} and $\abs{z_\text{vtx}} < 24$\unit{cm} in the transverse
and longitudinal direction, respectively. The fit of the vertex must
have at least 4 degrees of freedom.  For each of these vertices, the
impact parameters are measured for tracks with:
\begin{itemize}
\item more than six hits in the tracker, of which at least two are in the pixel detector,
\item at least one hit in the first layer of the BPIX or the first disk of the FPIX,
\item $\pt>1\GeVc$,
\item $\chi^{2}/N_\mathrm{dof}$ of the track smaller than 5.
\end{itemize}

The vertex position is recalculated excluding the track under scrutiny
from the track collection. A deterministic annealing clustering
algorithm \cite{Trk:DAVertexing} is used in order to make the method
robust against pileup, as in the default reconstruction
sequence.

\begin{figure*}[hbtp]
  \centering
    \includegraphics[width=0.45\textwidth]{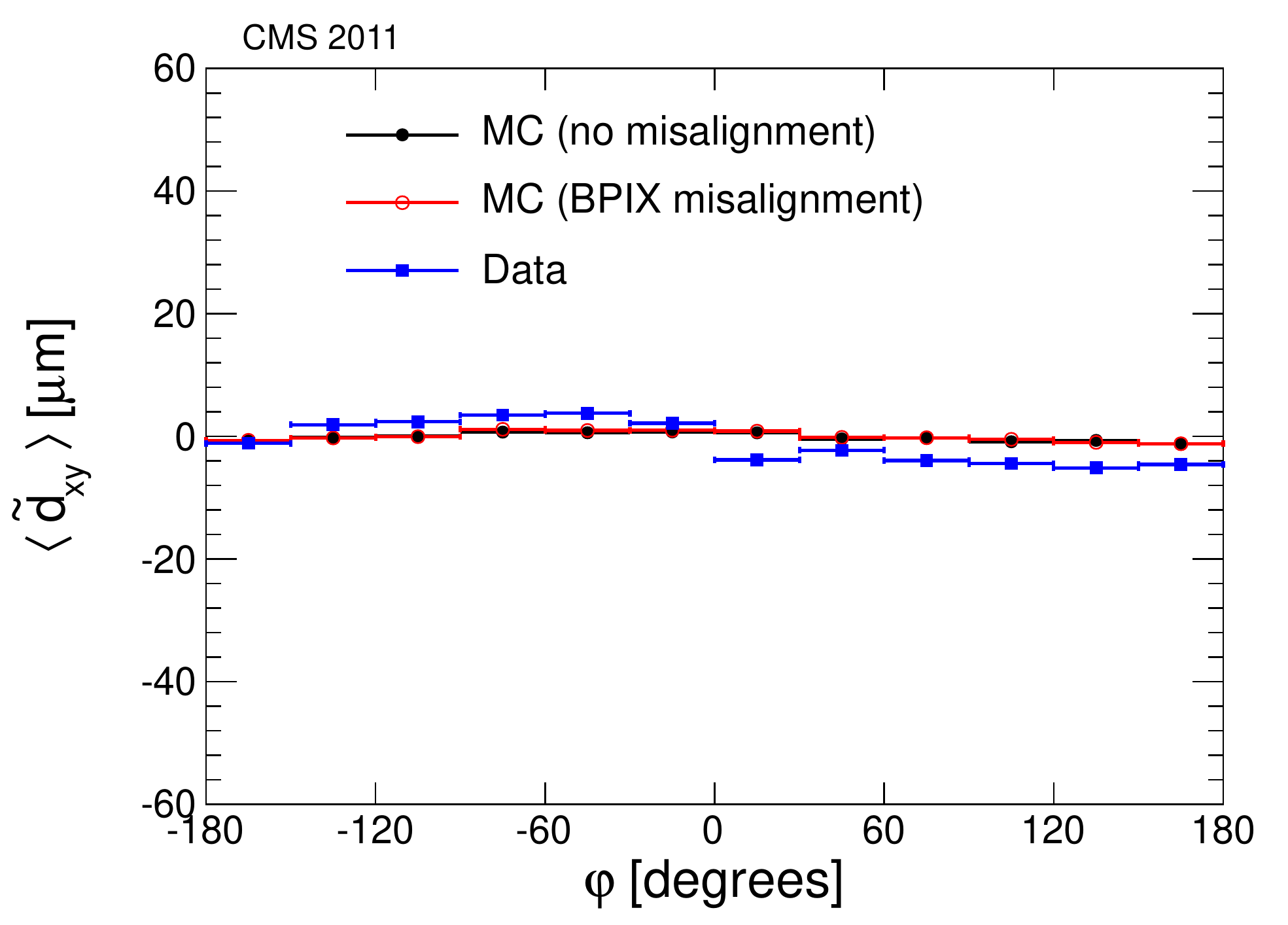}
    \includegraphics[width=0.45\textwidth]{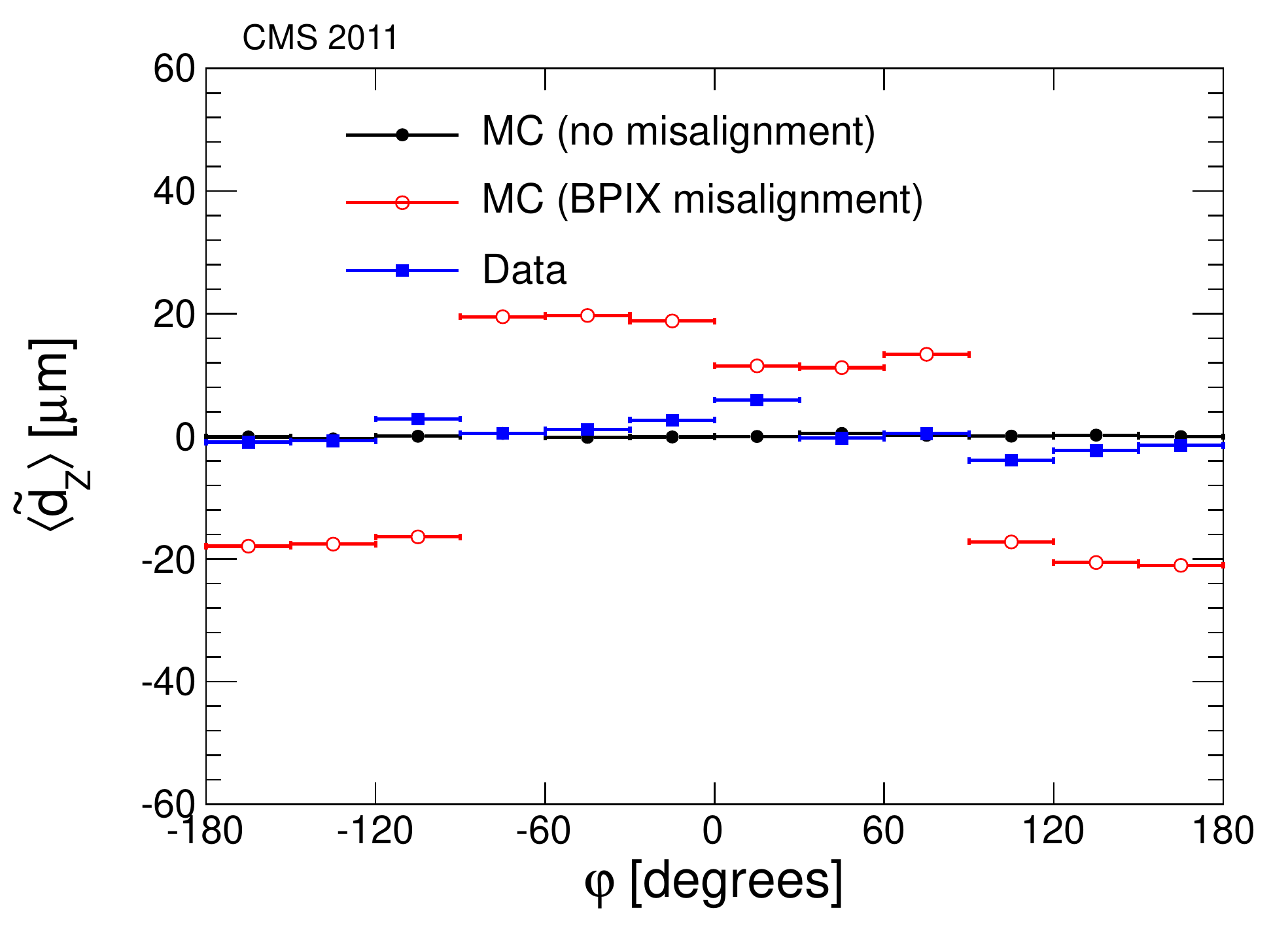}
    \caption{Means of the distributions of the unbiased transverse
      (left) and longitudinal (right) track-vertex residuals as a
      function of the azimuthal angle of the track. Blue squares show
      the distribution obtained from about ten thousand minimum bias events recorded
      in 2011. Full circles show the prediction by using a simulation with
      perfect alignment. Open red circles show the same prediction by using
      a
      geometry with the two BPIX
      half-shells shifted by 20\mum in opposite
      $z$-directions in the simulation.}
    \label{fig:PVValid_DzVsPhi}

\end{figure*}

\begin{figure*}[hbtp]
  \centering
    \includegraphics[width=0.5\textwidth]{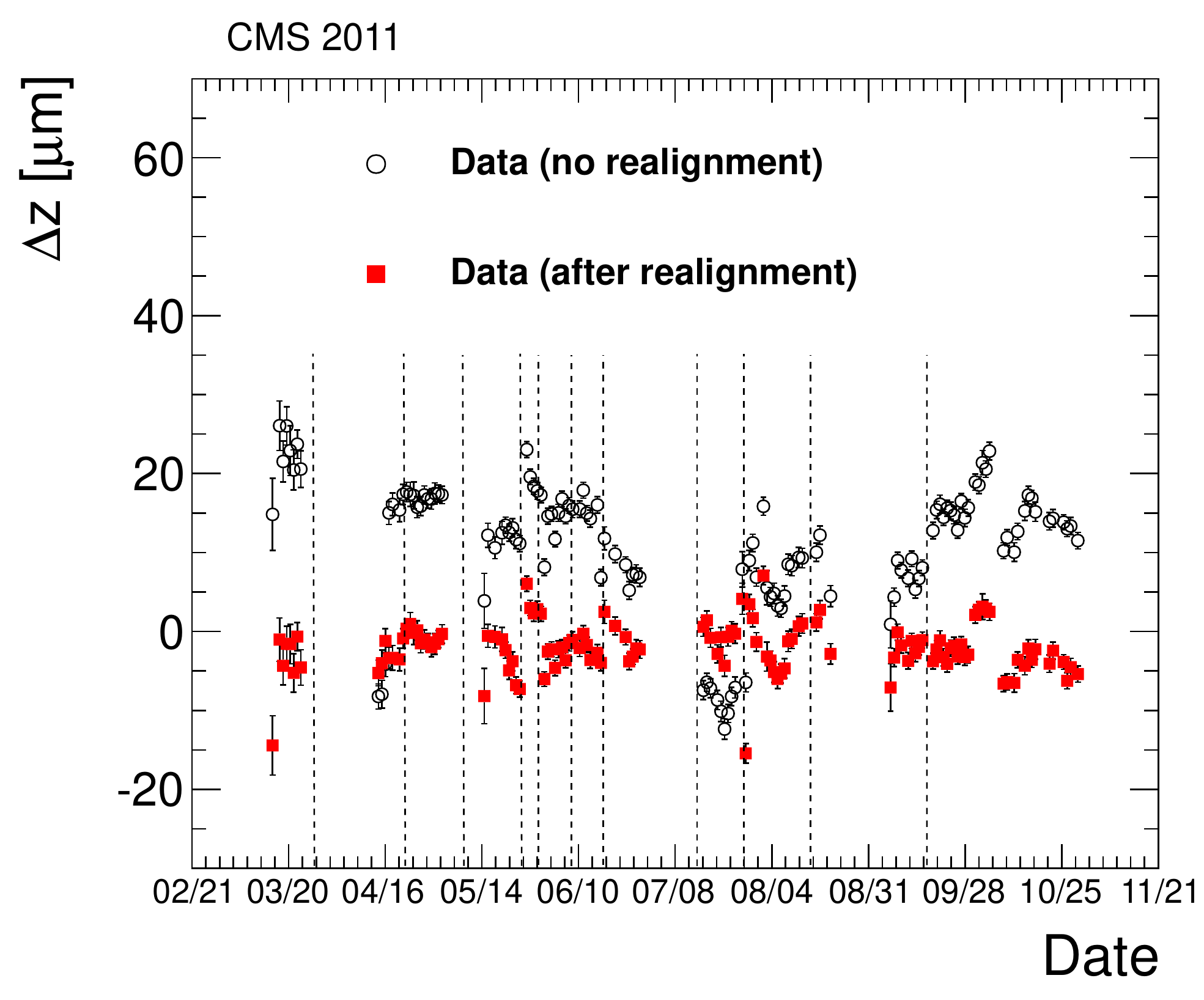}
    \caption{Daily evolution of the relative longitudinal shift
      between the two half-shells of the BPIX as measured with the
      primary-vertex residuals. The open circles show the shift
      observed by using prompt reconstruction data in
      2011. The same events were reconstructed again after the 2011
      alignment campaign, which accounts for the major changes in the
      positions of the half-shells (shown as filled squares). Dashed
      vertical lines indicate the chosen IOVs boundaries where a
      different alignment of the pixel layers has been performed.}
    \label{fig:PVValid_DzVsTime_2011}
\end{figure*}

The distributions of the unbiased track-vertex residuals in the
transverse plane, $\tilde{d}_{xy}$, and in the longitudinal direction,
$\tilde{d}_{z}$, are studied in bins of $\eta$ and $\varphi$ of the
track. Random misalignments of the modules affect only the resolution
of the unbiased track-vertex residual, increasing the width of the
distributions, but without biasing their mean. Systematic movements of
the modules will bias the distributions in a way that depends on the
nature and size of the misalignment and the $\eta$ and $\varphi$ of
the selected tracks. As an example, the dependence of the means of the
$\tilde{d}_{xy}$ and $\tilde{d}_{z}$ distributions as a function of
the azimuthal angle of the track is shown in figure
\ref{fig:PVValid_DzVsPhi}. The focus on the $\varphi$-dependence is
motivated by the design of the BPIX, which is divided into one
half-shell with modules at
$\varphi \in [-\pi/2, \pi/2]$
and another with modules at
$\varphi \in [\pi/2, \pi] \cup [-\pi,-\pi/2]$.  Small movements of the
two half-shells are mechanically allowed by the mechanical design of
the pixel detector. The observed movements have not been associated to
a specific cause, although thermal cycles executed on the pixel
detector increase the chances that such movements will happen. As an example,
the impact of a movement of one half-shell with respect to the other
in the longitudinal direction is shown by the open circles in figure
\ref{fig:PVValid_DzVsPhi} for a simulated sample of minimum bias
events. Such a movement is reflected in a very distinctive feature in
the dependence of the mean of the $\tilde{d}_{z}$ distribution as a
function of $\varphi$.  The size of the movement can be estimated as
the average bias in the two halves of the BPIX. The time dependence of
this quantity in the 2011 data is illustrated in figure
\ref{fig:PVValid_DzVsTime_2011}, which shows some discontinuities.
Studies carried out on simulated data show that the b-tagging
performance is visibly degraded in the case of uncorrected shifts with
amplitude $\abs{\Delta z} > 20\mum$ \cite{CMS:BTV-12-001}. For this
reason, IOVs with different alignments of the pixel layers are
conservatively defined according to the boundaries of periods with
steps of $\abs{\Delta z}$ larger than 10\mum.  The time-dependent
alignment parameters of BPIX layers and FPIX half-disks during the
first eight IOVs (until end of June 2011) were determined in a single
global fit.  Within each time interval, the positions of the modules
with respect to the structure were found not to need any further
correction. Because of this, the positions of the pixel layers and
half-disks were determined by a dedicated alignment procedure keeping
the other hierarchies of the geometry unchanged.  The aligned geometry
performs well over the entire data-taking period, reducing the
observed jumps in the expected way. Residual variations can be
attributed to small misalignments with negligible impact on physics
performance and to the resolution of the validation method itself.

\section{Statistical accuracy of the alignment}\label{Sec:StatAccuracy}
\begin{figure*}[btp]
  \centering
    \includegraphics[width=0.45\textwidth]{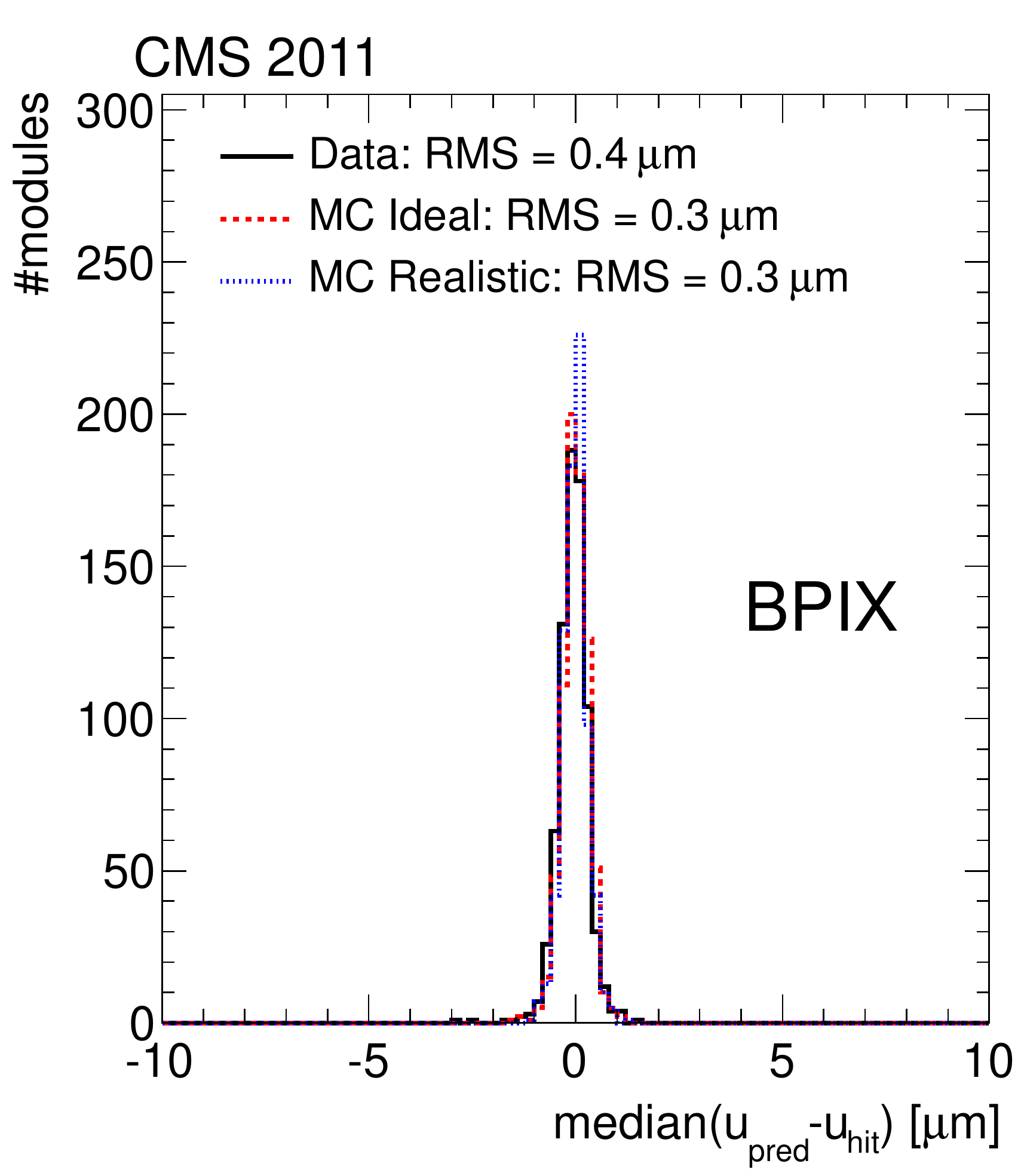}\hspace{1cm}\includegraphics[width=0.45\textwidth]{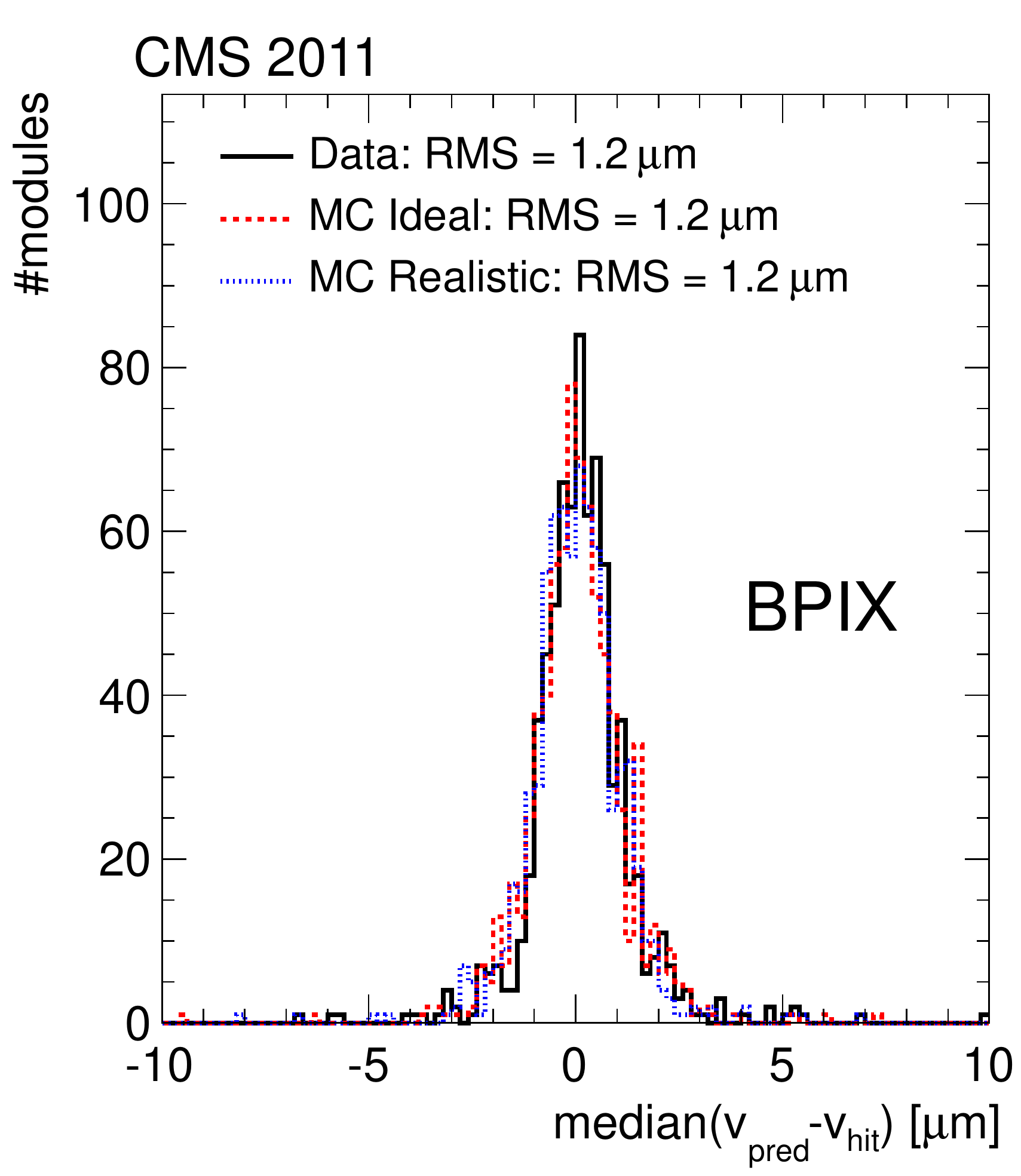}\\
    \includegraphics[width=0.45\textwidth]{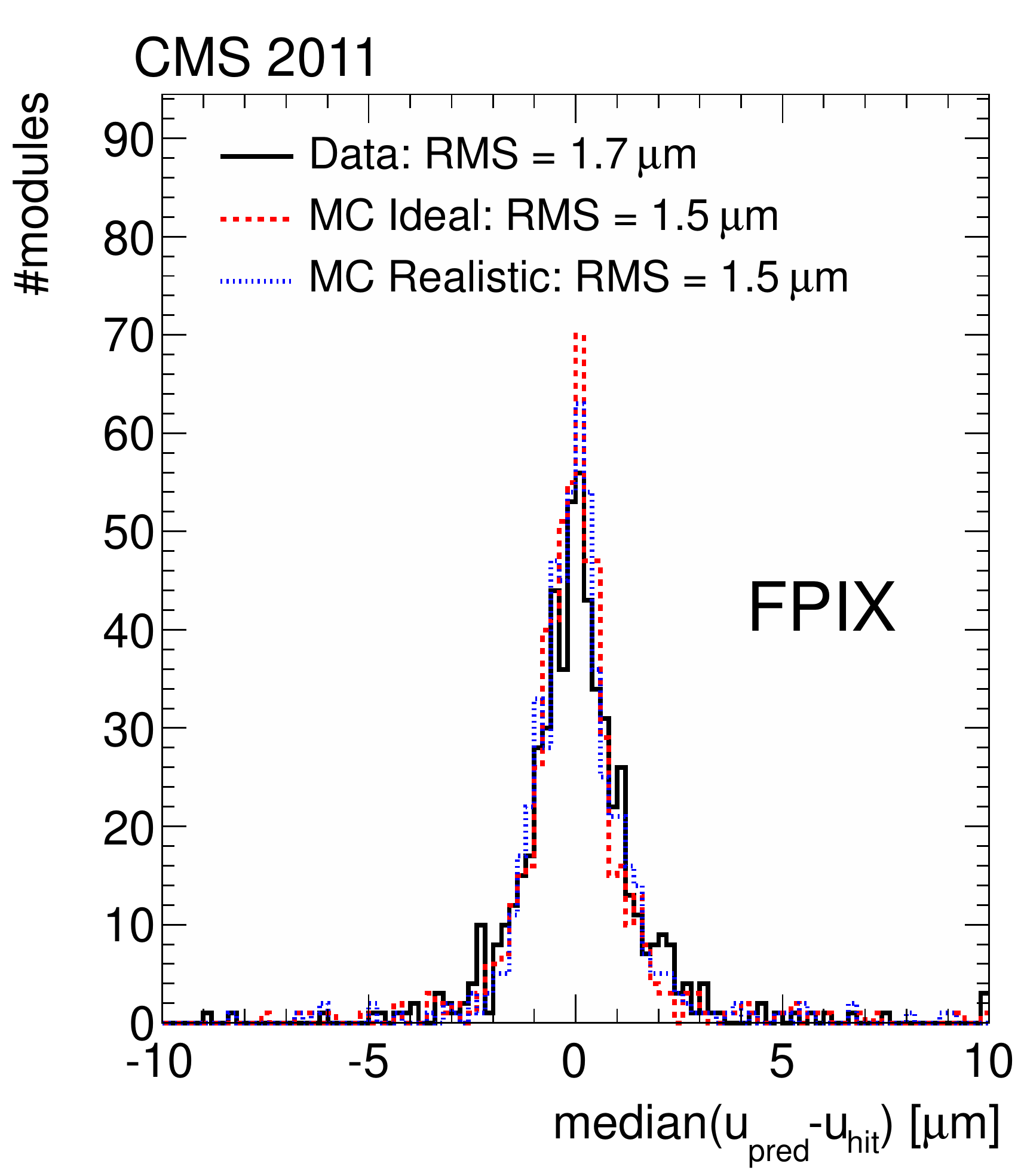}\hspace{1cm}\includegraphics[width=0.45\textwidth]{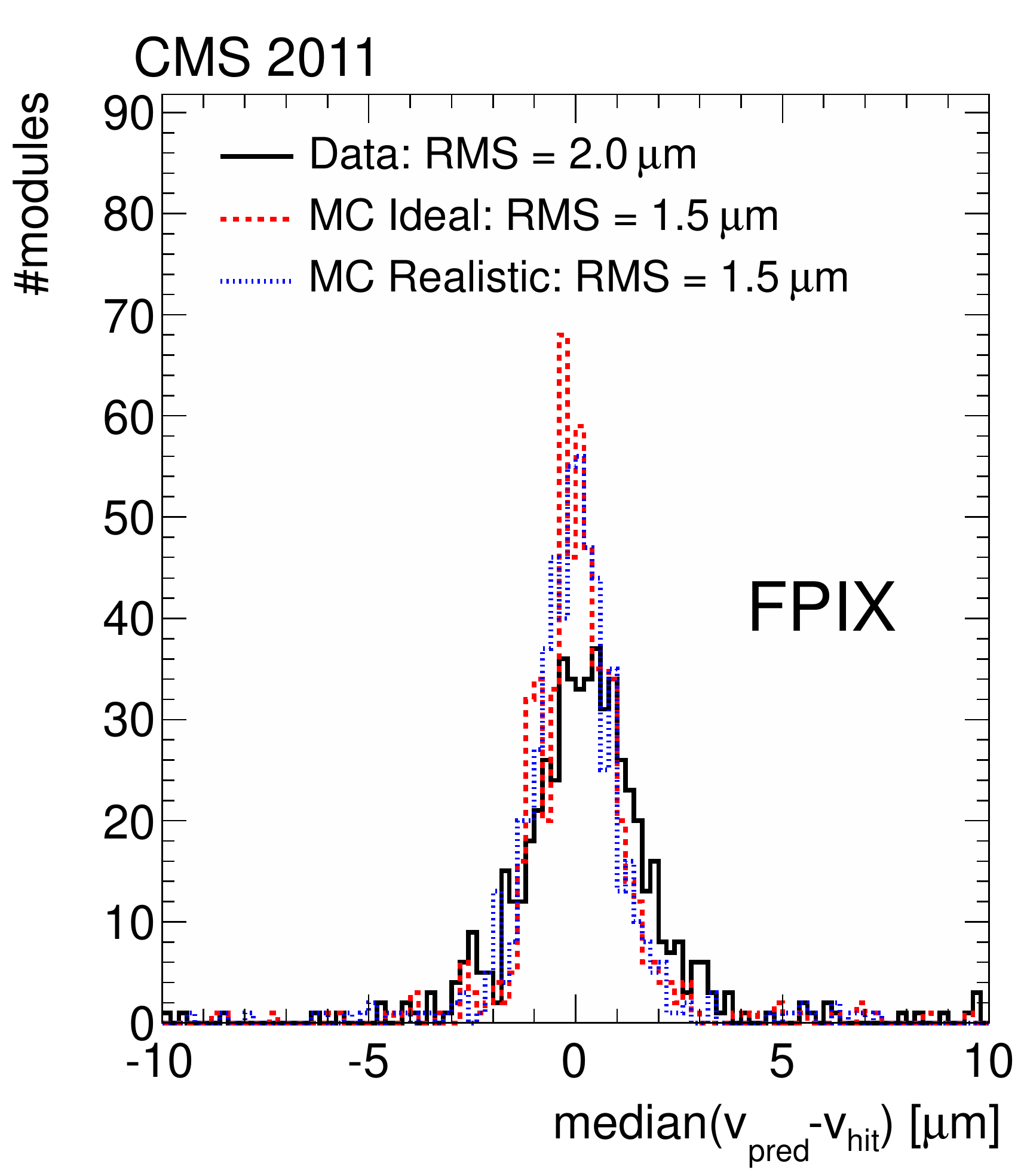}
    \caption{Distributions of the medians of the
    residuals, for the pixel tracker barrel (top) and endcap modules
    (bottom) in $u$ (left) and $v$ (right) coordinates. Shown in each
    case are the distributions after alignment with 2011 data (solid
    line), in comparison with simulations without any
    misalignment (dashed line) and with realistic misalignment (dotted line).}
    \label{fig:dmrTP}
\end{figure*}

\begin{figure*}[btp]
\centering
      \includegraphics[height=0.28\textheight]{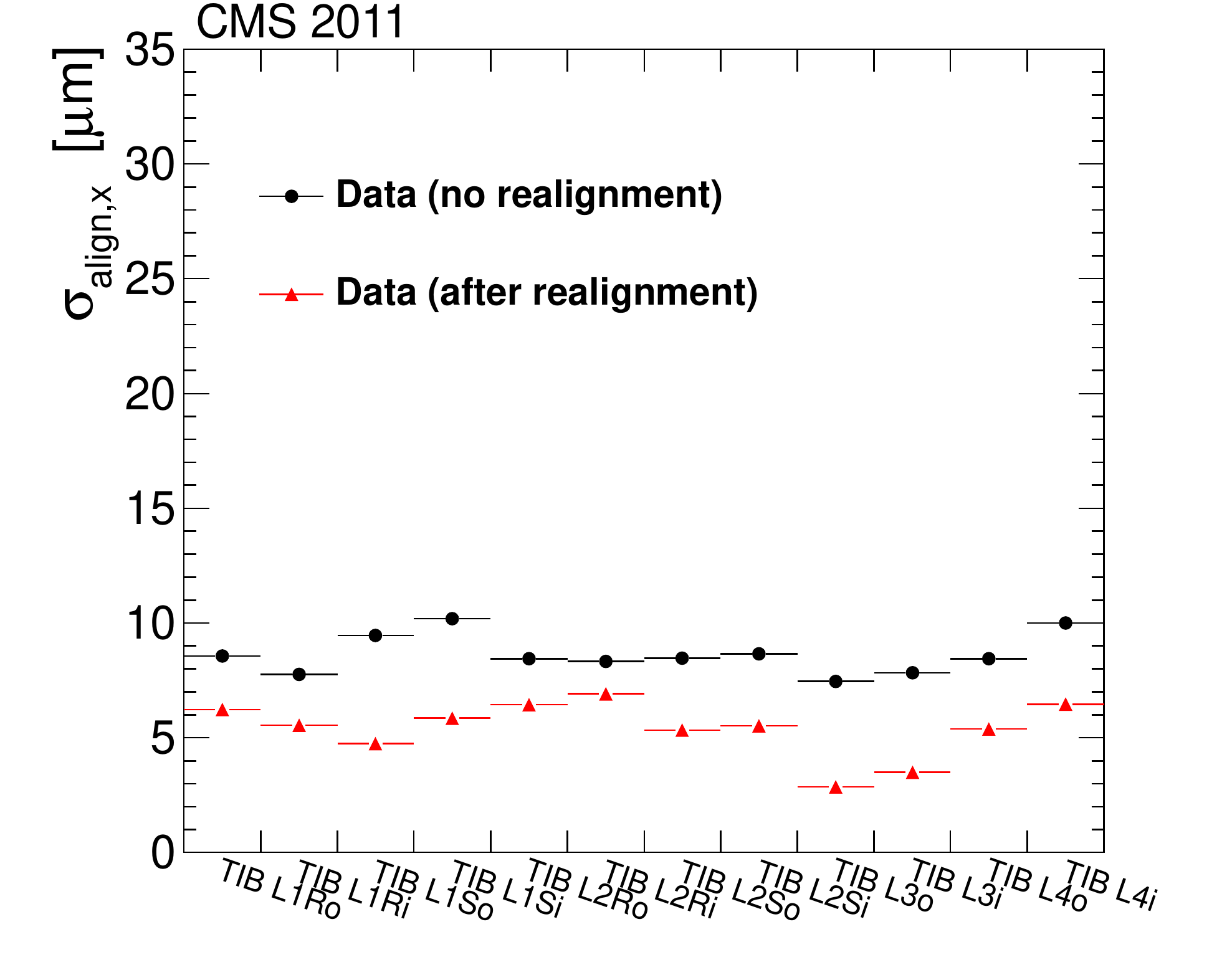}\includegraphics[height=0.28\textheight]{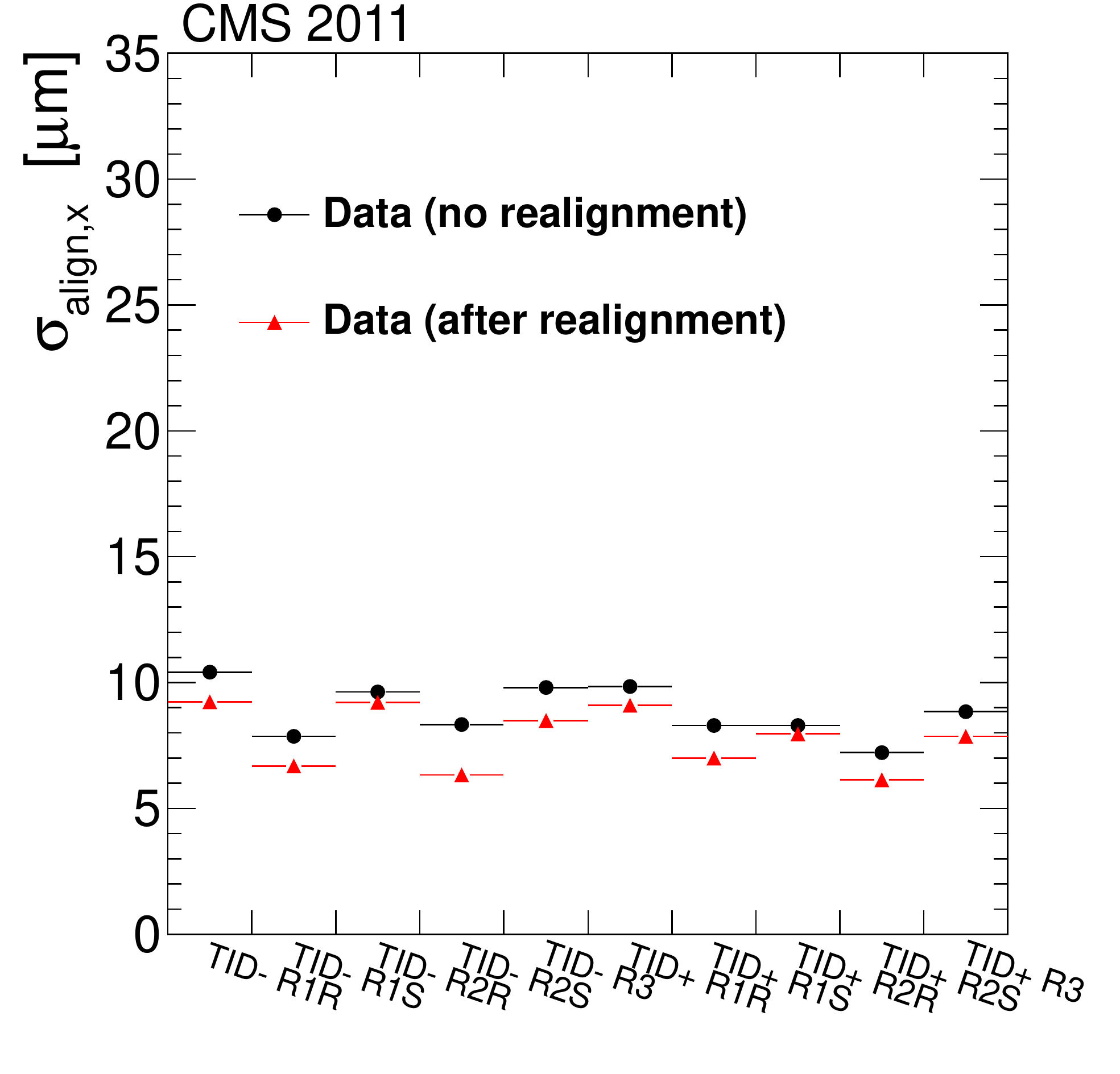}\\
      \includegraphics[height=0.28\textheight]{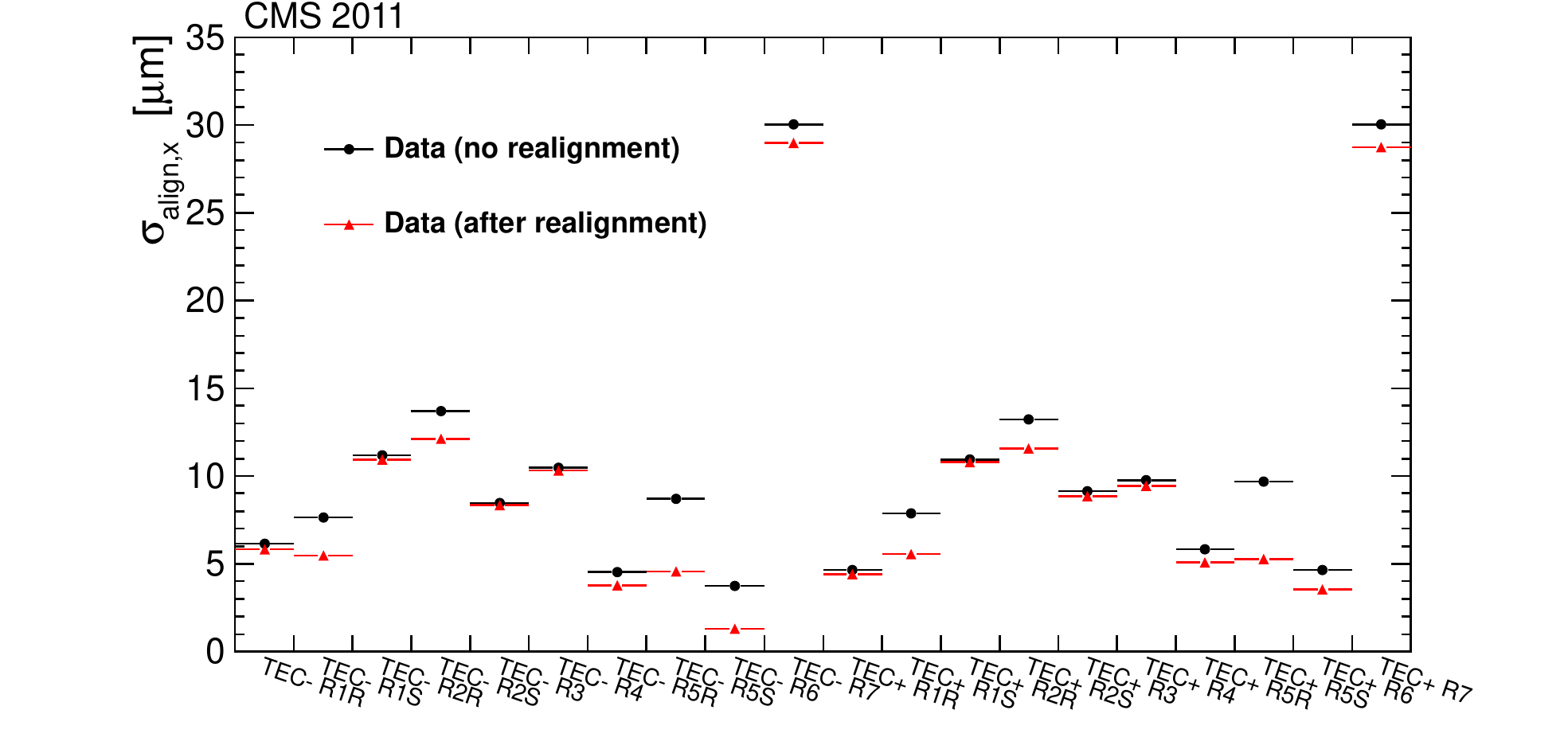}
    \caption{Alignment accuracy in the subdetectors TIB,
    TID and TEC, determined for each sensor group, with the normalised residuals method. The black dots show
    the alignment accuracy before the dedicated alignment with the
    2011 data, with the older alignment constants used in the prompt
    reconstruction, while the red triangles are obtained with the
    dedicated alignment applied in the reprocessing of the data.}
    \label{fig:APEResults}
\end{figure*}
A method for assessing the achieved statistical precision of the aligned
positions in the sensitive direction of the modules has been
successfully explored and adopted in the alignment of the CMS tracker
during commissioning with cosmic rays, described
in~\cite{ALI:TkAlCraft08}. The results from the validation are based
on isolated muon tracks with a transverse momentum of $\pt > 40$\GeVc
and at least ten hits in the tracker. The tracks are
refitted with the alignment constants under study. Hit residuals are
determined with respect to the track prediction, which is obtained
without using the hit in question to avoid any correlation between hit
and track. From the distribution of the unbiased hit
residuals in each module, the median
is taken and histogrammed for all modules in a detector subsystem. The
median is relatively robust against stochastic effects from multiple
scattering, and thus the distribution of the medians of residuals (DMR) is
a measure of the alignment accuracy. Only modules
with at least 30 entries in their distribution of residuals are
considered.

The addition of proton-proton collision events leads to a huge
increase of the number of tracks available for the alignment,
especially for the innermost parts of the tracker. Compared to the
alignment with cosmic rays alone~\cite{ALI:TkAlCraft08}, considerable
improvements are consequently observed in the pixel tracker,
especially in the endcaps.  The corresponding DMR are shown in the
figure~\ref{fig:dmrTP}, separately for the $u$ and $v$
coordinates; for both pixel tracker barrel (BPIX) and endcap (FPIX)
detectors, the RMS is well below 3\mum in both directions, compared to
about 13\mum for the endcaps in the cosmic ray-only alignment. These
numbers are identical or at most only slightly larger than those
obtained in simulation without any misalignment, which are below 2\mum and
thus far below the expected hit resolution. In the case of no
misalignment, the remaining DMR width is non-zero because of statistical
fluctuations reflecting the limited size of the track sample. Thus,
the DMR width of the no-misalignment case indicates the intrinsic
residual uncertainty of the DMR method itself. The remaining
uncertainty after alignment determined from recorded data is close to
the sensitivity limit of the DMR method. The DMR obtained with the
realistic misalignment scenario (see section~\ref{Sec:Strategy}) are
also shown in figure~\ref{fig:dmrTP}. The distributions are very close
to the ideal case.

The alignment accuracy of the strip detector is investigated in
smaller groups of sensors with a different method by using normalised
residuals~\cite{ALI:haukThesis2012}.  Each group consists of sensors
that are expected to have similar alignment accuracy. The distinction
is by layer (``L'') or ring (``R'') number, by longitudinal hemisphere
(``$+$'' and ``$-$'' for positive and negative $z$ coordinate,
respectively), and according to whether the surface of a barrel module
points inwards (``i''), \ie towards the beamline, or outwards
(``o''). The letter ``S'' indicates a stereo module. The method
applied here is based on the widths of the
distributions of normalised unbiased residuals of each sensor group. Since the misalignment
dilutes the apparent hit resolution, its degree can be derived from
the widening of these distributions of normalised residuals.
The residual resolution
is the square
root of the quadratic sum of the resolutions of the intrinsic hit reconstruction
and of the track prediction, excluding the hit
in question. The alignment uncertainty is added in quadrature to the
intrinsic hit resolution of the cluster and adjusted such that the
width of the distribution of normalised residuals matches the
ideal one, which is determined from simulated events.
In this way, misalignment in all degrees of freedom of the modules is
contributing to the measurement.

Since the
width of each distribution of normalised residuals is also influenced by the
alignment uncertainties of the surrounding detector areas, these
parameters are highly coupled, and iteration
is required. Each track is refitted with the current
estimate of all alignment uncertainties taken into account, and the
procedure is repeated. In each iteration, a damping factor of 0.6 is
applied to the correction
to mitigate oscillations. Convergence is achieved after 15 iterations.

The resulting alignment accuracy per sensor group, $\sigma_{\text{align, x}}$, is shown in
figure~\ref{fig:APEResults} for the TIB, TID and TEC subsystems. In
all cases, a significant improvement resulting from the alignment procedure is
observed. The alignment accuracy is between 3--8\mum for TIB,
between 6--10\mum for TID, and better than 13\mum for the TEC. The
only exception in ring 7 of the TEC is well understood;
it is due to a systematic misplacement of these sensors in the (almost
insensitive) radial direction by about 1.3\mm, which has been corrected
for in further
alignment procedures. For large parts of the TOB and the pixel detector, the
remaining misalignment cannot be distinguished from zero within the
systematic limitations of this method.

Overall, the statistical accuracy of the alignment is such that its
effect is small compared to the intrinsic measurement precision of the
sensors.  It should be noted, however, that quality estimators based
on track-residual distributions have little or no sensitivity to weak
modes; these will be addressed
in section~\ref{Sec:WeakModes}.

\section{Sensor and module shape parameters}\label{Sec:SensorShape}

\begin{figure}[tp]
  \centering
    \includegraphics[width=0.45\textwidth]{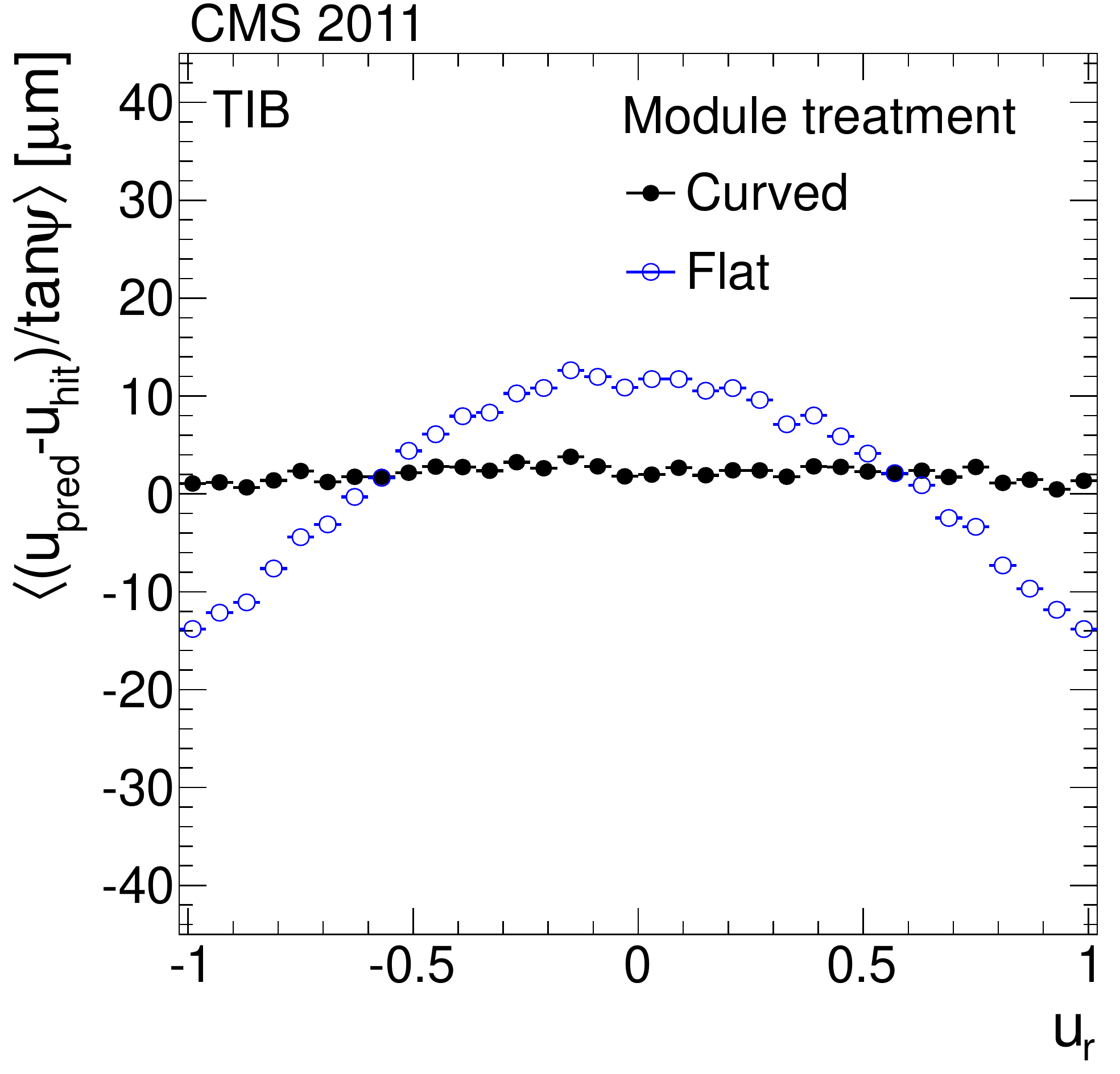}\hfill
    \includegraphics[width=0.45\textwidth]{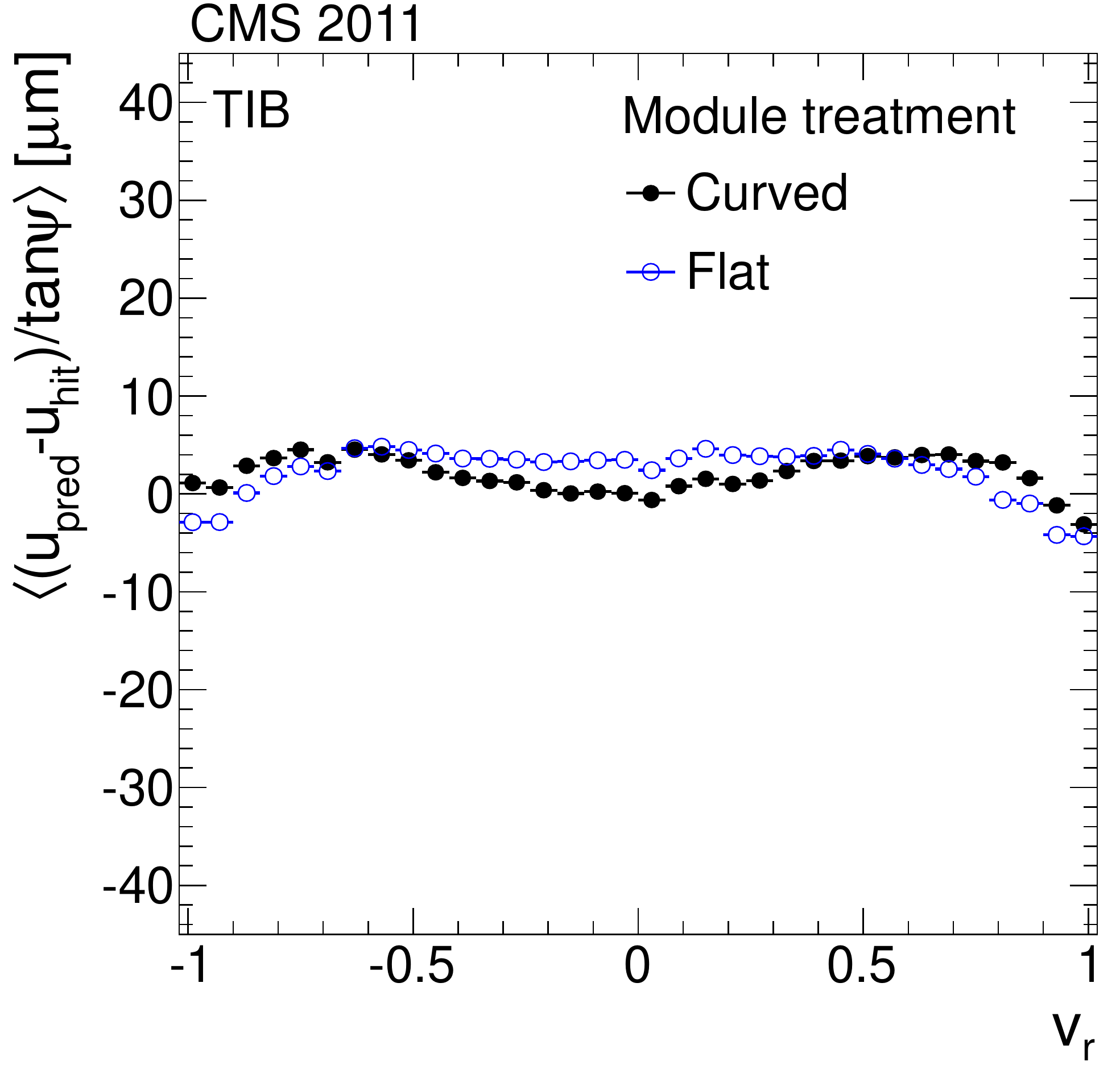}

    \includegraphics[width=0.45\textwidth]{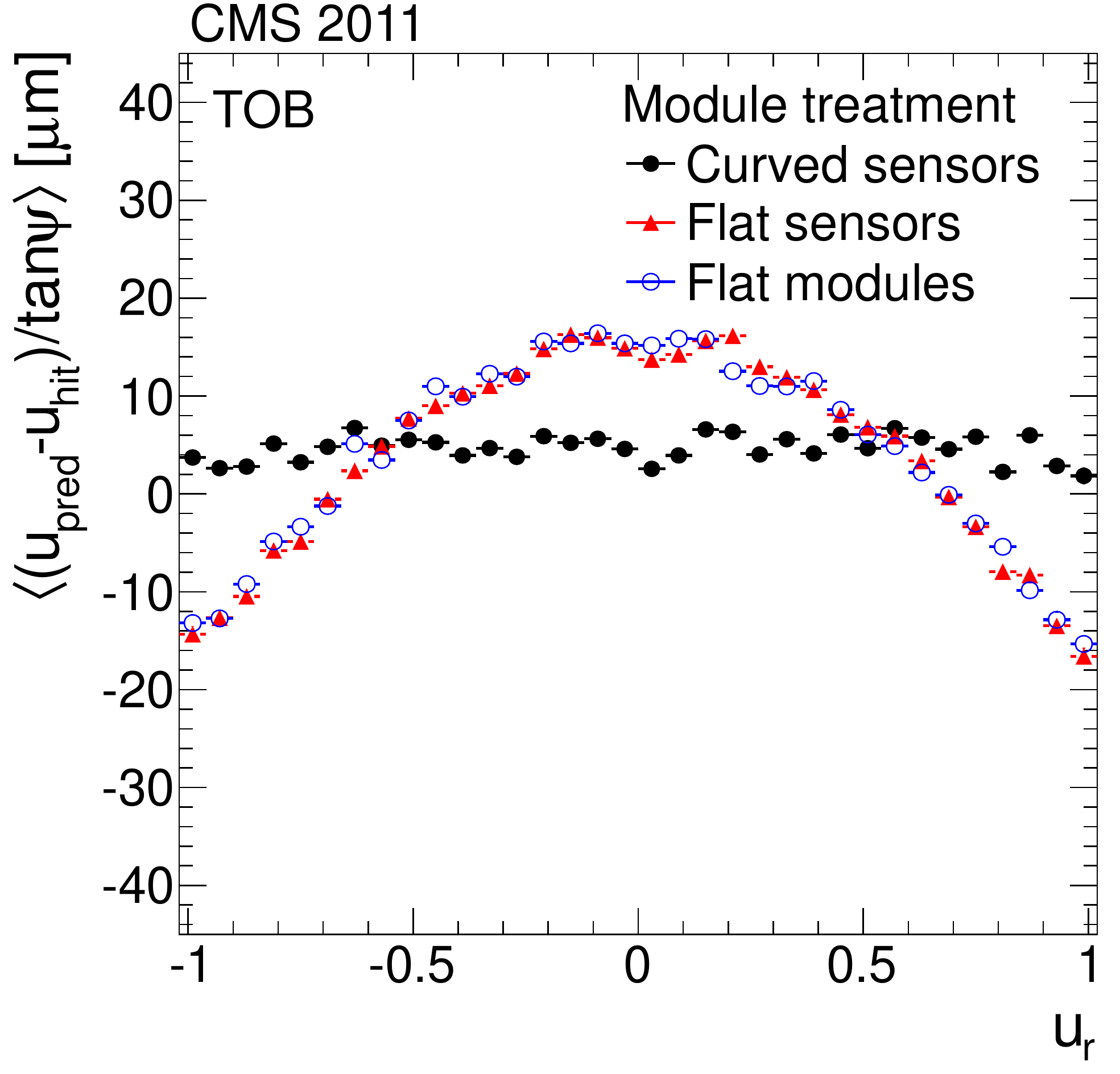}\hfill
    \includegraphics[width=0.45\textwidth]{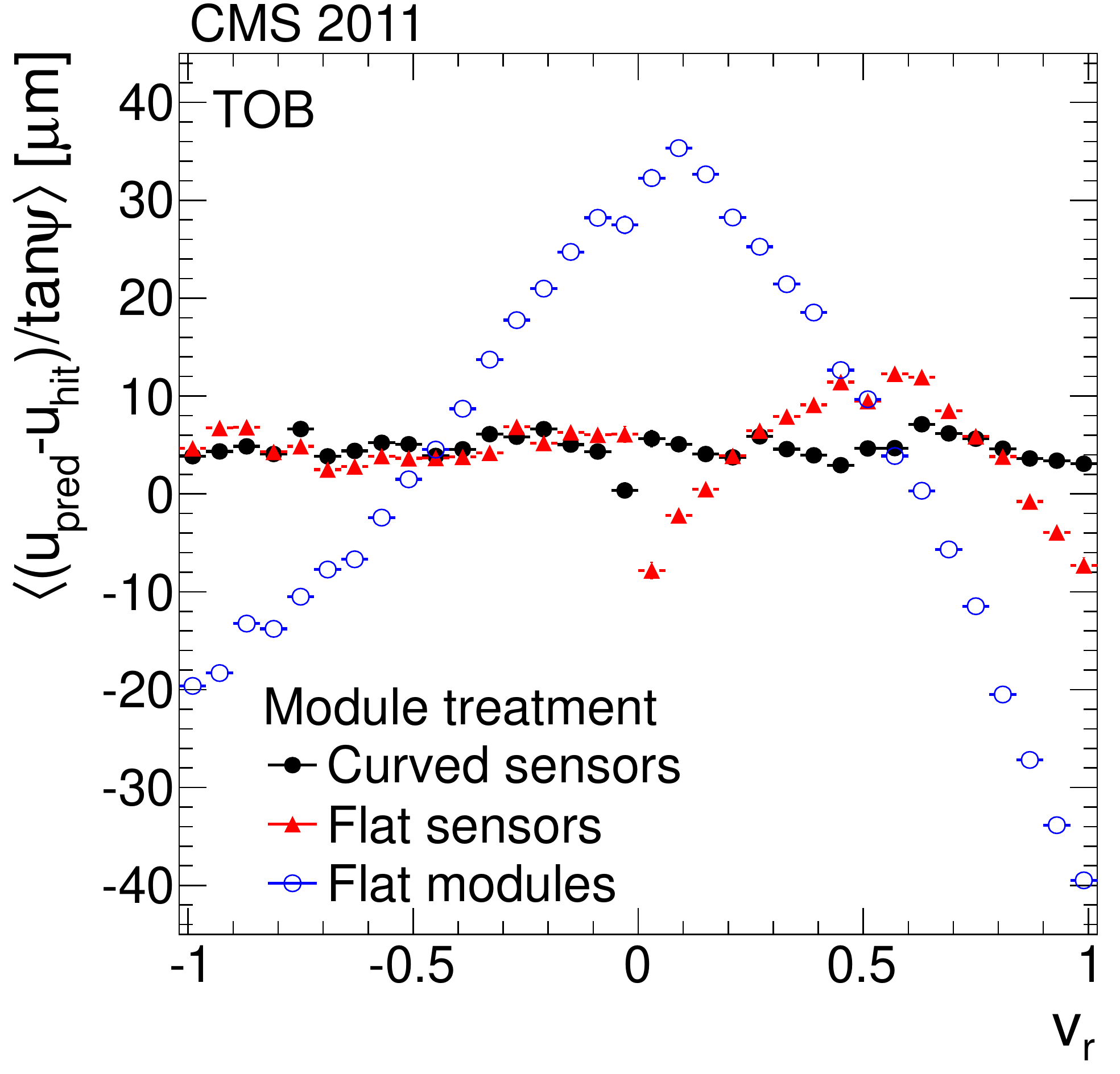}

    \includegraphics[width=0.45\textwidth]{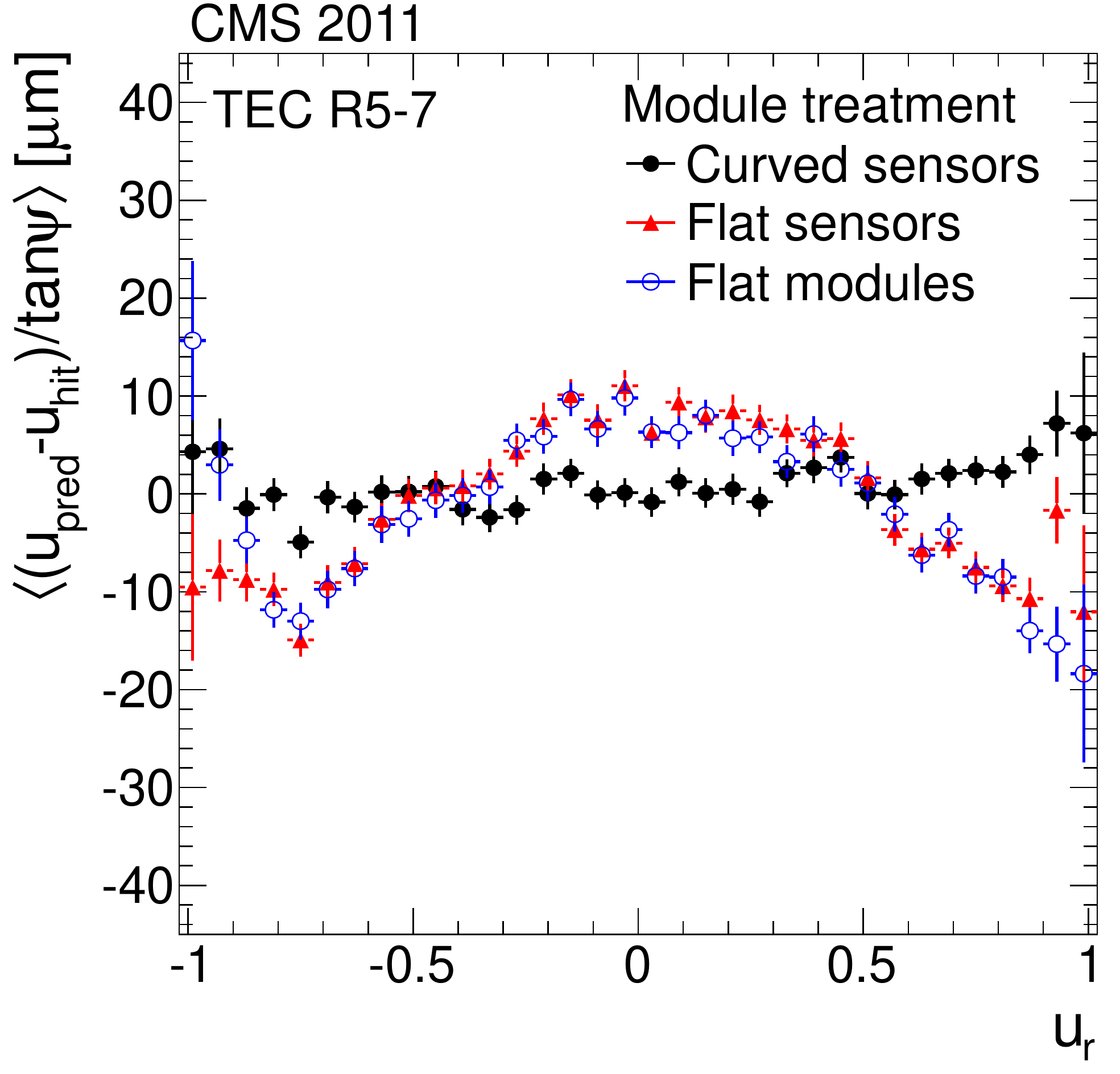}\hfill
    \includegraphics[width=0.45\textwidth]{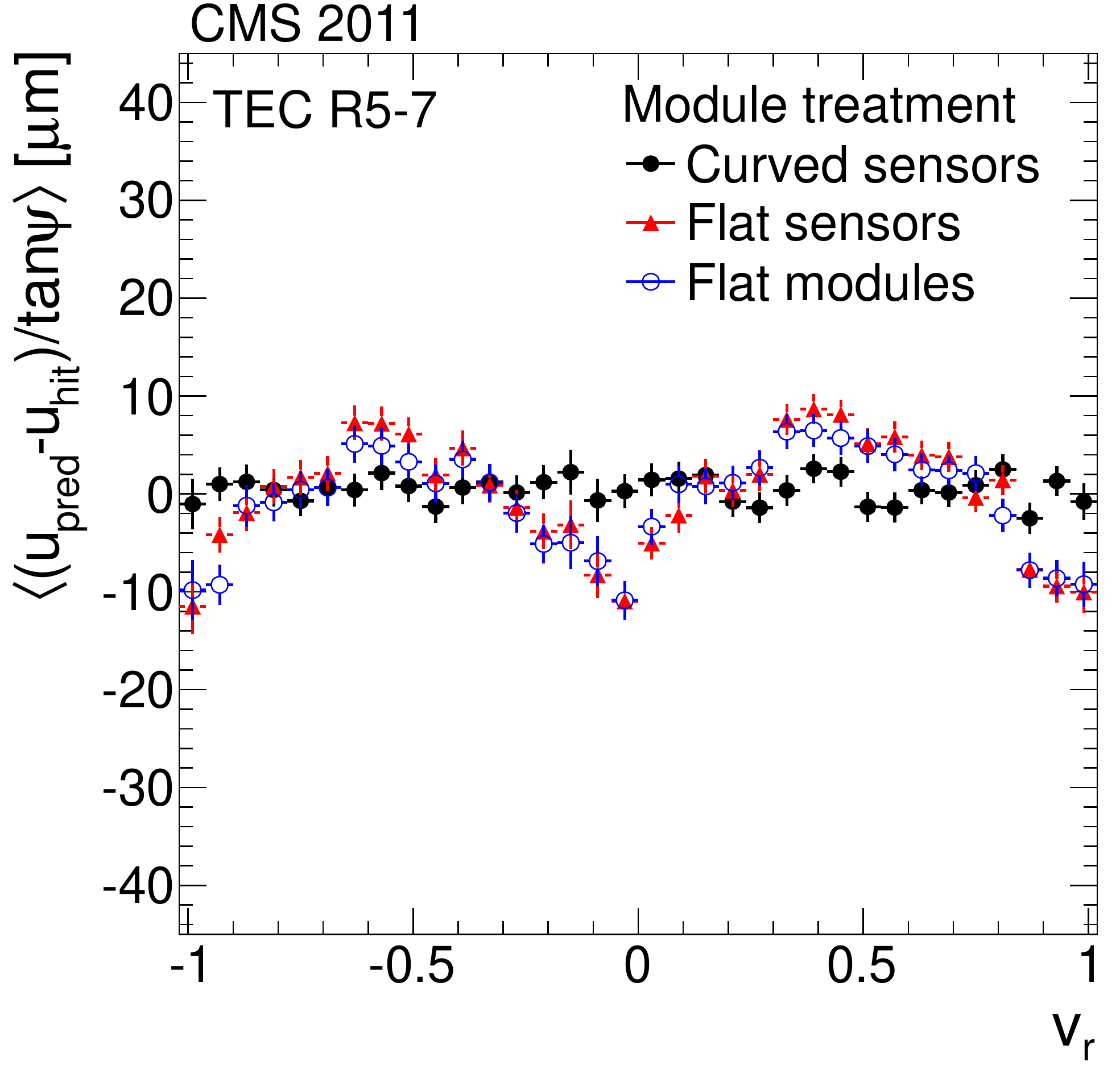}
  \caption{\label{fig:bowVal_strip}
    Distributions of the average weighted means of the
    $\Delta w = \Delta u/\!\tan\psi$
    track-hit residuals as a function of the relative positions of cosmic ray
    tracks on the modules along the local $u$- (left) and $v$-axis
    (right) for different approaches to parameterise the module shape.
    The first two rows show the average for all the TIB and TOB modules,
    respectively,
    and the last row shows the double-sensor modules of the rings
    5--7 of the TEC.
    Each residual is weighted by $\tan^2\psi$ of the track.
  }
\end{figure}

\begin{figure}[tb] 
  \centering
    \includegraphics[width=0.45\textwidth]{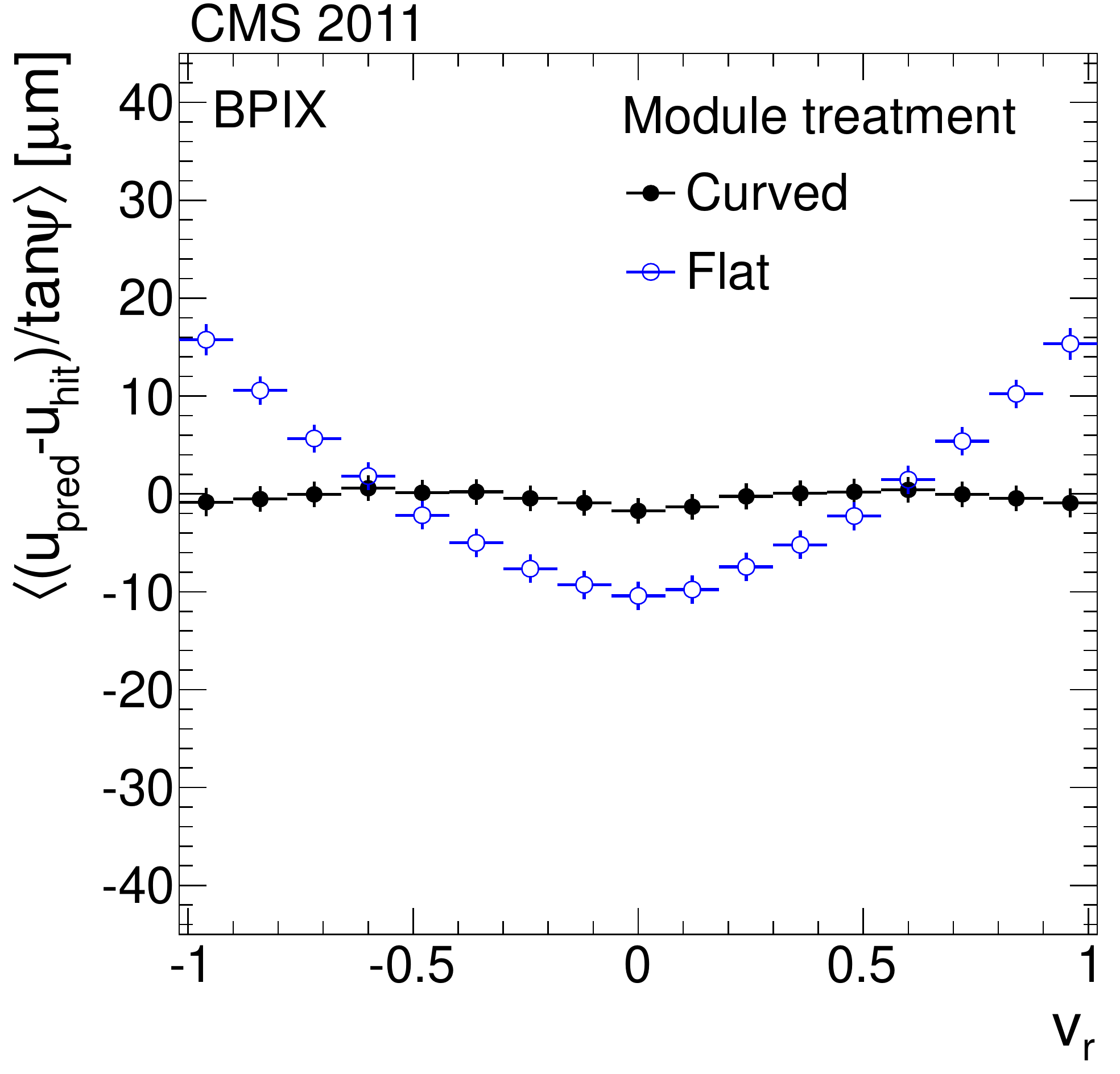}\hfill
    \includegraphics[width=0.45\textwidth]{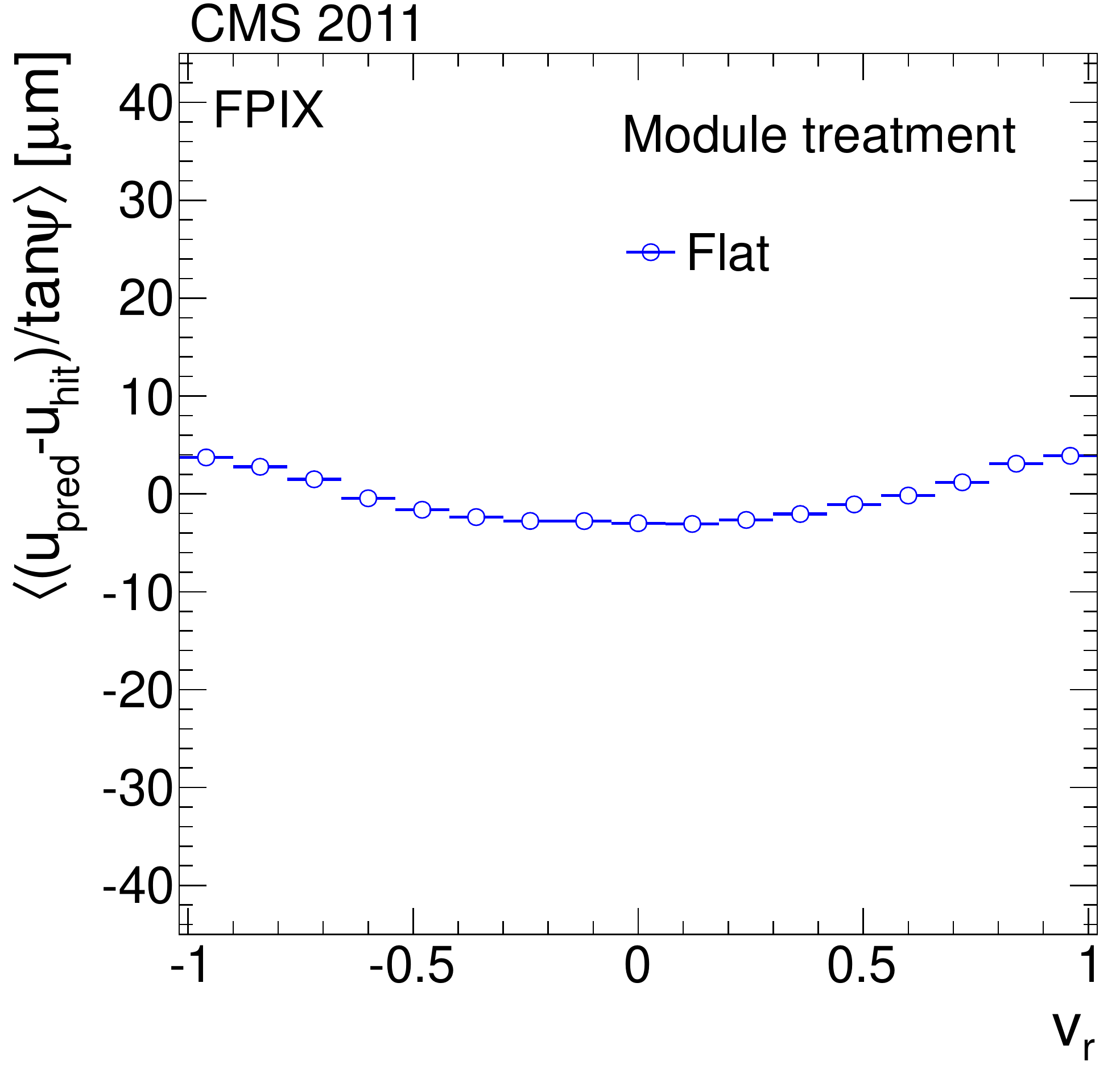}
  \caption{\label{fig:bowVal_pixel}
    Distributions of the average weighted means of the
    $\Delta w = \Delta u/\!\tan\psi$
    track-hit residuals as a function of the relative positions of
    tracks from proton-proton collisions  
    on the modules along the local $v$-axis for different approaches to
    parameterise the module shapes.
    The left shows the BPIX and the right the FPIX.
    Each residual is weighted by $\tan^2\psi$ of the track.
  }
\end{figure}

As discussed in section~\ref{sec:method_alignmentparametrisation},
the tracker modules are
not expected to be absolutely flat. If a silicon module is not flat,
the local $w$-coordinate of the point where a track intersects the
sensor (see figure~\ref{fig:LocalCoord}) depends on the
relative position ($u_r$, $v_r$).
The module shape can thus be investigated by track-hit residuals in $w$
as a function of the track position on the module
\cite{ALI:Babar}.
These residuals can be calculated from the one in the $u$-direction and
the track angle $\psi$ (figure~\ref{fig:LocalCoord}),
$
  \Delta w = \Delta u/\!\tan\psi.
$ 
The mean values of these residuals
are shown in
figures~\ref{fig:bowVal_strip} and~\ref{fig:bowVal_pixel} as a function of
the relative local track coordinates
$u_r$ and $v_r$, averaged over many modules of the strip and pixel
subdetectors, respectively. Since tracks
with a large angle $\psi$ relative to the surface normal are most sensitive to any deviation from flatness, each residual in the average is weighted by $\tan^2\psi$.

Several module shape parameterisations are investigated in the alignment
procedure as shown in figures~\ref{fig:bowVal_strip}
and~\ref{fig:bowVal_pixel}.
The blue open circles
are deduced by using alignment constants
obtained in a procedure similar to the one described in
section~\ref{Sec:Strategy}, but without taking any module shape
parameters into account. Clear deviations from zero are observed in almost all cases,
indicating that the modules are systematically not flat.
The red filled triangles
are obtained with the same alignment procedure
except that
the two sensors
of the modules in the TOB and in the TEC rings 5--7 are aligned
independently.
Finally, the black points
represent the full alignment.

For the distributions along $u_r$ for ``flat modules'', a parabolic shape
of the sensors of all subdetectors is clearly observed. These structures are correctly
compensated when curvatures are taken into account as
for the ``curved sensors''.
At the largest values of $\abs{u_r}$ in the TEC there are few tracks since
the modules are wedge-like in shape and their widths $l_u$ are defined by
their longer edges.
The distributions along the strip direction ($v_r$) show more varied
features. In the TIB, a structure remains that could be corrected
by a fourth order polynomial, but the amplitude is only a few
$\!\mum$ and
thus negligible for tracking purposes.
For the TOB, the V-shaped curve of the ``flat modules'' parameterisation
indicates a systematic kink between the two sensors of the modules.
After correcting for the relative misalignment of the sensors by means
of the ``flat sensors'' parameterisation, a parabolic shape can be seen for the
sensors at positive $v_r$.
Only the use of the ``curved sensors'' correction
results in a flat dependence.
For the double-sensor modules in the TEC rings 5--7, no systematic 
kink between the sensors is visible. However, both sensors are clearly curved,
which is corrected for with the ``curved sensors'' parameterisation.

In the pixel detectors, no systematic structure is observed along the
$u$-direction. The mean $w$-residual distributions along the
$v$-direction,
 determined with tracks from proton-proton collisions, is
shown in figure~\ref{fig:bowVal_pixel}.
An average curvature of the BPIX modules
can clearly be seen. The FPIX
modules also show curvatures, but with smaller amplitude. No corrections
are necessary for this subtle effect.

\begin{figure}[tb]
  \centering
    \includegraphics[width=0.55\textwidth]{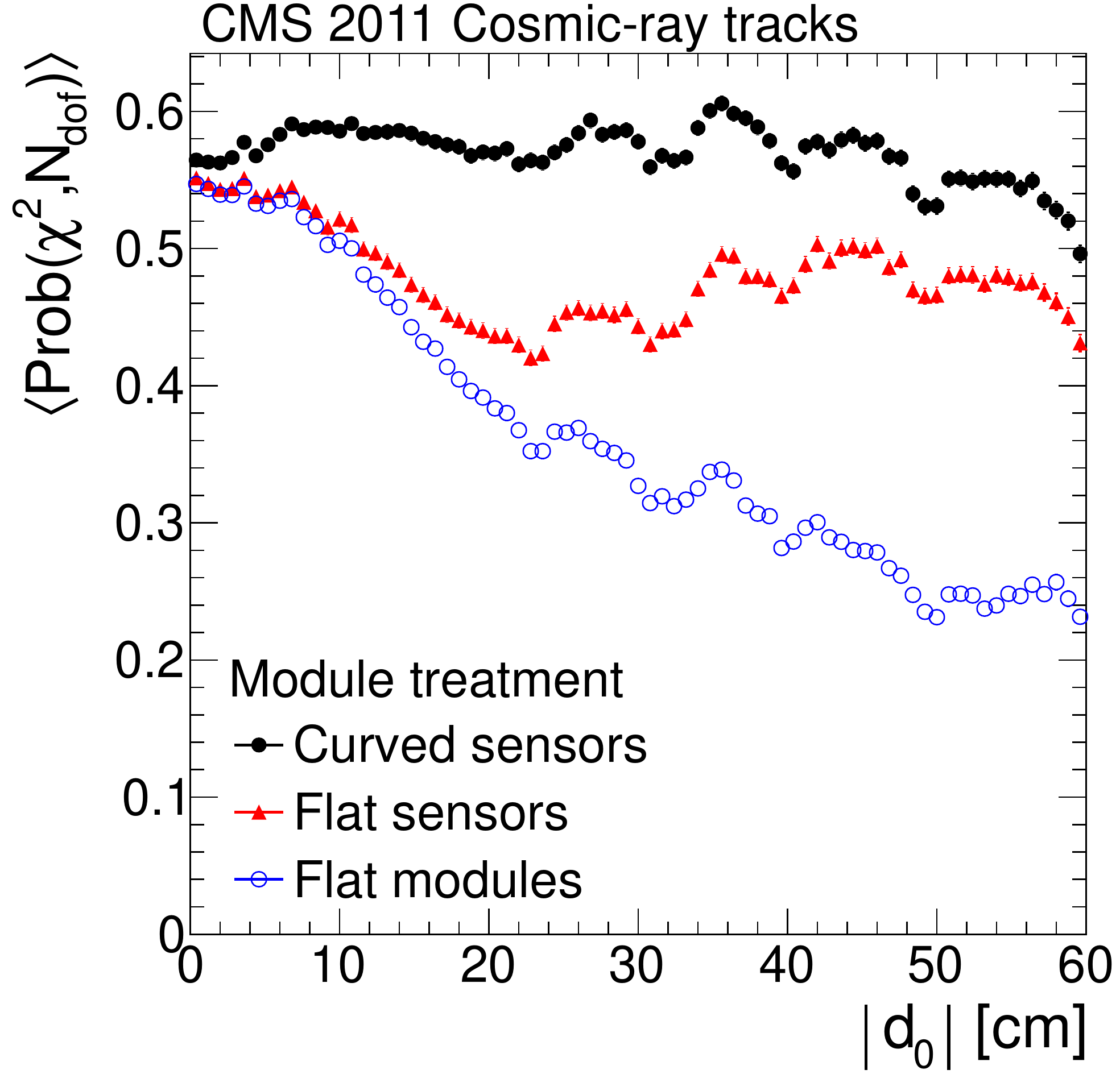}
  \caption{\label{fig:probChi2_vs_d0}
    Mean probability $\langle\text{Prob}(\chi^2,{N_\mathrm{dof}})\rangle$
    of cosmic ray track fits as a function of their distance
    of closest approach to the nominal beam line for 
    the different approaches to parameterise the module shapes.
  }

\end{figure}

The ``curved sensor'' parameterisation leads to a sizable improvement
of the quality of the tracks that cross modules with large angles relative to
the module normal; specifically, cosmic ray tracks crossing the barrel of the
tracker from the top to the bottom with large distances of
closest approach to the beam line $d_0$ predominantly have these large
track angles.
Figure~\ref{fig:probChi2_vs_d0} shows the average fit probability
$\langle\text{Prob}(\chi^2,N_\mathrm{dof})\rangle$
as a function of $\abs{d_0}$ for cosmic ray tracks.
Tracks with small $\abs{d_0}$ cross the modules with modest angles relative to the
normal. With larger $\abs{d_0}$, the average angle of incidence of the
track with respect to the module also increases, resulting in a
significant degradation of
the average fit quality for the ``flat modules'' parameterisation.
For ``curved sensors'' the distribution is approximately flat for $\abs{d_0} < 50$\cm,
resulting in improved consistency of the important cosmic ray track sample
with tracks from proton-proton collisions.
The remaining features of the dependence are correlated with the radii of the barrel layers.
If tracks cross a layer tangentially, the treatment of multiple
scattering effects by using thin scatterers only is an oversimplified approximation.

\begin{figure}[tbhp]
  \centering
  \includegraphics[height=0.35\textwidth]{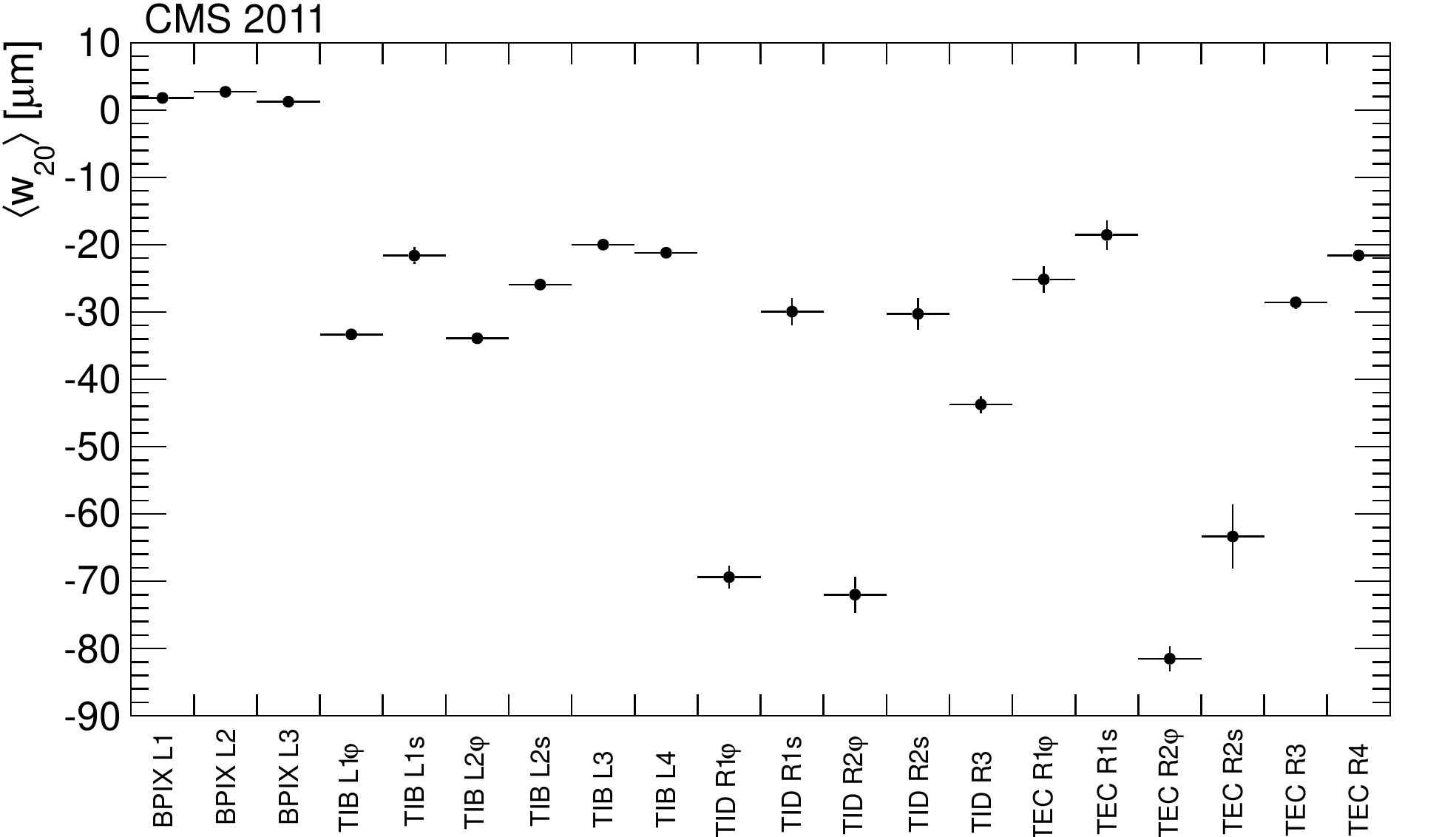}
  \includegraphics[height=0.35\textwidth]{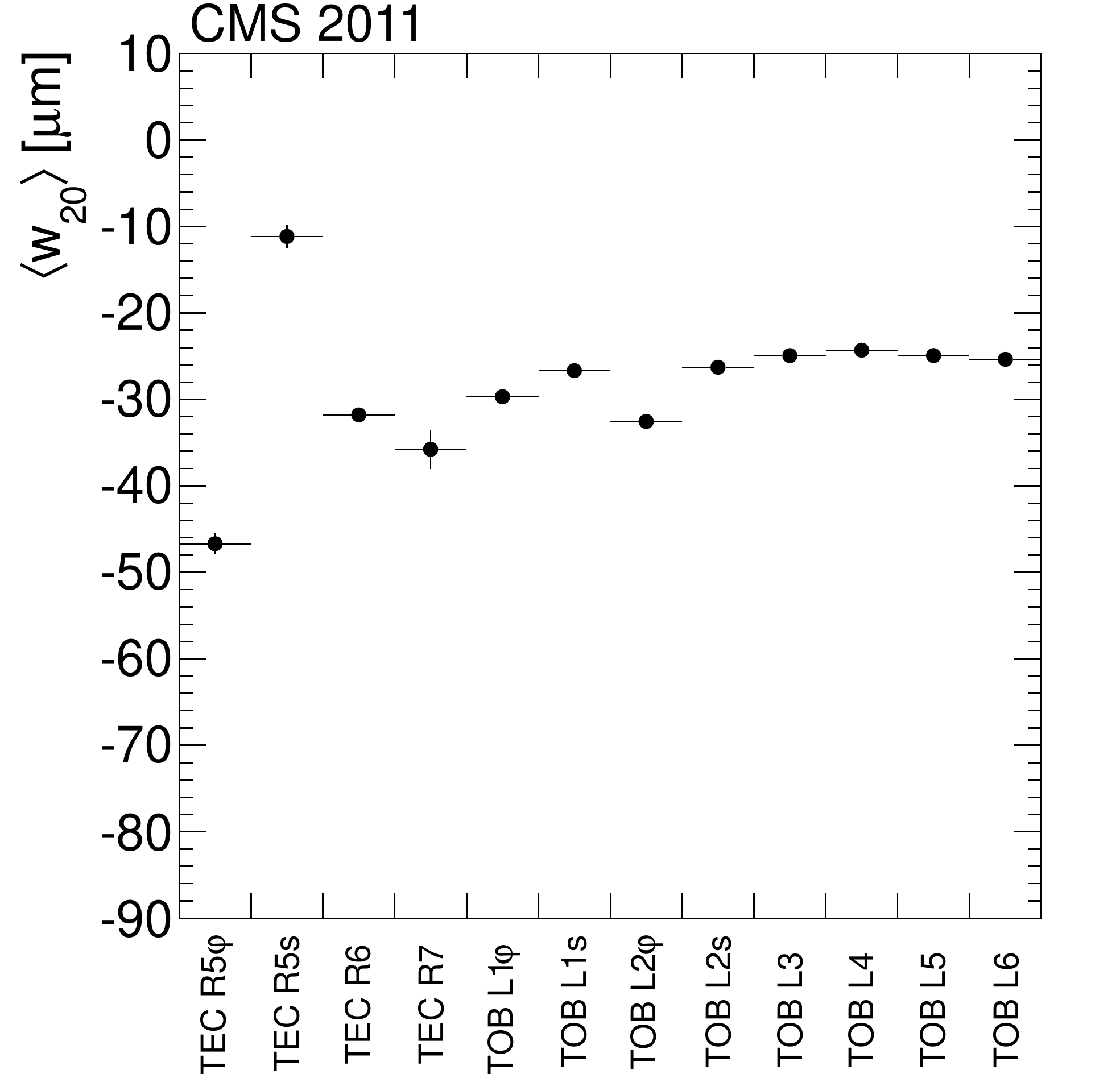}

  \includegraphics[height=0.35\textwidth]{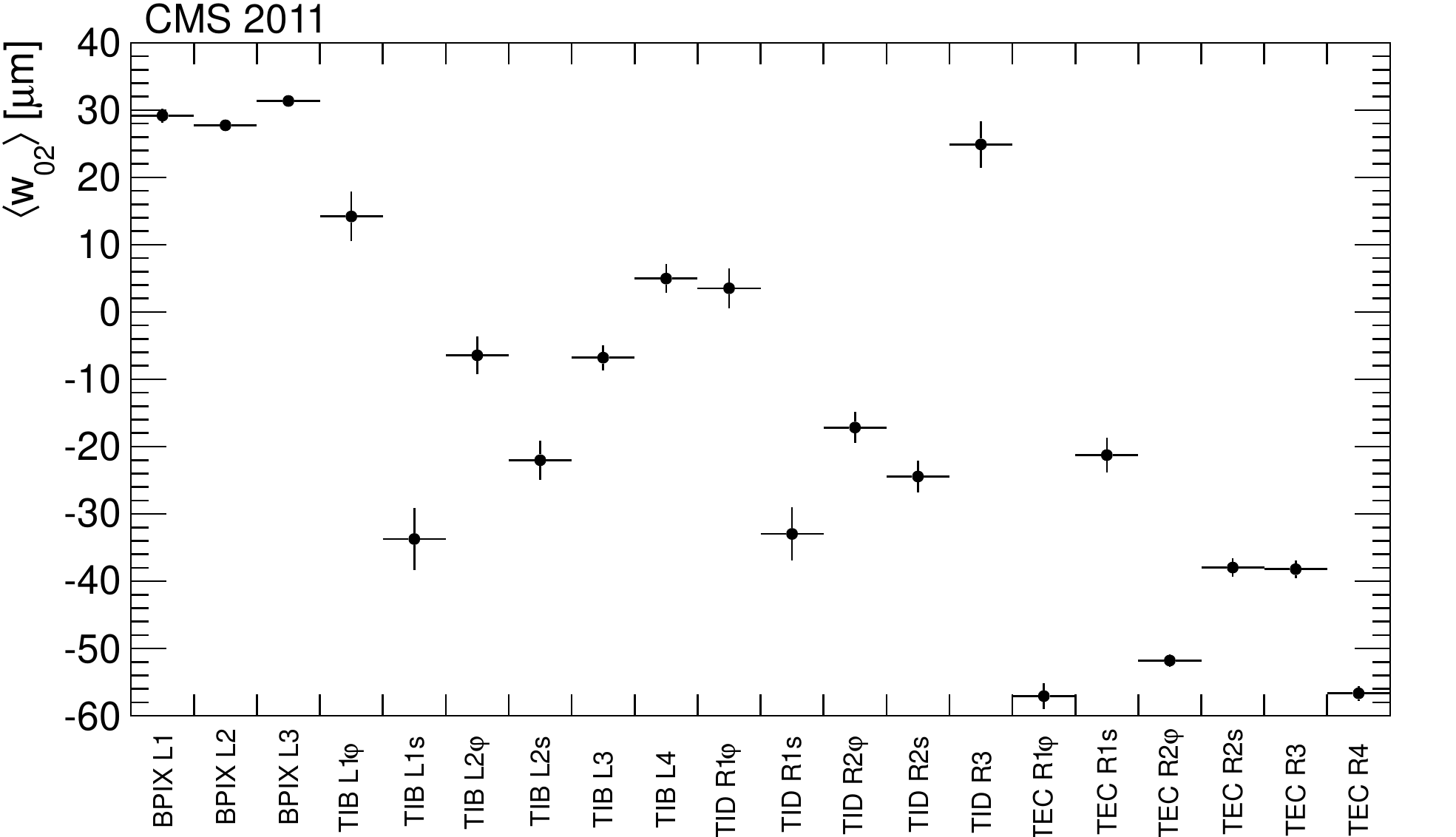}
  \includegraphics[height=0.35\textwidth]{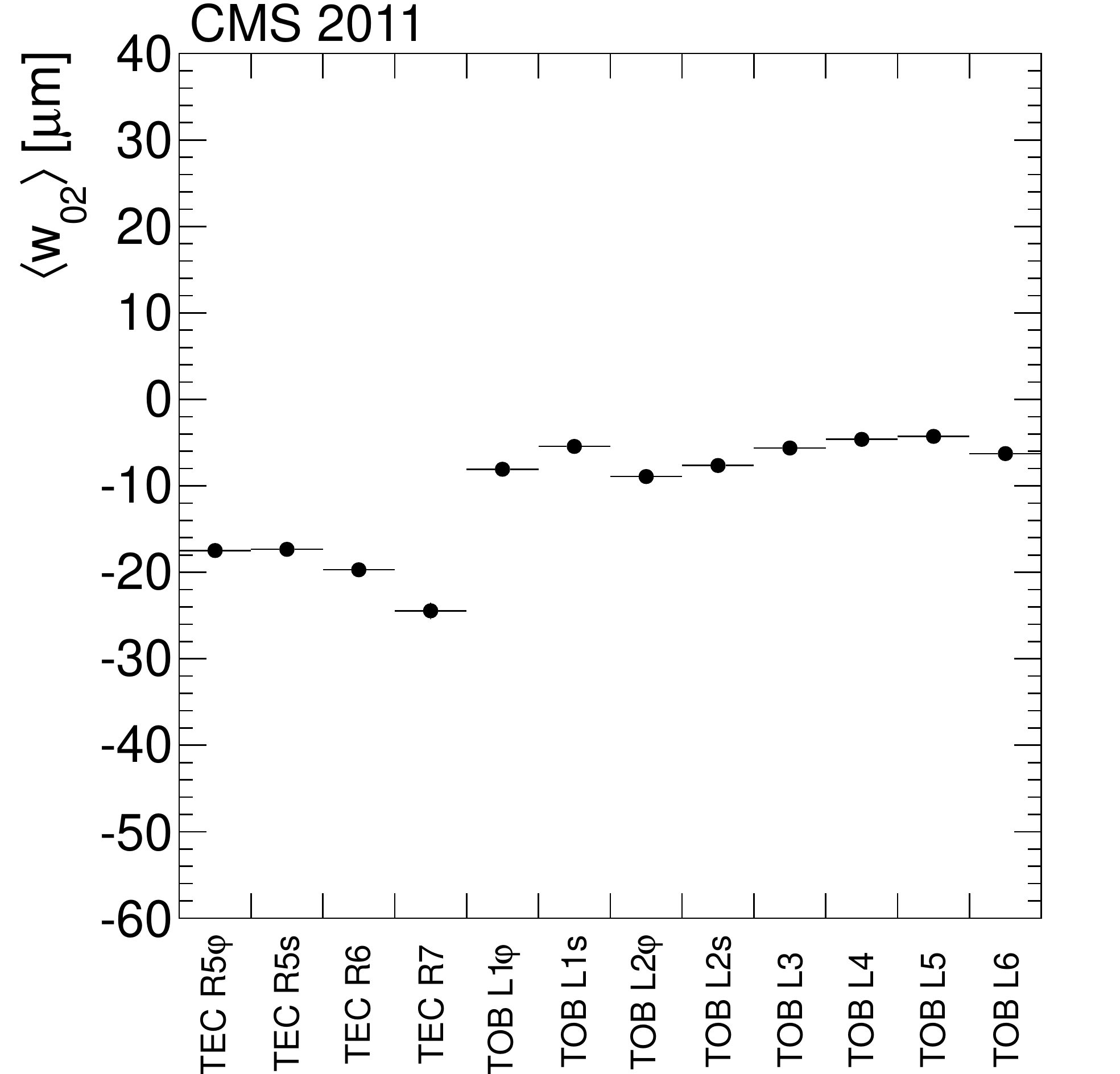}

  \caption{\label{fig:bowsU_V}
    Sensor curvatures along the local $u$ (upper row) and $v$ (lower row)
    coordinate for single- (left column) and double-sensor (right columns)
    modules, averaged over
    layers (L) and rings (R), respectively.
    Stereo (s) and \rphi ($\varphi$) modules within a layer or ring are separated.
  }
\end{figure}

The average sizes of the sagittae of the sensor curvatures along the $u$-
($w_{20}$) and $v$-direction ($w_{02}$),
as determined sensor-by-sensor, are shown in figure~\ref{fig:bowsU_V}
for the different layers and rings, further differentiating for stereo and \rphi
modules, \ie grouping modules with similar sensors and mounting.
While the average sagitta $\langle w_{20}\rangle$ for BPIX sensors is almost zero,
it is usually around $-30\mum$ in the strip subdetectors. This matches well the
sagittae of the average module shapes along $u$ seen for the ``flat module''
distributions on the left pane of figure~\ref{fig:bowVal_strip}.
Stronger curvatures with average sagittae up to
$\langle w_{20}\rangle < -80\mum$ are observed for specific
sensor types and mounting positions, \eg the \rphi modules in TEC ring 2.
The average sagitta $\langle w_{02}\rangle$ shows variations from $+30\mum$ for
BPIX modules (matching with the left pane of figure~\ref{fig:bowVal_pixel}) to almost
$-60$\mum for some sensor types in the TEC.
While the average sensor sagittae shown here are clearly below the construction
specifications of 100\mum, the tails of the distributions extend to
$\abs{w_{ij}} > 200$\mum and even beyond. This could be due to stresses
induced by the mechanics.

\begin{figure}[tbhp]
  \begin{minipage}[t]{0.495\textwidth}
    \centering
      \includegraphics[height=.72\textwidth]{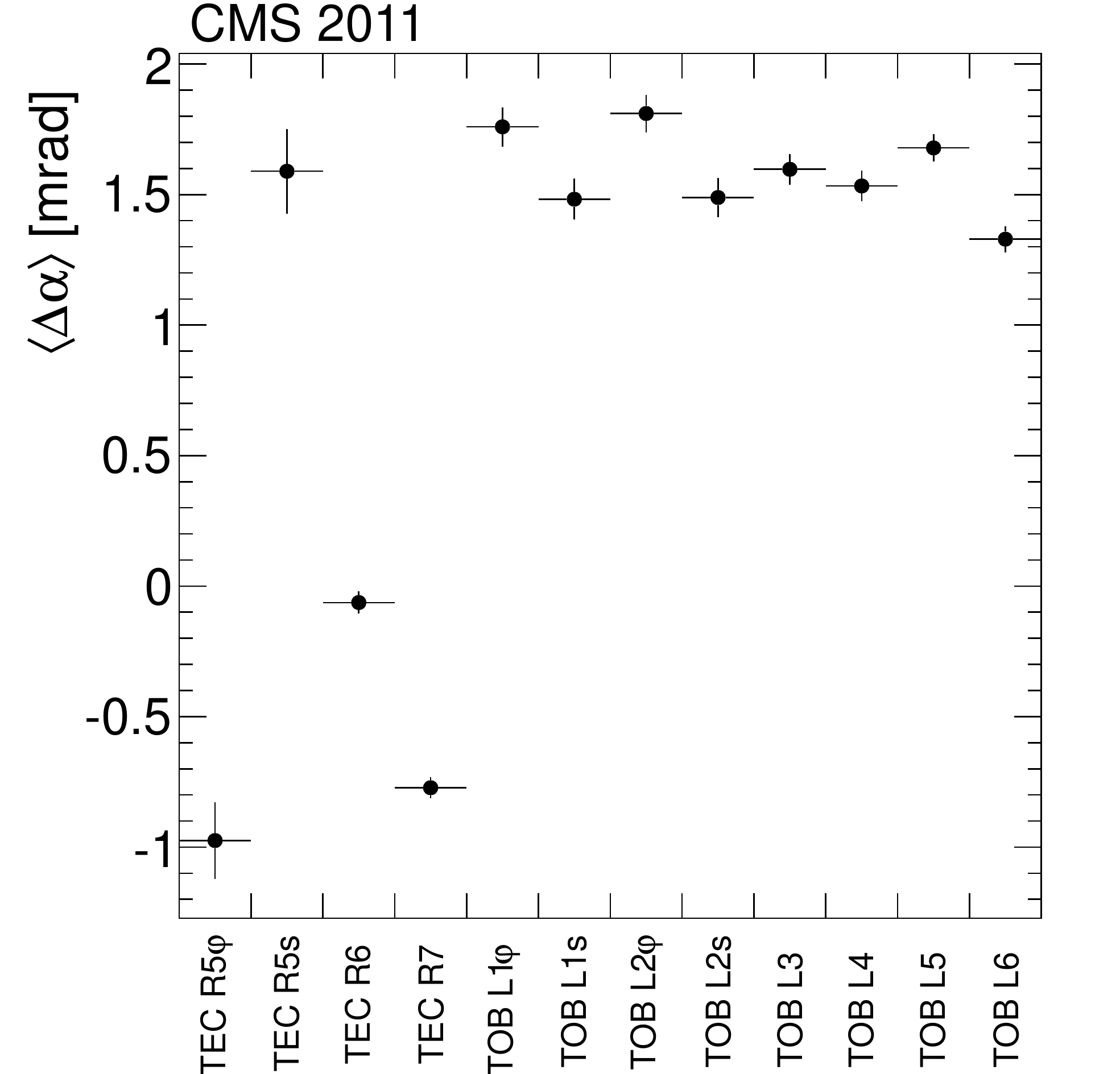}
    \caption{\label{fig:meankinks}
      Kink angle $\Delta\alpha$ for double-sensor modules,
      averaged over
      layers (L) and rings (R), respectively.
      Stereo (s) and \rphi ($\varphi$) modules within the same layer or ring are
      separated.
    }
  \end{minipage}
  \hfill
  \begin{minipage}[t]{0.45\textwidth}
    \centering
      \includegraphics[height=.8\textwidth]{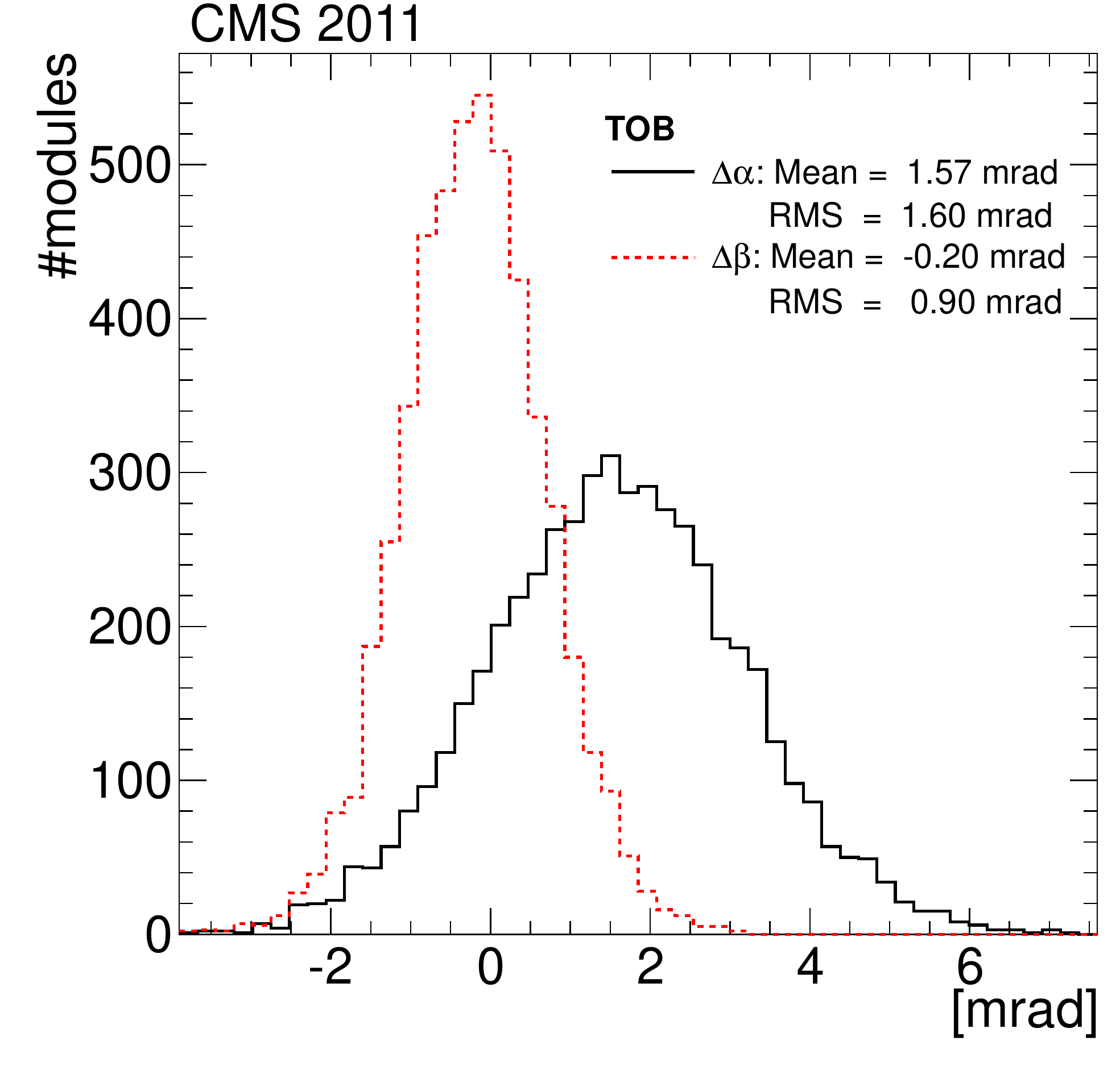}
    \caption{\label{fig:kinkDistributions}
      Distributions of the kink angles $\Delta\alpha$ and $\Delta\beta$
      of TOB modules.
    }
  \end{minipage}
\end{figure}

Aligning both sensors of the modules in the TOB and the TEC rings 5--7
independently indicates that the sensors were slightly misaligned with respect
to each other during module assembly.
As an example, figure~\ref{fig:meankinks} shows
their average differences of the rotation angles around the local $u$-axis,
$\Delta \alpha = \alpha_1 - \alpha_2$, where the first sensor is the one
closer to the readout electronics at $v < 0$.
The TOB modules show an average kink of
$\langle\Delta\alpha\rangle \approx 1.6$\unit{mrad}, matching well the kink seen
in the TOB graph \vs $v$ of figure~\ref{fig:bowVal_strip}.
The value varies significantly among different modules in the TEC. This variation has
averaged out the kink effect in the TEC distribution \vs $v$.
The angular misalignment of the two sensors 
exhibits a significant spread as shown for $\Delta\alpha$ and,
for the rotation around $v$, $\Delta\beta$ of the
TOB modules in figure~\ref{fig:kinkDistributions}. Values of
$\Delta\alpha = 6$\unit{mrad} are reached, corresponding to
$\Delta w \approx 150\mum$ at the edges of the 10\cm long sensors.

In summary, the module shapes can be described with polynomials
up to the second order for each sensor, and their coefficients are
successfully determined sensor-by-sensor within the alignment procedure.
Applying corrections to the hit positions that depend on the module
parameters determined and on the track parameters on the module surface
significantly improves the overall track description, especially for the important
cosmic ray tracks.
The sensor parameterisation used here is valid as long as the effect of
the curvatures can be approximated by a change of the local $w$ coordinate
only, neglecting changes in $u$ and $v$. Within these boundaries,
the construction criterion of sensor bows below 100\mum
could have
been relaxed
since the alignment successfully takes
care of this effect.

\section{Control of systematic misalignment}\label{Sec:WeakModes}

The monitoring of standard physics references and the comparison with
other subdetectors of CMS provides a direct check of the robustness of
the alignment procedure and potentially indicates the presence of
systematic misalignments. This information can be included in the
alignment algorithm in order to better constrain the systematic
misalignments, as described in section~\ref{sec:resonanceTreatment}.
The sensitivity to weak modes of the alignment procedure followed in this analysis is
discussed in section~\ref{subsec:systmis} following the approach presented in
\cite{ALI:TkAlCraft08, ALI:Babar}.

\subsection{Monitoring of the tracker geometry with \texorpdfstring{$\Z \rightarrow \mu\mu$}{Z to mu mu} events}\label{subsec:zmumu}

As described in sections \ref{sec:resonanceTreatment} and
\ref{Sec:Strategy}, muonic decays of \Z bosons provide a standard
reference that can be used for validating the aligned geometry.  The
selection of well-reconstructed $\Z \to \mu^+\mu^-$ candidates
requires two muons reconstructed with both the tracker and the muon
system (\textit{global muons}), where at least one of them passes the
tight quality selections as defined in \cite{Muo:MUO-Performance}.
The muons must pass the following kinematic selections:
\begin{itemize}
\item $\pt > 20$\GeVc,
\item $\abs{\eta}<2.4$,
\item $80 < M_{\mu\mu} < 120$\GeVcc .

\end{itemize}

The distribution of the mass of the \Z candidates is then fitted with
a Voigtian function to model the reconstructed \Z-boson lineshape and an exponential
function to model the background. The width of the Breit--Wigner
component of the Voigtian function is fixed to the decay width of the
\Z boson. The mass of the \Z candidates is estimated with the mean
of the fitted Voigtian. The reconstructed mass is
slightly below the nominal mass of 91.2\GeVcc, at about 90.8\GeVcc,
mostly because of the presence of QED final-state
radiation~\cite{Muo:MUO-Performance}. The mass of the \Z
candidates is measured as a function of $\eta$ and $\varphi$ of
the positively charged muon. A twist-like weak mode would bias the
curvature measurement of each muon depending on its polar angle,
manifesting itself as a strong dependence of the \Z-boson mass on the muon
pseudorapidity (with opposite signs for the two muon charges).

\begin{figure*}[hbtp]
  \centering
    \includegraphics[width=0.49\textwidth]{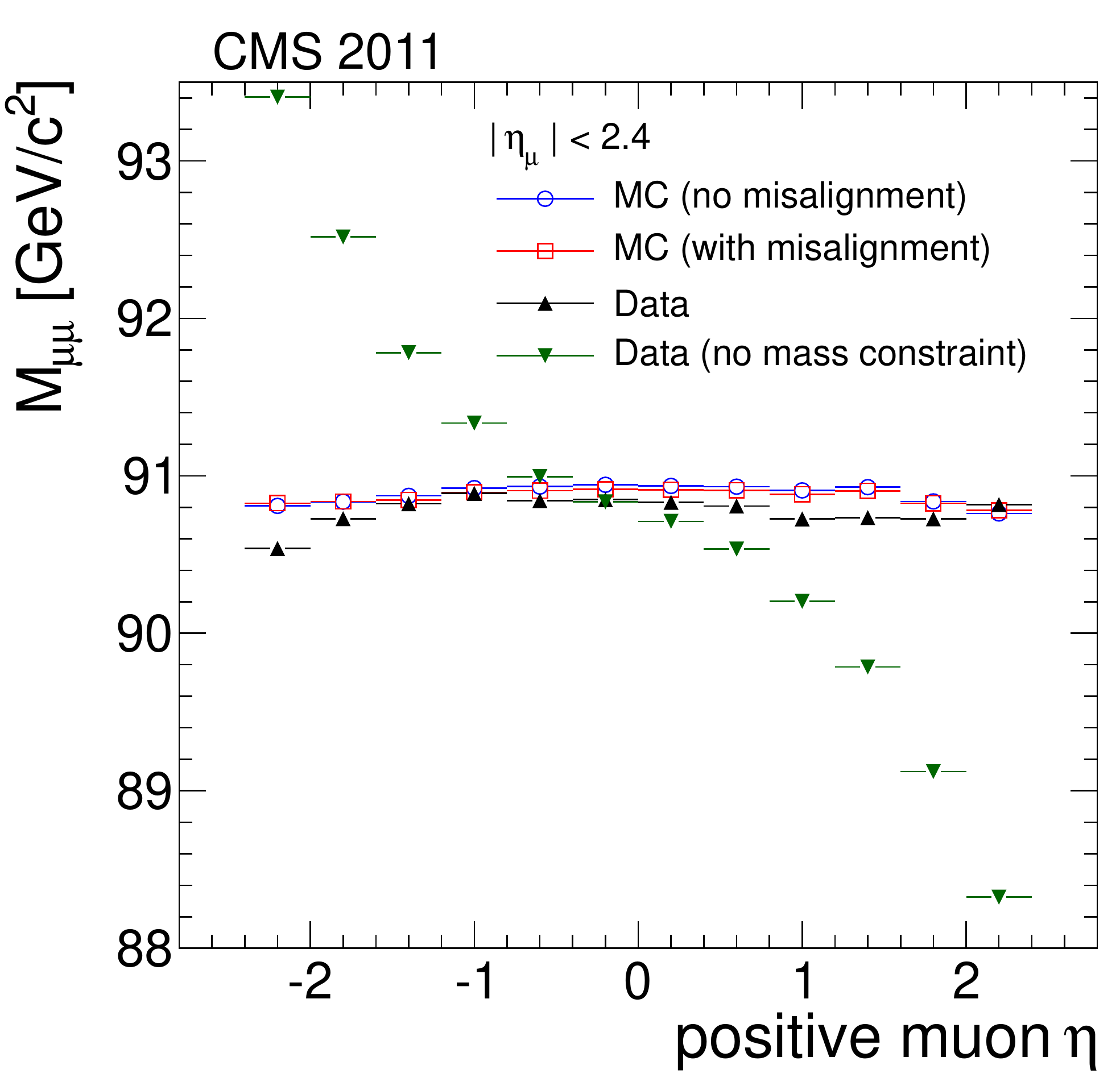}
    \hfill
    \includegraphics[width=0.49\textwidth]{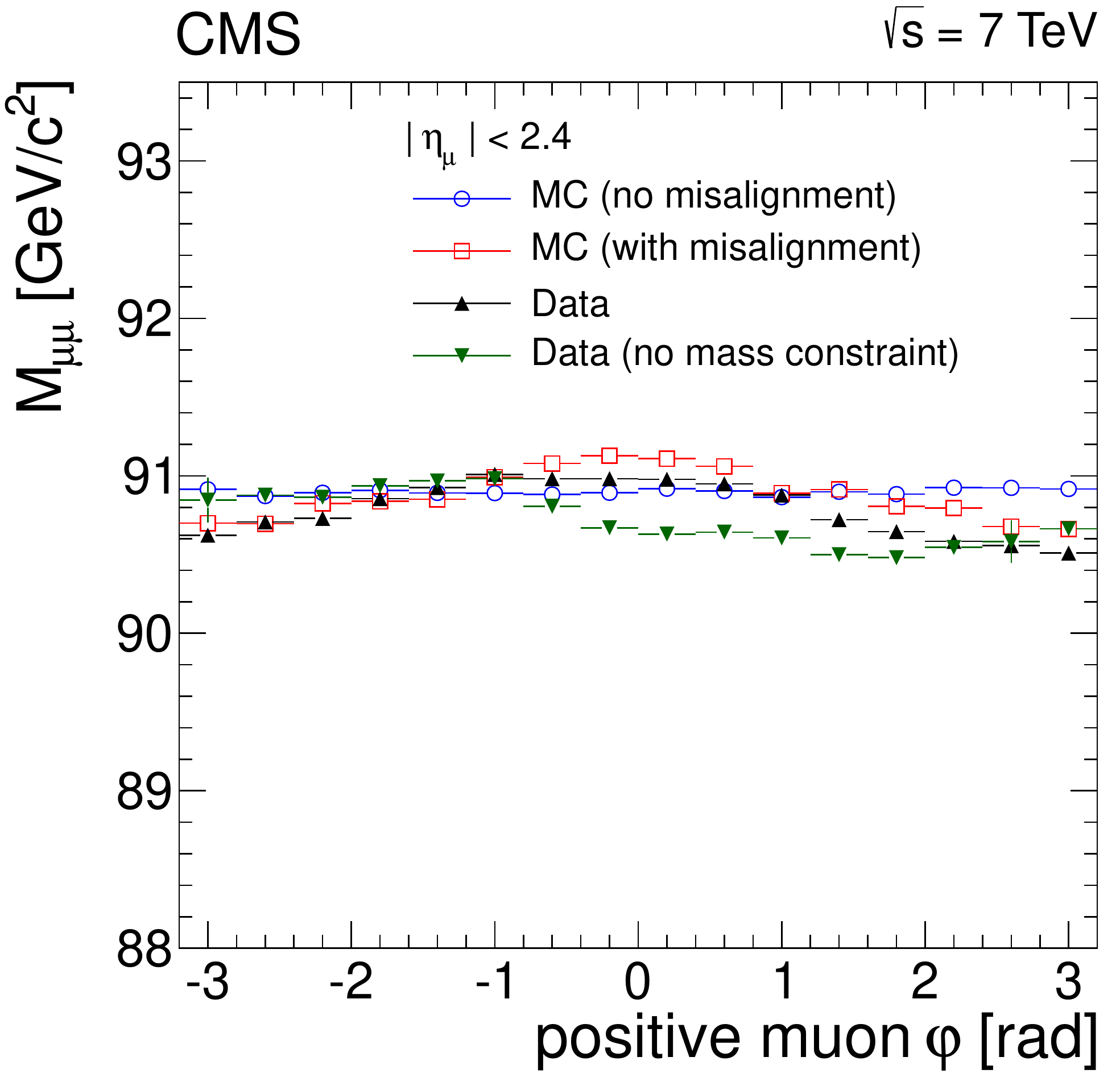}
    \caption{Invariant mass of $\Z \to\mu^+\mu^-$ candidates
    as a function of $\eta$ (left) and $\varphi$ (right) of the
    positively charged muon.  Distributions from aligned data are
    shown as black upward-pointing triangles. Distributions from a
    simulation without misalignment and with realistic misalignment (see section~\ref{Sec:Strategy}) are
    presented as blue hollow circles and red hollow markers,
    respectively. The same distribution with the data but with a
    geometry produced without using the \Z-boson mass information is presented with
    green downward-pointing triangles.}
    \label{fig:ZMassVsEtaPhi}

\end{figure*}

\begin{figure*}[hbtp]
  \centering
    \includegraphics[width=0.5\textwidth]{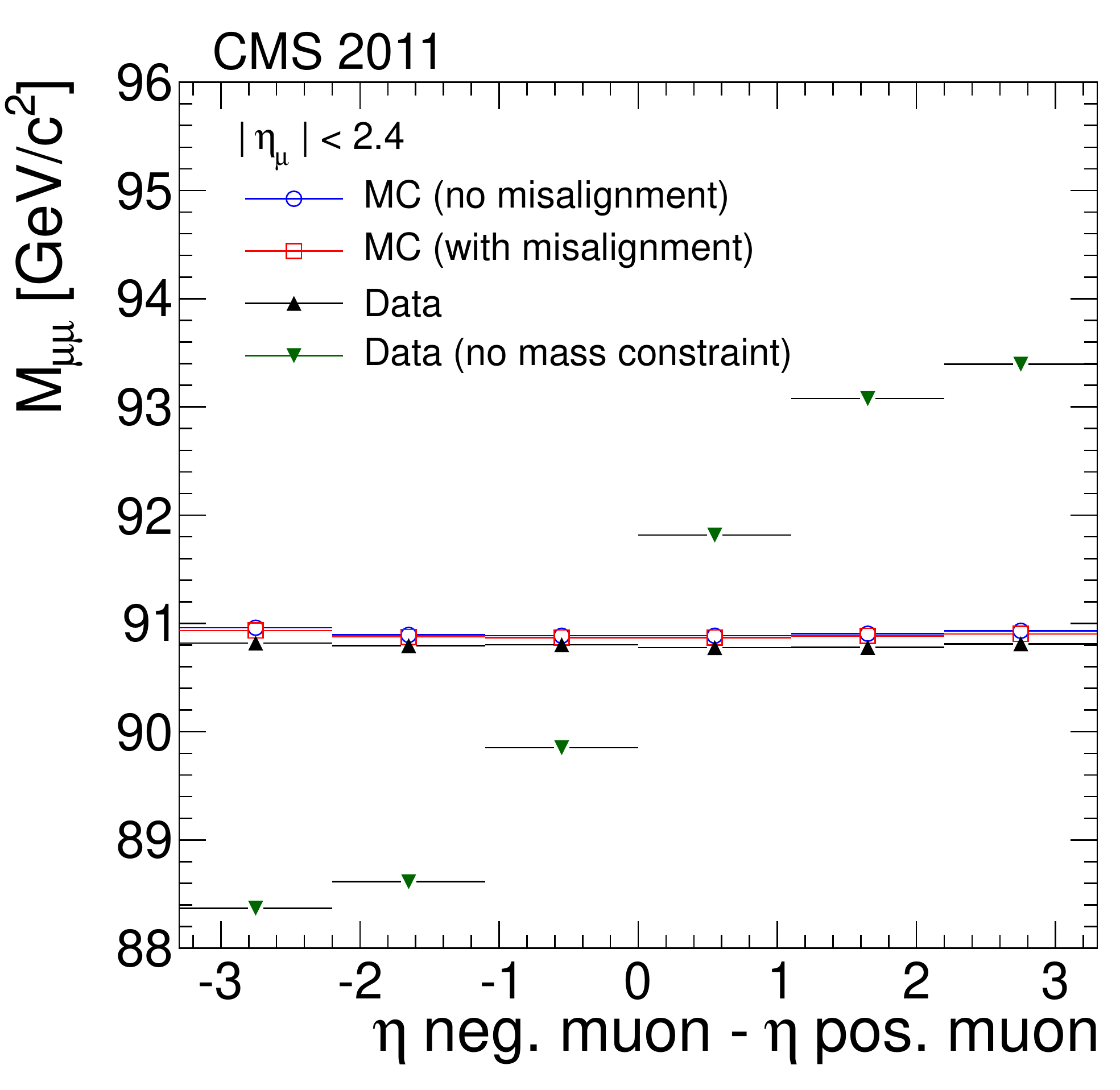}
    \caption{Invariant mass of $\Z \to \mu^+\mu^-$ candidates as
      a function of the $\eta$ separation of the two muons.
      Distributions from aligned data are shown as black upward-pointing
      triangles. Distributions from a simulation without misalignment and with
       realistic misalignment are presented as blue
      hollow circles and red hollow markers, respectively. The same
      distribution from the data but with a geometry produced without
      using the \Z-boson mass information is presented with green
      downward-pointing triangles.  }
    \label{fig:ZMassVsDeltaEta}
\end{figure*}

The result of this study is presented in figure
\ref{fig:ZMassVsEtaPhi} for both the 2011 data and the simulation, and
the corresponding dependence of the \Z-boson mass on the difference
in pseudorapidity of the two muons, $\Delta \eta = \eta^{+} -
\eta^{-}$, is shown in figure~\ref{fig:ZMassVsDeltaEta}.  By using the
nominal geometry, the estimated value of the \Z-boson mass in the
simulation is 90.8\GeVcc, as expected after final-state
radiation. Without using the \Z-boson mass information (downward-pointing
triangles), a pronounced $\eta$ and $\Delta \eta$ dependence of the
reconstructed invariant mass is observed, which extends over a range
of more than 5\GeVcc and is attributed to a twist weak mode as seen in
section~\ref{subsec:eoverp}.  The inclusion of the \Z-boson mass
information (upward-pointing triangles) removes this bias and leads to an
almost flat dependence on $\eta$ and $\Delta \eta$.

The agreement between the data and the simulation is good. A
small offset of about 100\MeVcc in the simulation with respect to the
data is visible. The size of any remaining bias, also as a function of
the azimuthal angle, is of the order of a few per mil and thus small
compared to the \pt resolution targeted for muons in the typical
momentum range of the \Z-boson decays, which is at best 1\%
\cite{Muo:MUO-Performance}. This suggests that in terms of performance the
aligned geometry in data is very close to a perfectly aligned tracker,
with a beneficial impact on the physics measurements of CMS. Offline
corrections to the muon momentum can be applied after the
reconstruction level, further improving the momentum scale and
resolution of the muons \cite{Muo:MUO-Performance}.

\subsection{Monitoring of the tracker geometry with the CMS calorimeter}\label{subsec:eoverp}

The measurements of the CMS electromagnetic and hadron calorimeters
can be exploited to study systematic effects in the momentum
measurement.  This check is valuable since it is an alternative to the
$\Z \rightarrow \mu^+\mu^-$ decays, which are already used in the
alignment procedure. Weak modes altering the true azimuthal angle of
the modules would modify the track curvature in the opposite way for
positively and negatively charged tracks.
If $\phi$ denotes the change of the azimuthal position at the radius $r$
of a track due to bending in the magnetic field, the following
relation holds for the reconstructed \pt of the track in case of a twist-like
deformation:
\begin{equation}
\pt^{\pm} = \frac{0.57\GeVc \cdot r [\text{m}]}{\sin(\phi \mp \Delta \phi)}\, .
  \label{eq:EoverP1}
\end{equation}
A longitudinal magnetic field strength of 3.8~T is assumed. The $\pm$
indicates the electric charge of the particle, $r[\text{m}]$ is the
radius (measured in metres) at which the particle leaves the tracker
volume, and $\Delta \phi$ is the azimuthal misplacement of the hits
due to the deformation.  So the relative angle $\Delta \phi$ is
related to the asymmetry in the \pt measurement of oppositely charged
tracks with the same true \pt and same $\theta$. An external
measurement of the energy of the charged particle, $E$, is provided by
the ECAL and the HCAL \cite{ALI:enderleThesis2012}.  At a given value
of $\langle E\!/\!p\rangle$, the average ratio between the energy and
momentum of a charged track at fixed \pt, $\Delta \phi$ is measured as
a function of the asymmetry between positively and negatively charged
tracks, $(\langle E\!/\!p^{-}\rangle - \langle E\!/\!p^{+}\rangle)$:

\begin{equation}
  \Delta\phi=\frac{1}{2}\left[
    \arcsin {\left(\frac{0.57\cdot r\text{[m]}}{\left\langle E\cdot\sin\theta\right\rangle\text{[}\!\GeV\text{]}}\left\langle\frac{E}{p^{-}}\right\rangle\right)}
    -\arcsin{\left(\frac{0.57\cdot r\text{[m]}}{\left\langle E\cdot\sin\theta\right\rangle\text{[}\!\GeV\text{]}}\left\langle\frac{E}{p^{+}}\right\rangle\right)}
    \right]\, ,
\label{eq:EoverP2}
\end{equation}

which for large \pt ($\pt \gtrsim 10\GeVc$) and small misalignments
($\Delta \phi \ll 1$) approximates to

\begin{equation}
\Delta \phi =  \frac{(0.57 \cdot r)}{2}\,\frac{ \left\langle\frac{E}{p^{-}}\right\rangle  -  \left\langle\frac{E}{p^{+}}\right\rangle   }{ \left\langle E \cdot \sin \theta \right\rangle}\,.
  \label{eq:EoverP3}
\end{equation}

\begin{figure*}[hbtp]
  \centering
    \includegraphics[width=0.46\textwidth]{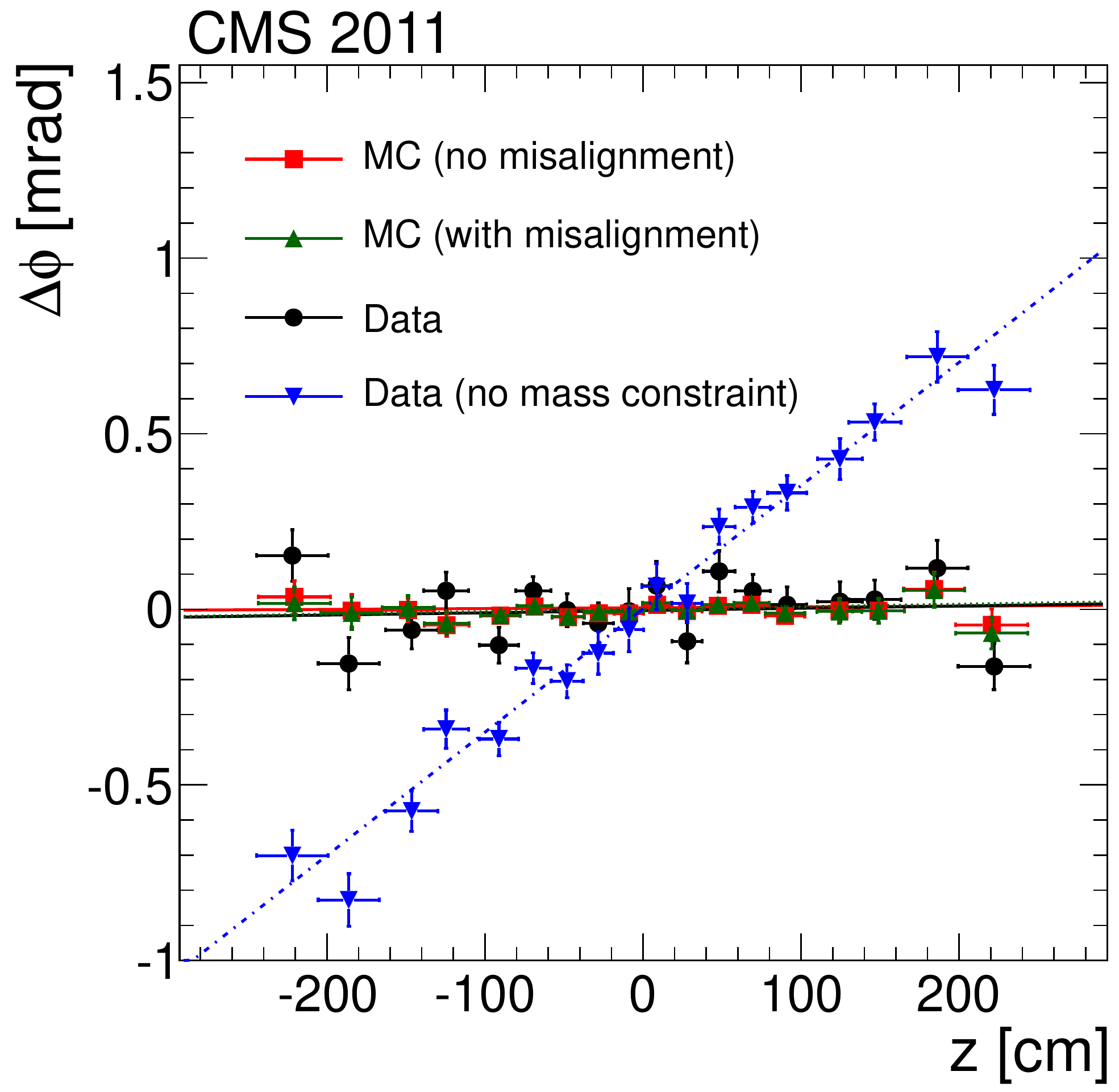}
    \hfill
    \includegraphics[width=0.46\textwidth]{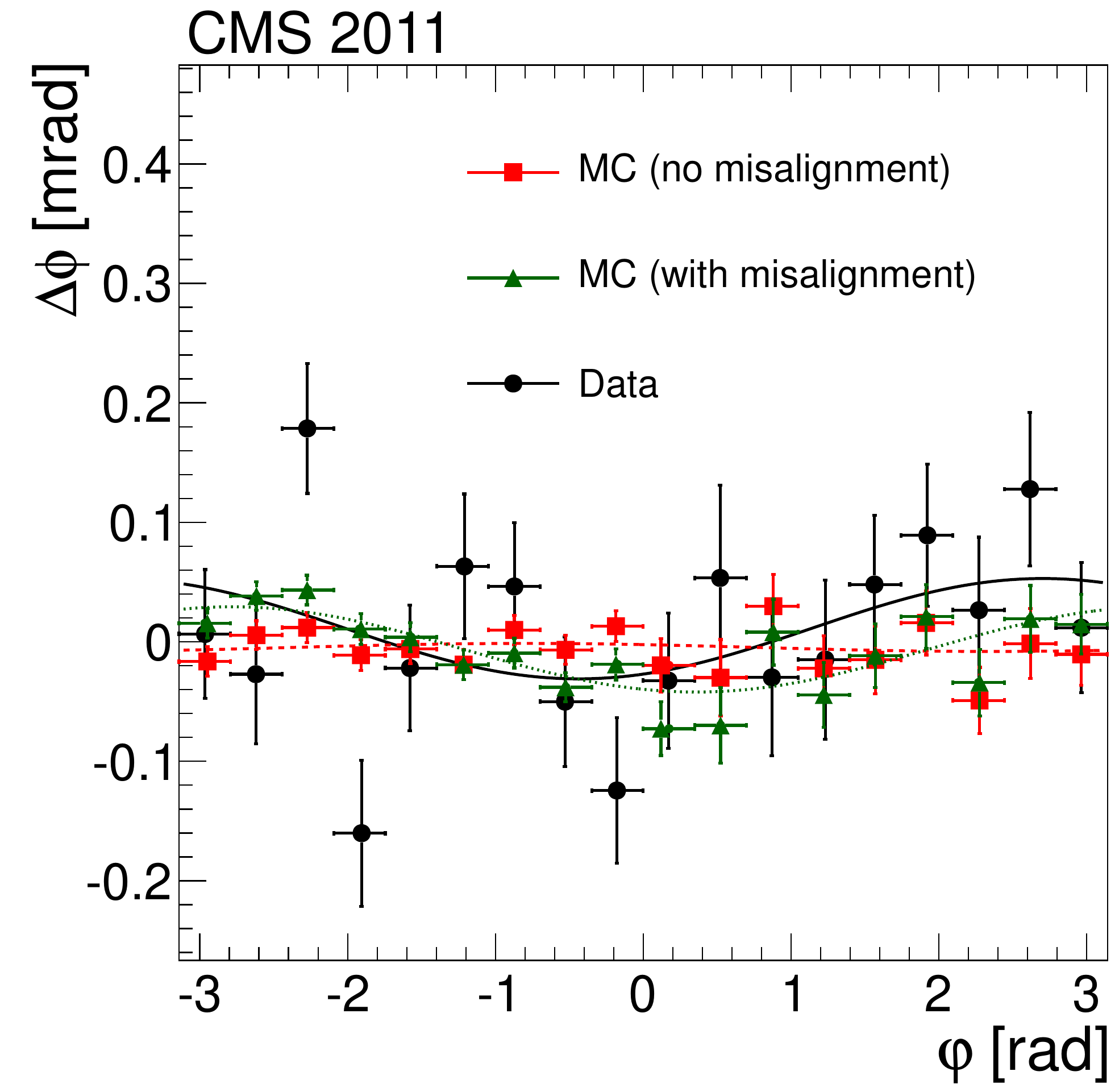}
    \caption{Rotational misalignment, $\Delta \phi$, as a function of
      the $z$-position at $r=1$\unit{m} (left) and $\varphi$ (right) of
      the track. Distributions from 2011 data with an aligned
      geometry are shown as black dots. The distributions from the
      simulation without misalignment and with realistic misalignment are
      presented as red squares and green upward-pointing triangles, respectively.
      The
      blue downward-pointing triangles in the left figure show the
      distribution by using an alignment obtained without using the
      information on the \Z-boson mass.  }
    \label{fig:EoverP}
\end{figure*}

The track sample used for the validation is selected from an input
data set of events triggered by requiring a track with a total momentum
$p > 38\GeVc$ and matched to an HCAL cluster.  A charged-track
isolation requirement is applied at the trigger level ensuring that no
track with a transverse momentum of $\pt > 2\GeVc$ is allowed to be
in a circle with a radius of 40\cm around the impact point on the ECAL
surface of the track considered.  The distributions of
$\langle E\!/\!p^{-}\rangle$ and $\langle E\!/\!p^{+}\rangle$ for
particles with similar energy $E$ are fitted with a Gaussian
function. The means of the fits are used in
equation~(\ref{eq:EoverP3}) in order to measure the $\Delta \phi$ for
that specific bin. The results for different bins of the calorimeter
energy are finally averaged.  This method uses the calorimetric
information only to identify tracks with the same energy, improving
its robustness against miscalibrations of the absolute energy scale of
the calorimeters. The dependence of $\Delta \phi$ on the $z$-component of the
impact point at the radial position $r=1$\unit{m} and on the $\varphi$ of
the track is shown in figure \ref{fig:EoverP} for different
geometries. A twist deformation would show as a linear trend in the
$z$ dependence.  From the observed dependence, no significant systematic
distortion in the aligned geometry is visible within the current
uncertainties of the validation method. A clear improvement with
respect to a geometry not exploiting the mass information is visible.
A linear fit to the distributions is performed in order to quantify
the bias. In the absence of the mass information in the alignment
procedure, the linear fit exhibits a slope significantly different
from zero, $(0.351 \pm 0.012)\unit{mrad}\!/\!\text{m}$. In the case of
the baseline alignment on data, the slope is
$0.002 \pm 0.012\unit{mrad}\!/\!\text{m}$.
Slopes compatible with zero are observed also in
the simulation case, both without misalignment and in the realistic
misalignment scenario. A layer rotation, \ie a systematic rotation of
the layers with an amplitude proportional to the radius ($\Delta
\varphi=\rho \cdot (r-r_0)$), would appear as a constant offset.  The
$\varphi$-dependence displays the same trend as already seen in the
validation with $\Z \rightarrow \mu^+\mu^-$ decays of figure
\ref{fig:ZMassVsEtaPhi}. The trend of $\Delta \phi$ is quantitatively
characterised by means of a fit to the binned distribution with a
sinusoidal function. The fit returns a parameterisation of $[(0.036
\pm 0.005) \cdot \sin(\varphi + (2.8 \pm 0.2)\,)]\unit{mrad}$.

\subsection{Sensitivity to systematic misalignment}\label{subsec:systmis}

Beyond the validation of the aligned geometry, the sensitivity to weak
modes of the alignment procedure has been studied.  Following the
approach in \cite{ALI:TkAlCraft08,ALI:Babar}, a set of basic
deformations were applied on top of the aligned tracker geometry. The
full alignment procedure was then repeated starting from the
misaligned scenario, obtaining a set of "realigned'' geometries.  Nine
systematic misalignment scenarios were studied, giving a matrix of
deformations expressed with $\Delta r$, $\Delta z$, and $\Delta
\varphi$ as a function of the same three variables.  These
misalignments were applied only to the directions to which the silicon
modules are effectively sensitive. Movements along non-measurement
directions (\eg the longitudinal direction for strip barrel modules)
are irrelevant, since the alignment procedure does not associate any
parameters with them.

The ability of the alignment procedure to compensate for
misalignments indicates its robustness against systematic distortions
of this or similar type. A close match between the realigned and
initial geometry means that the procedure is fully sensitive to this
specific deformation and able to keep it under control. On
the other hand, few or no changes compared to the misaligned
scenario indicate poor sensitivity.

\begin{figure*}[hbtp]
  \centering
    \begin{tabular}{c}
      \includegraphics[width=0.4\textwidth]{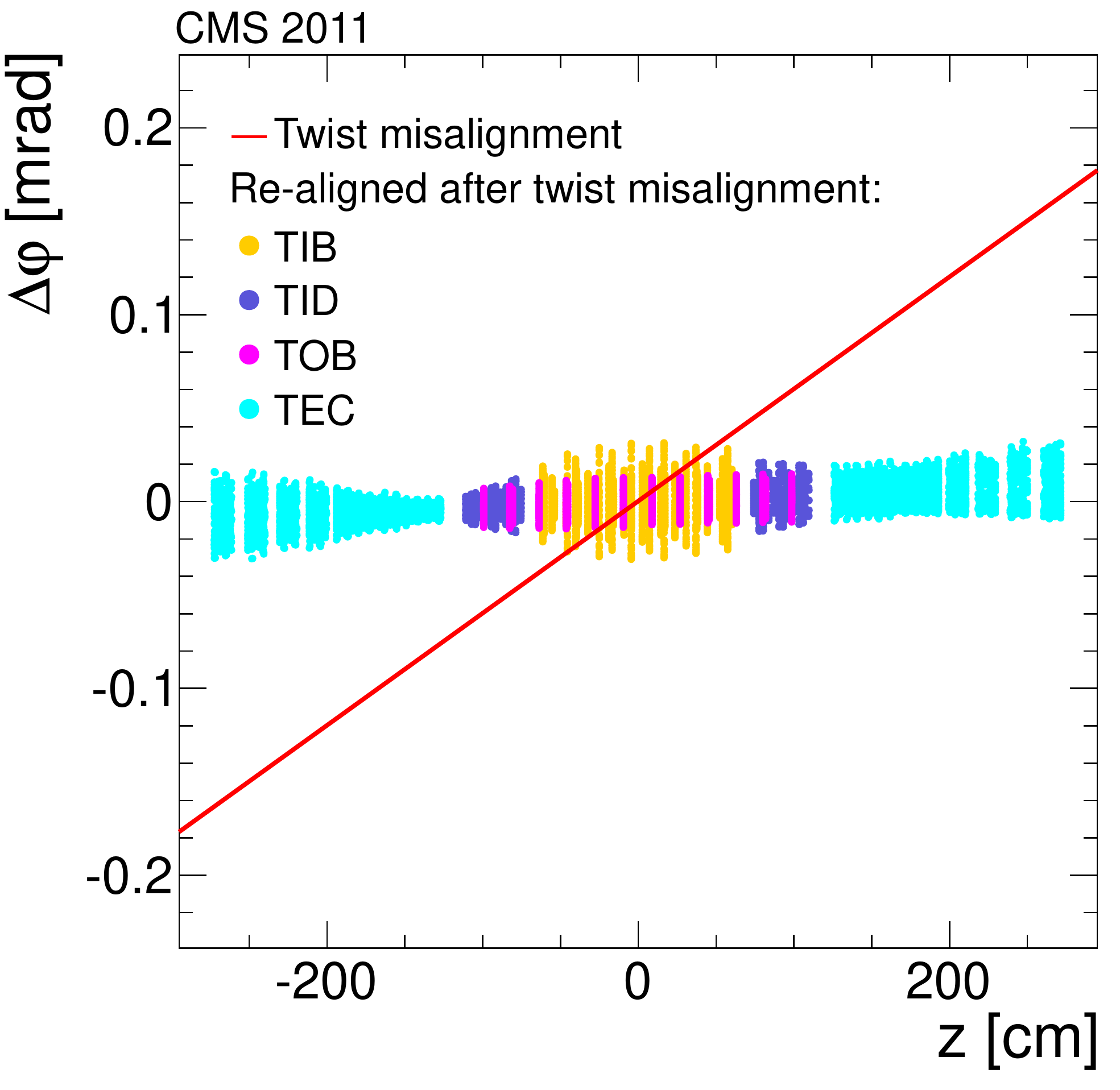}
      \includegraphics[width=0.4\textwidth]{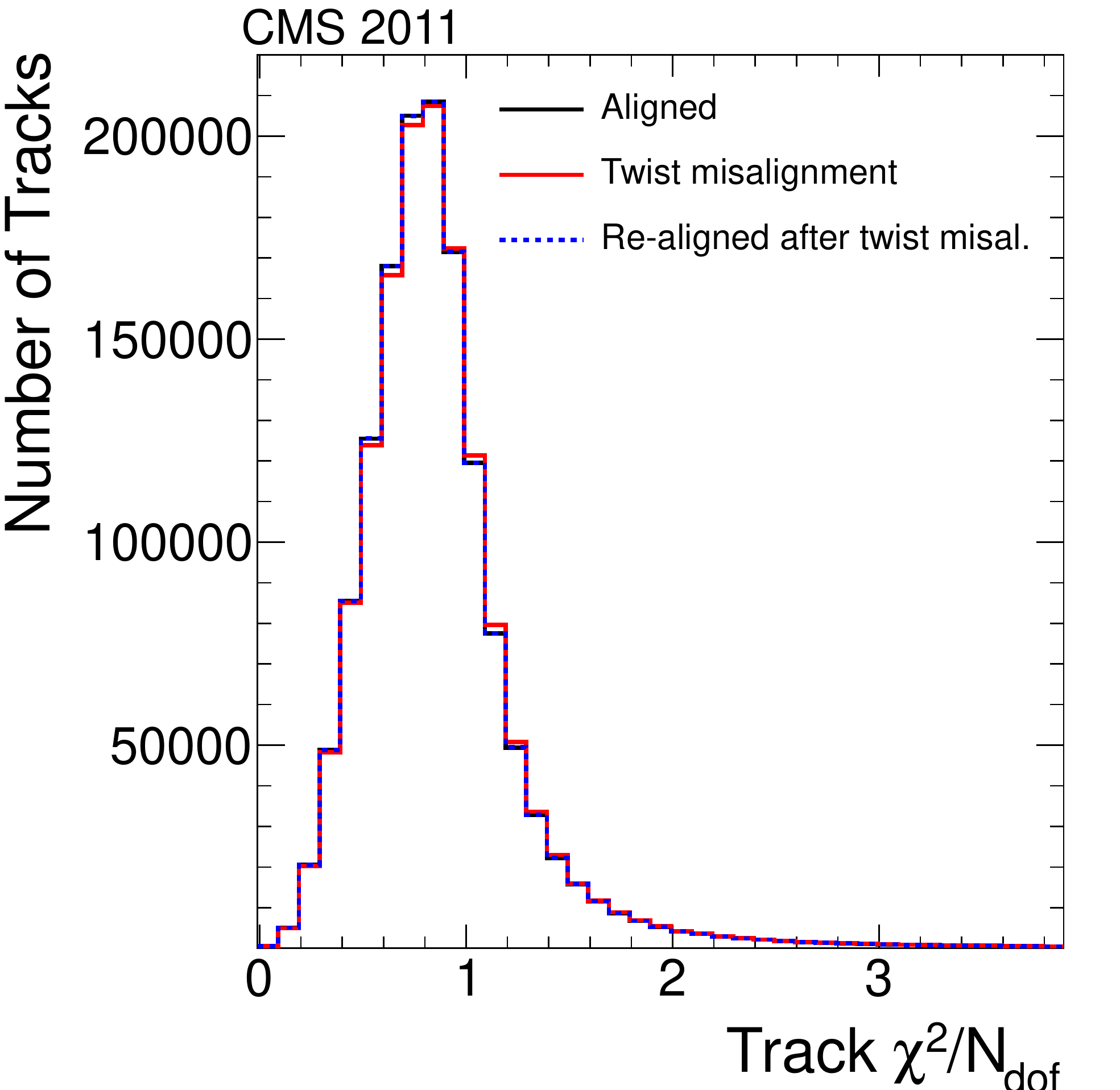} \\
      \includegraphics[width=0.4\textwidth]{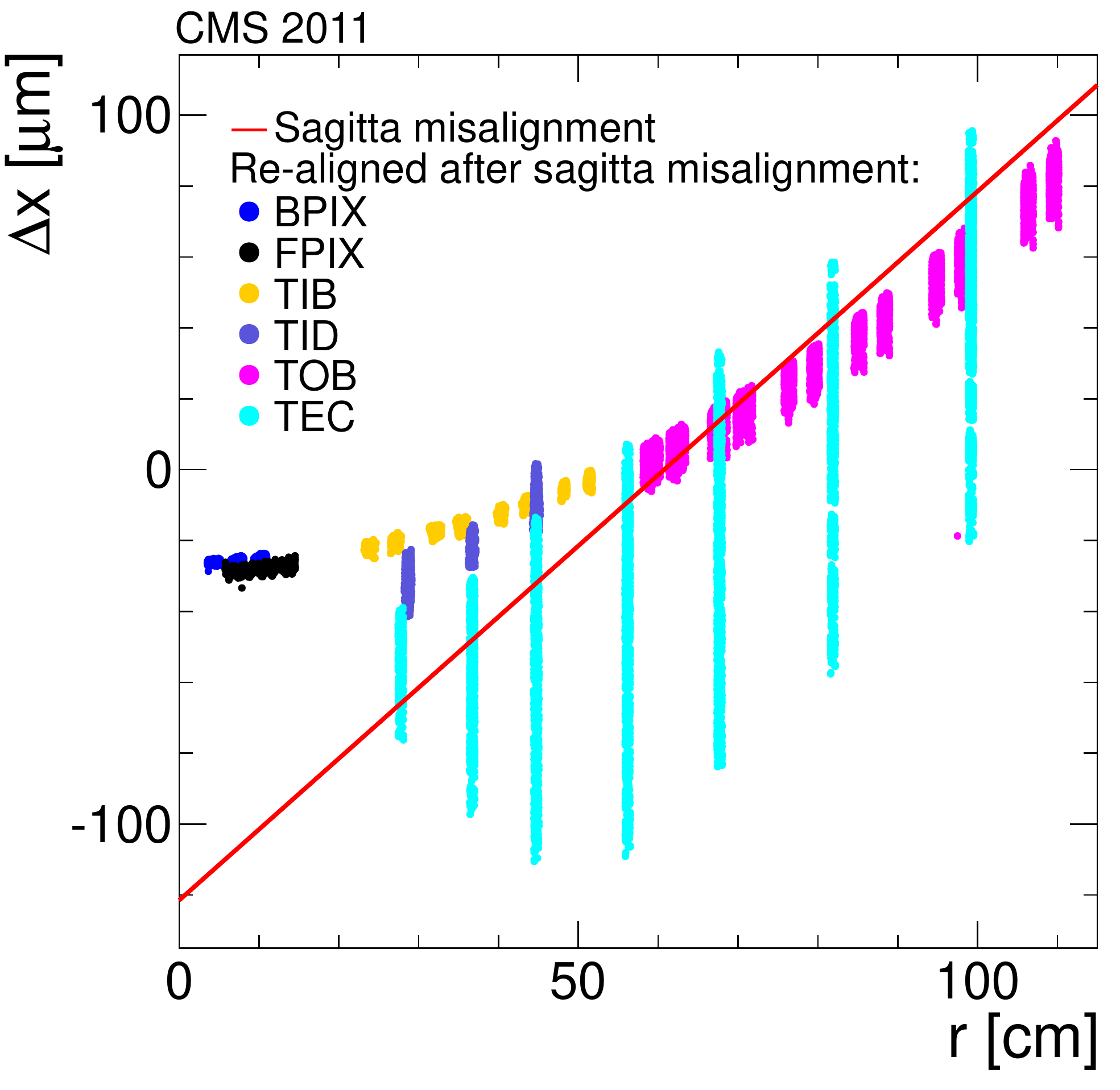}
      \includegraphics[width=0.4\textwidth]{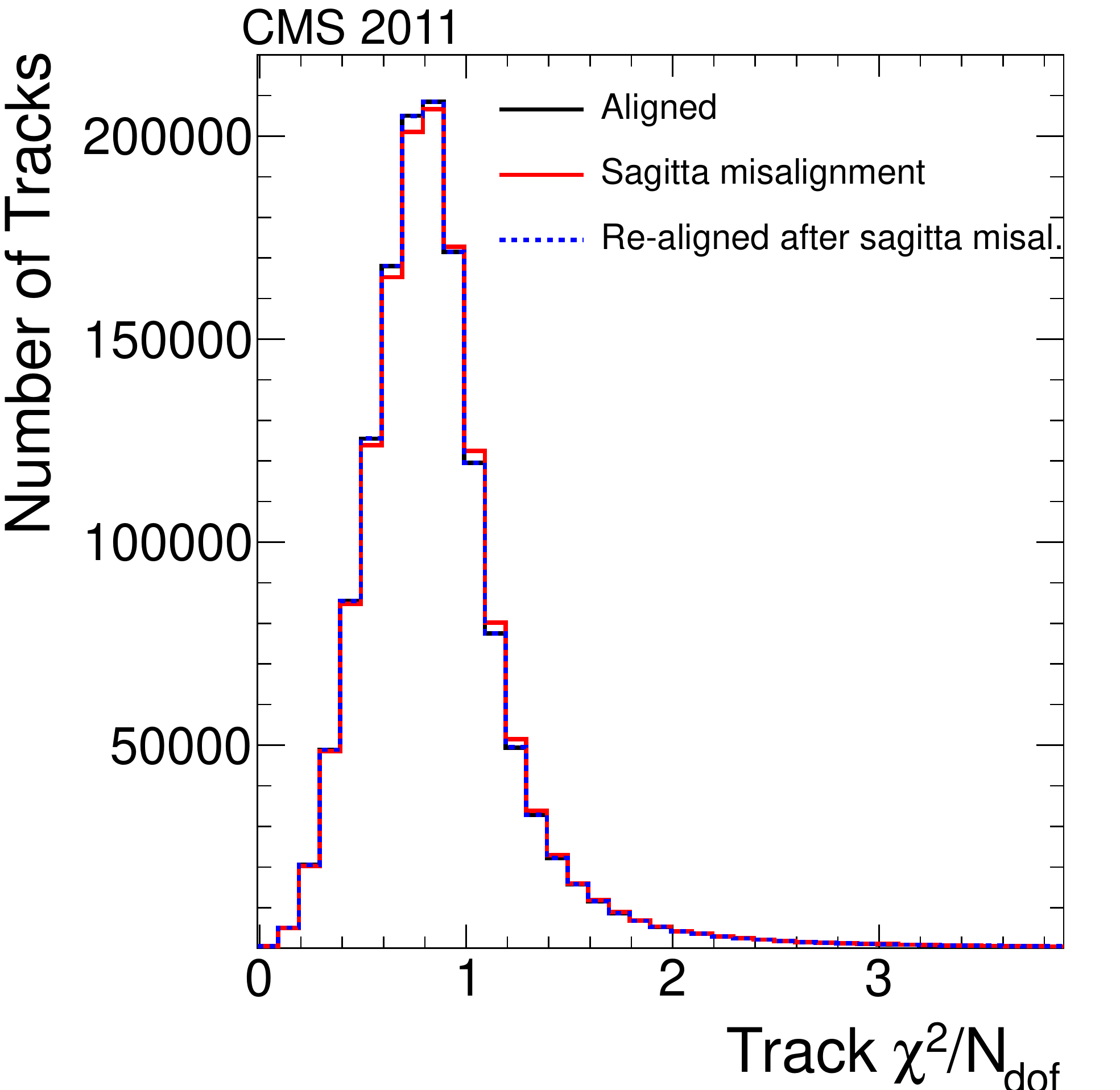}\\
      \includegraphics[width=0.4\textwidth]{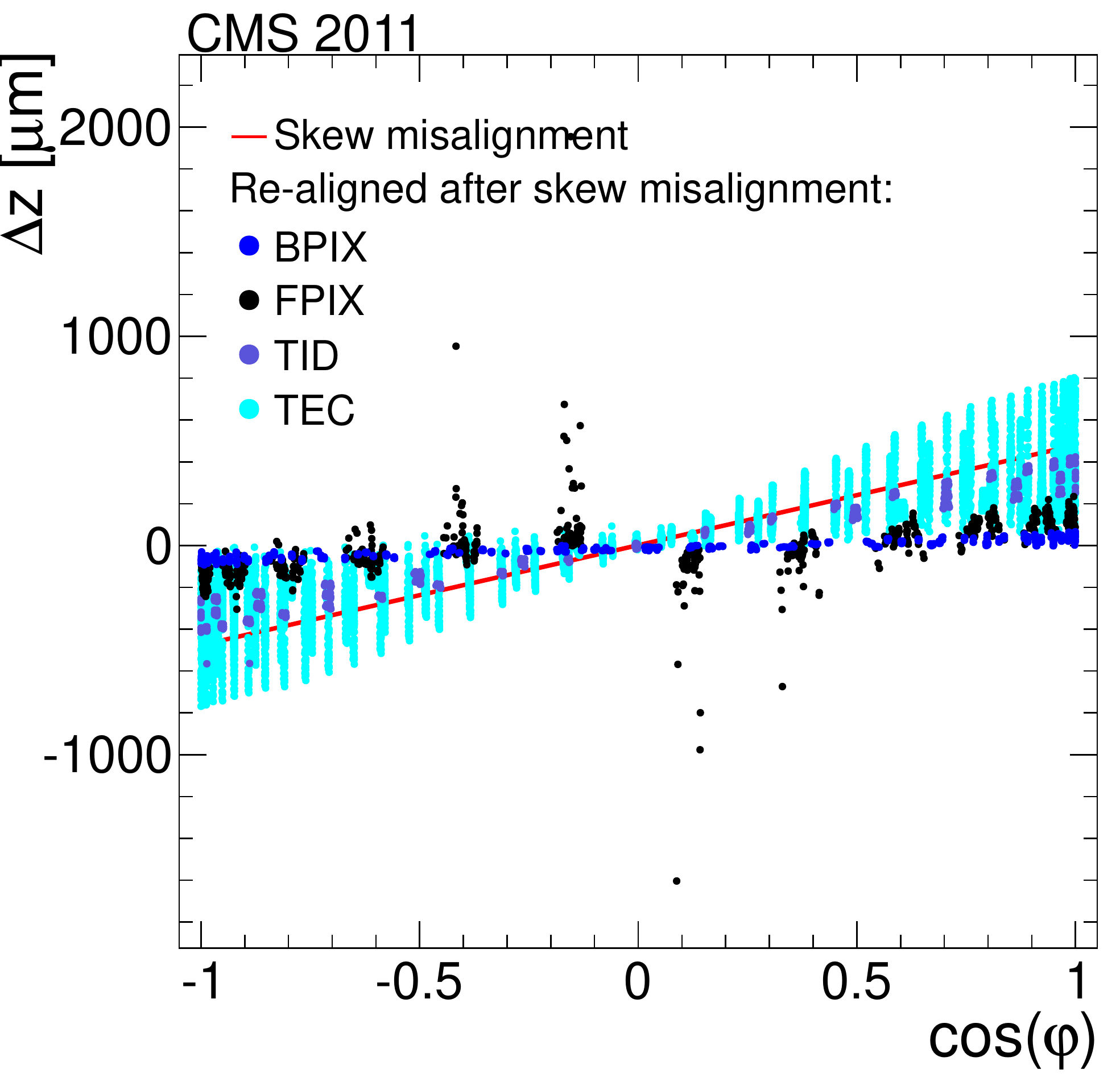}
      \includegraphics[width=0.4\textwidth]{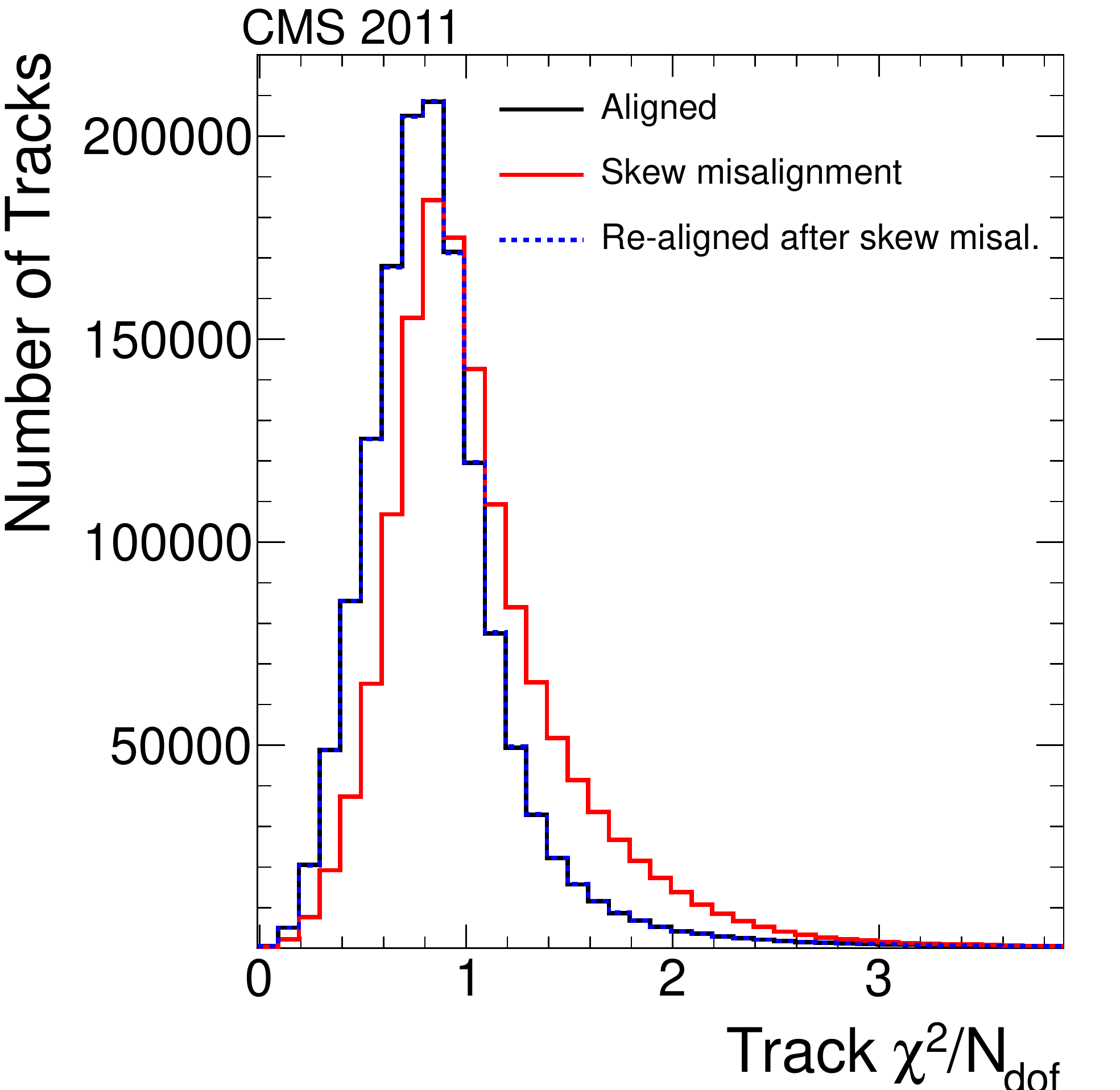}
    \end{tabular}
    \caption{
      Impact of intentional application of a twist (top row), sagitta
      (middle row) and skew systematic misalignment (bottom
      row). In the left columns, the red line shows the size of the applied
      misalignment, and the coloured dots show the difference of selected
      alignment parameters, module by module after realignment, to the
      initial values prior to misalignment. The plots in the right
      column show the distributions of goodness-of-fit for
      loosely selected isolated muons from an independent data sample,
      with transverse momentum $\pt >5$\GeVc.
    }
    \label{fig:weakmode_study}

\end{figure*}

The left-hand side of figure \ref{fig:weakmode_study} displays the
difference between the positions of the modules in the initial aligned
and the deliberately misaligned geometry for some benchmark
misalignment scenarios. The behaviour of the distribution of the
normalised $\chi^{2}$ for the three stages of the study (initially
aligned, misaligned and realigned) is presented in the right-hand side
of figure \ref{fig:weakmode_study}, by using a sample of isolated
muons. The upper, middle and bottom rows present the cases of twist,
skew and sagitta deformations as introduced in
section~\ref{subsec:weakmodes}.  The results show that the alignment
procedure has very good control over twist-like deformations.  The
normalised $\chi^{2}$ of the isolated muons does not change
significantly when introducing the twist-like deformation because this
type of tracks does not have sensitivity to it. The control over
twists comes largely from the constraining power of the muonic
\Z-boson decays.  Skew and sagitta are interesting because they are the
systematic distortions most difficult to control. The skew
misalignment is fully recovered in the pixel barrel, but not in the
other subdetectors. The alignment procedure is only partially
resilient against sagitta distortion, with the best recovery observed
at small radii.  The other six misalignment scenarios considered
prove to be well controlled.  These results represent a significant
improvement with respect to \cite{ALI:TkAlCraft08}, thanks to the
inclusion of tracks from proton-proton collisions with several
different topologies and the usage of vertex constraint and mass
information with muons coming from \Z-boson decays.

\section{Summary}
The alignment procedure for the CMS tracker and its results for the
first high-luminosity data-taking period during the year 2011 have been presented. Among the most
prominent features are the successful handling of the large degree of complexity of a highly
granular silicon detector, the simultaneous determination of shape
parameters at the sensor level, the use of the Z resonance
signature to control systematic effects, and the parallelised
implementation of the whole procedure resulting in a fast execution of the workflow.

The alignment is
based on global minimisation of track-to-hit residuals. The internal
alignment is performed with the \mptwo algorithm, which is enhanced
compared to its predecessor to handle about 200\,000 alignment
parameters simultaneously. A dedicated track parameterisation is
included, based on the general broken lines method, which allows
rigorous and execution-time-efficient
treatment of multiple scattering 
in the
global fit. The execution time of the fit is considerably shortened by
parallelisation on a multi-core architecture.

The time dependence of the tracker alignment is monitored with laser beams and
tracks. The tracker geometry is found to be very
stable with time. The most important movements are observed between the
half-shells of the barrel pixel detector, whose longitudinal separation
varies by up to 40\mum. The alignment procedure corrects for these
movements such that the residual variation after alignment is kept
below 10\mum.

The overall tilt angles of the tracker with respect to the magnetic
field are determined to be at the sub-mrad level. The statistical
accuracy of the alignment is found to be generally---often significantly---better than
10\mum, with the exception of some rings in the tracker endcap
disks. Sensor- and module-shape parameters are determined at the module level
simultaneously with other alignment parameters. Curvatures of
individual sensors and kink angles of adjacent sensors in modules are
observed and measured;
sensor curvature amplitudes vary according to
subsystem, and their averages per layer and ring
range up to about 80\mum in the endcap
systems. Kink angles of up to several mrad are observed.

Besides cosmic ray tracks, reconstructed $\Z\to\mu^+\mu^-$ decays
play an essential role in constraining systematic deformations of the
aligned geometry with small leverage to the track $\chi^2$, also known as
weak modes. The remaining variation of the \Z-boson mass peak is less
than 0.5\% and thus small compared to other resolution effects in the
corresponding momentum range. The successful control of weak modes
is confirmed by studies involving the energy measured in the
hadronic calorimeter.

The stability of the alignment with respect to weak modes is further
investigated by the study of the effect of deliberately added
distortions and subsequent re-alignment. The procedure is found to have very good
control over twist modes, while strong sagitta and skew misalignments
are at least partially recovered.

In summary, this article describes the comprehensive alignment procedure
for the largest and most complex silicon detector ever
built. The achieved alignment accuracy enables the tracking to take full benefit of the high intrinsic resolution of the silicon modules.
The quality of the alignment is thus an essential building block
for the excellent physics performance of the CMS detector.

\section*{Acknowledgements}
\hyphenation{Bundes-ministerium Forschungs-gemeinschaft Forschungs-zentren} We congratulate our colleagues in the CERN accelerator departments for the excellent performance of the LHC and thank the technical and administrative staffs at CERN and at other CMS institutes for their contributions to the success of the CMS effort. In addition, we gratefully acknowledge the computing centres and personnel of the Worldwide LHC Computing Grid for delivering so effectively the computing infrastructure essential to our analyses. Finally, we acknowledge the enduring support for the construction and operation of the LHC and the CMS detector provided by the following funding agencies: the Austrian Federal Ministry of Science and Research and the Austrian Science Fund; the Belgian Fonds de la Recherche Scientifique, and Fonds voor Wetenschappelijk Onderzoek; the Brazilian Funding Agencies (CNPq, CAPES, FAPERJ, and FAPESP); the Bulgarian Ministry of Education and Science; CERN; the Chinese Academy of Sciences, Ministry of Science and Technology, and National Natural Science Foundation of China; the Colombian Funding Agency (COLCIENCIAS); the Croatian Ministry of Science, Education and Sport, and the Croatian Science Foundation; the Research Promotion Foundation, Cyprus; the Ministry of Education and Research, Recurrent financing contract SF0690030s09 and European Regional Development Fund, Estonia; the Academy of Finland, Finnish Ministry of Education and Culture, and Helsinki Institute of Physics; the Institut National de Physique Nucl\'eaire et de Physique des Particules~/~CNRS, and Commissariat \`a l'\'Energie Atomique et aux \'Energies Alternatives~/~CEA, France; the Bundesministerium f\"ur Bildung und Forschung, Deutsche Forschungsgemeinschaft, and Helmholtz-Gemeinschaft Deutscher Forschungszentren, Germany; the General Secretariat for Research and Technology, Greece; the National Scientific Research Foundation, and National Innovation Office, Hungary; the Department of Atomic Energy and the Department of Science and Technology, India; the Institute for Studies in Theoretical Physics and Mathematics, Iran; the Science Foundation, Ireland; the Istituto Nazionale di Fisica Nucleare, Italy; the Korean Ministry of Education, Science and Technology and the World Class University program of NRF, Republic of Korea; the Lithuanian Academy of Sciences; the Ministry of Education, and University of Malaya (Malaysia); the Mexican Funding Agencies (CINVESTAV, CONACYT, SEP, and UASLP-FAI); the Ministry of Business, Innovation and Employment, New Zealand; the Pakistan Atomic Energy Commission; the Ministry of Science and Higher Education and the National Science Centre, Poland; the Funda\c{c}\~ao para a Ci\^encia e a Tecnologia, Portugal; JINR, Dubna; the Ministry of Education and Science of the Russian Federation, the Federal Agency of Atomic Energy of the Russian Federation, Russian Academy of Sciences, and the Russian Foundation for Basic Research; the Ministry of Education, Science and Technological Development of Serbia; the Secretar\'{\i}a de Estado de Investigaci\'on, Desarrollo e Innovaci\'on and Programa Consolider-Ingenio 2010, Spain; the Swiss Funding Agencies (ETH Board, ETH Zurich, PSI, SNF, UniZH, Canton Zurich, and SER); the National Science Council, Taipei; the Thailand Center of Excellence in Physics, the Institute for the Promotion of Teaching Science and Technology of Thailand, Special Task Force for Activating Research and the National Science and Technology Development Agency of Thailand; the Scientific and Technical Research Council of Turkey, and Turkish Atomic Energy Authority; the National Academy of Sciences of Ukraine, and State Fund for Fundamental Researches, Ukraine; the Science and Technology Facilities Council, UK; the US Department of Energy, and the US National Science Foundation.

Individuals have received support from the Marie-Curie programme and the European Research Council and EPLANET (European Union); the Leventis Foundation; the A. P. Sloan Foundation; the Alexander von Humboldt Foundation; the Belgian Federal Science Policy Office; the Fonds pour la Formation \`a la Recherche dans l'Industrie et dans l'Agriculture (FRIA-Belgium); the Agentschap voor Innovatie door Wetenschap en Technologie (IWT-Belgium); the Ministry of Education, Youth and Sports (MEYS) of Czech Republic; the Council of Science and Industrial Research, India; the Compagnia di San Paolo (Torino); the HOMING PLUS programme of Foundation for Polish Science, cofinanced by EU, Regional Development Fund; and the Thalis and Aristeia programmes cofinanced by EU-ESF and the Greek NSRF.

\bibliography{auto_generated}   

\cleardoublepage \appendix\section{The CMS Collaboration \label{app:collab}}\begin{sloppypar}\hyphenpenalty=5000\widowpenalty=500\clubpenalty=5000\textbf{Yerevan Physics Institute,  Yerevan,  Armenia}\\*[0pt]
S.~Chatrchyan, V.~Khachatryan, A.M.~Sirunyan, A.~Tumasyan
\vskip\cmsinstskip
\textbf{Institut f\"{u}r Hochenergiephysik der OeAW,  Wien,  Austria}\\*[0pt]
W.~Adam, T.~Bergauer, M.~Dragicevic, J.~Er\"{o}, C.~Fabjan\cmsAuthorMark{1}, M.~Friedl, R.~Fr\"{u}hwirth\cmsAuthorMark{1}, V.M.~Ghete, C.~Hartl, N.~H\"{o}rmann, J.~Hrubec, M.~Jeitler\cmsAuthorMark{1}, W.~Kiesenhofer, V.~Kn\"{u}nz, M.~Krammer\cmsAuthorMark{1}, I.~Kr\"{a}tschmer, D.~Liko, I.~Mikulec, D.~Rabady\cmsAuthorMark{2}, B.~Rahbaran, H.~Rohringer, R.~Sch\"{o}fbeck, J.~Strauss, A.~Taurok, W.~Treberer-Treberspurg, W.~Waltenberger, C.-E.~Wulz\cmsAuthorMark{1}
% ...
\vskip\cmsinstskip
% ...
1:~~Also at Vienna University of Technology, Vienna, Austria\\
2:~~Also at CERN, European Organization for Nuclear Research, Geneva, Switzerland\\
% ...

\end{sloppypar}
\end{document}